\documentclass[tightenlines,a4paper,11pt]{article}
\pdfoutput=1
\usepackage{jheppub,keyval,amsfonts,slashed,bm}
\DeclareMathAlphabet\mathbfcal{OMS}{cmsy}{b}{n}
\newcommand{\beq}{\begin{equation}} % Begin an equation.
\newcommand{\eeq}{\end{equation}} % End an equation.
\newcommand{\beqa}{\begin{eqnarray}} % Begin an array of equation.
\newcommand{\eeqa}{\end{eqnarray}} % End an array of equation.
\newcommand{\rmi}[1]{{\mbox{\scriptsize #1}}} % Subscript name.

\newcommand{\aE}[1]{\alpha_\rmi{E#1}}

\newcommand{\lmsb}{\bar{\Lambda}} % MSbar scheme, renormalization scale.
\def\intedk{\int_{\kern-0.25em \lower.5em \hbox{\mbox{\scriptsize $0$}}}^{\kern-0.25em \lower-.5em\hbox {\mbox{\scriptsize $\infty$}}}\kern-0.75em\mbox{\normalsize $\mathop{{\rm d}k}$}\ } % 1d momentum integration over k.
\def\inteddw{\int_{\kern-0.25em \lower.5em \hbox{\mbox{\scriptsize $0$}}}^{\kern-0.25em \lower-.5em \hbox{\mbox{\scriptsize $\infty$}}}\kern-0.75em\mbox{\normalsize $\mathop{{\rm d}\omega}$}\ } % 1d momentum integration over w (going with another 1d momentum integration over k).
\def\sumint{\hbox{$\sum$}\!\!\!\!\!\!\!\int} % Sum-Integral symbol.
\def\inteddk{\int_{\kern-0.25em \lower.5em \hbox{\mbox{\scriptsize $\omega$}}}^{\kern-0.25em \lower-.5em \hbox{\mbox{\scriptsize $\infty$}}}\kern-0.75em\mbox{\normalsize $\mathop{{\rm d}k}$}\ } % 1d momentum integration over k (going with another 1d momentum integration over w).
\def\intdwdk{\int_{\kern-0.25em \lower.5em\hbox{\mbox{\scriptsize $0$}}}^{\kern-0.25em \lower-.5em\hbox{\mbox{\scriptsize $\infty$}}}\kern-0.75em \mbox{\normalsize $\mathop{{\rm d}\omega}$}\kern-0.15em\int_{\kern-0.25em \lower.5em\hbox{\mbox{\scriptsize $\omega$}}}^{\kern-0.25em \lower-.5em \hbox{\mbox{\scriptsize $\infty$}}}\kern-0.75em\mbox{\normalsize $\mathop{{\rm d}k}$}\,} % combination of 1d momentum integration over k and 1d momentum integration over w, to save some space.
\newcommand\intec[3]{\oint_{\lower+.25em\hbox{$\mbox{\tiny $#1$}$}}\kern-#2 \mbox{\Large $\frac{\mathop{{\rm d}\omega}}{2\pi i}$}\kern-#3\mbox{}} % Contour integral along the path #1.
\def\inteddwbis{\int_{\kern-0.25em \lower.5em \hbox{\mbox{\scriptsize $0$}}}^{\kern-0.25em \lower-.5em\hbox{\mbox{\scriptsize $k$}}}\kern-0.50em\mbox{\normalsize $\mathop{{\rm d}\omega}$}\ } % Another 1d momentum integration over w (going with another 1d momentum integration over k).
\newcommand\inteddwters[2]{\int_{\kern-0.25em \lower.5em\hbox {\mbox{\scriptsize $#1$}}}^{\kern-0.25em \lower-.5em \hbox{\mbox{\scriptsize $#2$}}}\kern-0.75em\mbox{\normalsize $\mathop{{\rm d}\omega_\rmi{E}}$}} % Again another 1d momentum integration over w (going with another 1d momentum 
\newcommand\disc{{\rm Disc}\ } % Disc() symbol.
\newcommand\ImPart{\rm {Im}\ } % Imaginary part.
\newcommand\RePart{\rm {Re}\ } % Real part.
\newcommand\arctanh{\text{arctanh}} % Inverse hyperbolic tangent.
\title{Equation of State of hot and dense QCD: Resummed perturbation theory confronts lattice data}
\author[a]{Sylvain~Mogliacci,}
\author[b]{Jens~O.~Andersen,}
\author[c]{Michael~Strickland,}
\author[a]{Nan~Su}
\author[d]{and Aleksi~Vuorinen}
\affiliation[a]{Faculty of Physics, University of Bielefeld, D-33615 Bielefeld, Germany}
\affiliation[b]{Department of Physics, Norwegian University of Science and Technology,\\ N-7491 Trondheim, Norway}
\affiliation[c]{Department of Physics, Kent State University, Kent OH 44242 USA}
\affiliation[d]{Department of Physics and Helsinki Institute of Physics, P. O. Box 64, FI-00014 University of Helsinki, Finland}
\emailAdd{sylvain@physik.uni-bielefeld.de}
\emailAdd{andersen@tf.phys.ntnu.no}
\emailAdd{mstrick6@kent.edu}
\emailAdd{nansu@physik.uni-bielefeld.de}
\emailAdd{aleksi.vuorinen@helsinki.fi}

\preprint{BI-TP 2013/17}

\abstract{We perform a detailed analysis of the predictions of resummed perturbation theory for the pressure and the second-, fourth-, and sixth-order diagonal quark number susceptibilities in a hot and dense quark-gluon plasma. First, we present an exact one-loop calculation of the equation of state within hard-thermal-loop perturbation theory (HTLpt) and compare it to a previous one-loop HTLpt calculation that employed an expansion in the ratios of thermal masses and the temperature. We find that this expansion converges reasonably fast. We then perform a resummation of the existing four-loop weak coupling expression for the pressure, motivated by dimensional reduction. Finally, we compare the exact one-loop HTLpt and resummed dimensional reduction results with state-of-the-art lattice calculations and a recent mass-expanded three-loop HTLpt calculation.}

\arxivnumber{1307.8098}

\begin{document}
\maketitle
\flushbottom
%%%%%%%%%%%%%%%%%%%%%%%%%%%%%%%%%%%%%%%%%%%%%%%%%%%%%%%%%%%%%%%%%%%%%%%%%%%%%%%%%%%%

\section{Introduction}

Understanding the behavior of strongly interacting matter subject to extreme conditions is important in many physical contexts such as the study of the early universe and the interiors of compact stars. The determination of the QCD phase diagram has subsequently received considerable attention over the past few decades. On the experimental side, an enormous effort has gone to the creation of the quark-gluon plasma (QGP) in heavy ion collisions, most recently carried out at the Relativistic Heavy Ion Collider (RHIC) of Brookhaven~\cite{Tannenbaum}, and at the Large Hadron Collider (LHC) of CERN~\cite{Muller}. On the theoretical side, both lattice gauge theory and model calculations have been employed to map out the phase diagram of QCD and in particular locate a possible critical end point of a line of first order phase transitions (see e.g.~\cite{Satz} for a review).

Unfortunately, at nonzero quark chemical potentials $\mu_f$, lattice Monte Carlo simulations are hampered by the infamous sign problem related to the complex nature of the lattice action, rendering importance sampling techniques inapplicable. One of the proposed ways to circumvent this problem --- and thus enter the $\mu$--$T$ plane --- is by Taylor expanding various physical quantities in powers of the ratios $\mu_f/T$ and evaluating the coefficients at $\mu_f=0$. As these derivatives are evaluated at zero density, they can be computed on the lattice using standard techniques. In the case of the pressure, the coefficients are called quark number susceptibilities (QNS), which indeed carry information about the response of the system to nonzero baryon density as well as correlations between different quark flavors.

Let us denote by $\bm{\mu}$ an $N_\rmi{f}$\,-\,component vector consisting of the quark chemical potentials for the different flavors, $\bm{\mu}\equiv(\mu_1,\mu_2,...,\mu_{N_\rmi{f}})$, with $N_\rmi{f}$ being typically 2 or 3 at the relevant energies. We can then define the QNS, $\chi_{ijk}\left(T\right)$, simply as derivatives of the pressure $p\left(T,\bm{\mu}\right)$, according to
\beqa\label{Relation_chi_pressure}
\chi_{ijk}\left(T\right)&\equiv&\frac{\partial^{i+j+k+...}\; p\left(T,\bm{\mu}\right)}{\partial\mu_u^i\, \partial\mu_d^j \, \partial\mu_s^k\, ...}\bigg|_{\bm{\mu}=0} \, ,
\eeqa
where the indices $u,\,d,\,s\, ...$ refer to the quark flavors. The determination of the equation of state (EoS) in the $\mu$--$T$ plane is then limited only by the convergence of the corresponding series in powers of $\mu_f/T$, i.e.~ultimately by the magnitudes of the QNS. Recent studies of these quantities on the lattice can be found e.g.~in refs.~\cite{bieleHighT,biele1,biele2,bazavov12,wuppertalcharge,chi4}. In addition, lattice studies of two-color QCD with an even number of flavors, which does not suffer from the sign problem, have been performed at high density in refs.~\cite{HandsNc2_1,HandsNc2_2}.

Due to the difficulties in performing lattice simulations far above the pseudo-critical deconfinement transition temperature $T_\rmi{c}$ (recalling the crossover nature of the transition at $\mu_f=0$), it is important to have complementary techniques that can bridge the gap between low and high temperatures. In this context, analytic weak coupling techniques are clearly the method of choice, as they work optimally at asymptotically high temperatures and can be fairly easily continued to the vicinity of the transition region. In recent years, there have been numerous analytic calculations of the chemical potential dependence of the pressure as well as the QNS, using techniques such as unresummed perturbation theory~\cite{aleksi1,aleksi2,aleksi3,ipp3}, various hard-thermal-loop motivated approaches~\cite{blaizot1,blaizot2,mustafa1,mustafa2,mustafa3}, hard-thermal-loop perturbation theory (HTLpt)~\cite{baierredlich,JensMike1,sylvain1,mikehaque,Haque1,haque}, the large-$N_\rmi{f}$ limit of QCD~\cite{ipp1,ipp2}, and even the gauge/gravity duality in the context of strongly coupled large-$N_\rmi{c}\ $ ${\mathcal N}=4$ Super Yang-Mills theory~\cite{ads}. Finally, similar investigations have also been carried out using various field theoretical~\cite{PNJL1,ModelForQCD2,PNJL3} and holographic~\cite{Jinfeng} models of QCD.

In ref.~\cite{sylvain1}, four of the present authors applied both HTLpt and a resummation scheme motivated by dimensional reduction (DR) to determine the second- and fourth-order diagonal QNS of hot QCD. The calculation involving HTLpt was carried out only to one-loop order, and moreover utilized an expansion in the ratio $m/T$, where $m$ represents both the Debye and quark thermal masses of order $gT$. A similar technique was recently further employed to two-loop order in refs.~\cite{mikehaque,Haque1} as well as to three-loop order in~\cite{haque}, exhibiting sizable devitations from the results of~\cite{sylvain1}. Motivated by these developments, our aim in the current paper is to continue the work of~\cite{sylvain1} in three directions:
\begin{itemize}
\item Carry out an exact one-loop HTLpt calculation with no $m/T$-expansion, thereby analyzing the convergence of the expansion and the validity of the corresponding expanded results.
\item Generalize all of our previous results, both within one-loop HTLpt and four-loop DR, to the sixth order quark number susceptibility as well as the chemical potential dependence of the pressure itself, covering also the case of $N_\rmi{f}=2$.
\item Perform a careful explicit comparison of our results with state-of-the-art lattice data, and to the three-loop HTLpt results of~\cite{haque}, when possible.
\end{itemize}
The first of these points is particularly important, because it provides a highly nontrivial quantitative check of one of the crucial analytic approximations used in HTLpt calculations, namely the $m/T$-expansion, which in fact is an essential ingredient in all higher order computations within HTLpt.  Its convergence has been studied in scalar $\phi^4$ theory to three-loop order~\cite{motspt} with encouraging results; however, to our knowledge, our study is the first calculation to probe such an $m/T$-expansion within QCD.

Our paper is organized as follows. In section~\ref{sec:Section2_for_Intro}, we briefly describe hard-thermal-loop perturbation theory and dimensional reduction as two ways of reorganizing perturbative expansions within thermal QCD. We also outline the computation of the exact one-loop and partial four-loop pressures in these two setups, respectively. In section~\ref{sec:Section4_for_Intro}, we investigate the convergence of the $m/T$-\,expansion for the one-loop HTLpt results. In section~\ref{sec:Section3_for_Intro}, we compare our results with the three-loop HTLpt calculation~\cite{haque} as well as to lattice data. Finally, in section~\ref{sec:Section5_for_Intro} we draw our conclusions. Nearly all of the computational details have been relegated to the appendices. In appendix~\ref{sec:Notations}, we first explain our notation, in appendix~\ref{sec:DR_matching_para}, we list the matching coefficients of Electrostatic QCD (EQCD) needed in the derivation of the four-loop EoS within DR, and in appendix~\ref{Computing_Full_HTLpt} we present a detailed evaluation of the exact one-loop HTLpt pressure. Finally, in appendix~\ref{sec:BC_Disc}, we provide some computational details regarding the branch cut discontinuities that are encountered in exact HTLpt calculations, and in appendix~\ref{sec:Mass_expansion}, we go through the derivation of the $m/T$-\,expansion employed in ref.~\cite{sylvain1}.

We would like to point out that all calculations presented in this paper have been carried out in the limit of vanishing bare quark masses.  In the case of our DR calculation, we have explicitly checked that this only affects the results in any noticeable way at the very lowest temperatures shown, where the validity of perturbation theory is in any case questionable.  As for HTLpt, a study of the mass dependence of the one-loop quark self-energy and gluon polarization functions has been carried out in~\cite{Seipt1}, indicating that a similar conclusion should hold also in this case.

\section{Resummations in thermal QCD} \label{sec:Section2_for_Intro}

The (bare) Lagrangian density of massless QCD reads
\beqa\label{Lagran_Densi_QCD}
{\cal L}_{\rm QCD}&=&-\frac{1}{2}{\rm Tr}[G_{\mu\nu}\,G^{\mu\nu}]+i\bar{\psi}\gamma^{\mu}D_{\mu}\psi+{\cal L}_{\rm gf}+{\cal L}_{\rm gh}\,,
\eeqa
in which the gluon field strength tensor is defined by $G^{\mu\nu}=\partial^{\mu}A^{\nu}-\partial^{\nu}A^{\mu}-i g[A^{\mu},A^{\nu}]$ and the covariant derivative in the fundamental representation by $D^{\mu}=\partial^{\mu}-i gA^{\mu}$, while the term involving the quark fields $\psi$ contains an implicit sum over the $N_\rmi{f}$ quark flavors. The ghost term ${\cal L}_{\rm gh}$ depends on the gauge-fixing term ${\cal L}_{\rm gf}$; in the present work, we employ the general covariant gauge, where the latter reads
\beqa
{\cal L}_{\rm gf}&=&-\frac{1}{\xi}{\rm Tr}[(\partial_{\mu}A^{\mu})^2]\, ,
\eeqa
with $\xi$ standing for the gauge-fixing parameter.

Computing various thermodynamic quantities within entirely unresummed (`naive' or diagrammatic) perturbation theory amounts to expanding them in power series organized by even powers of the coupling constant $g$, letting the renormalization of the coupling (and possibly quark masses) cancel the $1/\epsilon$ ultraviolet (UV) divergences encountered.\footnote{We work within dimensional regularization, so all of our integrals are defined in $d=3-2\epsilon$ spatial dimensions.} This procedure, which amounts to an expansion around an ideal gas of massless quasiparticles, however typically runs into difficulties with physical infrared (IR) divergences already at low loop orders, necessitating some type of a physically motivated resummation of higher-loop Feynman diagrams to be carried out. This may be done in a way, in which one always expands the final $n$-loop result to order $g^{2(n-1)}$, dropping all ``extra'' terms from the result before evaluating it numerically. Such a procedure has, however, been seen to lead to rather poor convergence of different observables, and to this end, several different resummation schemes accounting for some higher order contributions have been proposed (see e.g.~\cite{blaizot3,kraemmer1,JensMike2} for reviews).

In the remainder of this section, we will introduce two ways of curing the IR problems encountered in thermal QCD calculations, involving the resummation of certain classes of higher order diagrams. These include firstly hard-thermal-loop perturbation theory, where the expansion point of perturbation theory is shifted to an ideal gas of massive quasiparticles, providing a marked improvement in the convergence properties of weak coupling expansion. At the same time, the physical idea of dimensional reduction can be seen to suggest a highly natural way of including certain higher order contributions to different equilibrium quantities, employing a weak coupling expansion within the effective theory Electrostatic QCD (EQCD). As our presentation will no doubt be rather superficial, we refer the interested reader to the original references~\cite{Jens2,Jens3,braaten2,Kajantie1} for more details about both of these topics.

\subsection{Hard-thermal-loop perturbation theory}

Within HTLpt, the reorganization of the perturbative expansion is achieved by writing the Lagrangian density of QCD in the form
\beqa
{\cal L_{\rm HTLpt}}&=&\left({\cal L}_{\rm QCD}+{\cal L}_{\rm HTL}\right)\Big|_{g\rightarrow\sqrt{\delta}g}+\Delta{\cal L}_{\rm HTL}\,,
\label{htlpttt}
\eeqa
where ${\cal L}_{\rm QCD}$ is given by (\ref{Lagran_Densi_QCD}), ${\cal L}_{\rm HTL}$ is an HTL improvement term, and $\delta$ is a formal expansion parameter introduced for bookkeeping purposes. The last term here, $\Delta{\cal L}_{\rm HTL}$, finally contains counterterms, which are necessary to cancel UV divergences introduced by the HTLpt reorganization. For QCD with dynamical massless quarks, the gauge invariant HTL improvement term reads
\beqa
{\cal L}_{\rm HTL}&=&-\frac{1}{2}(1-\delta)m_\rmi{D}^2{\rm Tr}\left(G_{\mu\alpha}\bigg\langle\frac{y^{\alpha}y^{\beta}}{(y\cdot D)^2}\bigg\rangle_{\!\!\!y}G^{\mu}_{\,\ \beta}\right) \nonumber \\
&+& (1-\delta)i\sum_f^{N_\rmi{f}} m^2_\rmi{q$_f$}\bar{\psi}_f\gamma^{\mu}\bigg\langle\frac{y_{\mu}}{y\cdot D}\bigg\rangle_{\!\!\!y}\psi_f, \ \ \ \ \ \ \ 
\eeqa
where $D^{\mu}=\partial^{\mu}-i gA^{\mu}$ now denotes covariant derivatives in both the adjoint and fundamental representations, $y=(1,\hat{\bf y})$ is a light-like four-vector, $\langle ... \rangle_y$ represents an average over the direction of $\hat{\bf y}$, and $m_\rmi{D}$ and $m_\rmi{q$_f$}$ are the Debye and quark thermal mass parameters. Note that $m_\rmi{q$_f$}$ carries dependence on the flavor index f, running from 1 to $N_\rmi{f}$.

In HTLpt, physical quantities are first expanded in power series in $\delta$, then truncated at some order, and finally evaluated after setting $\delta=1$. At leading order, this gives rise to dressed propagators that incorporate physical effects such as Debye screening and Landau damping. The starting point of HTLpt is thus an ideal gas of massive quasiparticles, which can be seen to be the main reason for its success. At two loops and beyond, the expansion in $\delta$ generates also dressed vertices as well as higher order interaction terms that ensure that there is no overcounting of Feynman diagrams.

If the expansion in $\delta$ is truncated at a finite order, then to complete the evaluation of a given physical observable, one needs a prescription for determining the values of the mass parameters $m_\rmi{D}$ and $m_\rmi{q$_f$}$, on which the result depends. In the case of the pressure, this can be achieved via a variational principle from two-loop order onwards, i.e.~extremizing the quantity as a function of $m_\rmi{D}$ and $m_\rmi{q$_f$}$. At one-loop order, the procedure however fails due to the absence of the coupling constant $g$ in the result~\cite{Jens1}. In line with earlier one-loop HTLpt calculations, we thus assign these parameters their leading order weak coupling values, keeping the number of colors $N_\rmi{c}$ and flavors $N_\rmi{f}$ arbitrary. Thus, our prescription reads
\beq\label{HTLpt_mD_mq_parameters}
m_\rmi{D}^2\equiv\frac{g^2}{3}\left[\left(N_\rmi{c}+\frac{N_\rmi{f}}{2}\right)\, T^2 +\frac{3}{2\pi^2}\sum_g \mu_g^2\right] \, ,\quad
m_\rmi{q$_f$}^2\equiv\frac{g^2}{16}\,\,\frac{N_\rmi{c}^2-1}{N_\rmi{c}}\,\Bigg(T^2+\frac{\mu_f^2}{\pi^2}\Bigg) \, ,
\eeq
which (together with the quark chemical potentials) we will frequently scale to be dimensionless via
\beq
\hat m_\rmi{D}\equiv \frac{m_\rmi{D}}{2\pi T} \, ,\quad \hat m_\rmi{q$_f$}\equiv \frac{m_\rmi{q$_f$}}{2\pi T} \, , \quad \hat \mu_f\equiv \frac{\mu_f}{2\pi T} \, .
\eeq

\subsubsection{One-loop HTLpt pressure}

Working out the pressure of QCD to one-loop order within the above HTLpt scheme, one arrives at the expression
\beq
p_\rmi{HTLpt}\left(T,\bm{\mu}\right) \equiv d_\rmi{A} \Big[(2-2\epsilon)\ p_\rmi{T}\left(T,\bm{\mu}\right) +\ p_\rmi{L}\left(T,\bm{\mu}\right) \Big] + N_\rmi{c} \sum_f\ p_\rmi{q$_f$}\left(T,\bm{\mu}\right) + \Delta p \, ,
\label{oneloophtl}
\eeq
where $d_\rmi{A}\equiv N_\rmi{c}^2-1$ and the contributions from transverse and longitudinal gluons as well as quarks read respectively
{\allowdisplaybreaks
\beqa
p_\rmi{T}\left(T,\bm{\mu}\right)&=&-\frac{1}{2}\,\sumint_{K}\log\Big[K^2+\Pi_\rmi{T}(i\omega_n,k)\Big] \label{Full_Transverse_Gluon}\, , \\
p_\rmi{L}\left(T,\bm{\mu}\right)&=&-\frac{1}{2}\,\sumint_{K}\log\Big[k^2+\Pi_\rmi{L}(i\omega_n,k)\Big] \label{Full_Longitudinal_Gluon}\, , \\
p_\rmi{q$_f$}\left(T,\bm{\mu}\right)&=&2\,\sumint_{\{K\}}\log\Big[A_\rmi{S}^2(i\widetilde{\omega}_n+\mu_f,k)-A_\rmi{0}^2(i\widetilde{\omega}_n+\mu_f,k)\Big] \label{Full_Quarks}\, ,
\eeqa}
\hspace{-0.12cm}and where $\Delta p$ stands for a counterterm necessary to cancel UV divergences (for our notation of the Matsubara frequencies, consult the appendix \ref{sec:Notations}). The transverse gluon self-energy $\Pi_\rmi{T}$, the longitudinal gluon self-energy $\Pi_\rmi{L}$, and the functions $A_\rmi{S}$ and $A_\rmi{0}$ are in turn given by the expressions
{\allowdisplaybreaks
\beqa
\Pi_\rmi{T}(i\omega_n,k) &\equiv& - \frac{m_\rmi{D}^2}{2-2\epsilon} \frac{\omega_n^2}{k^2} \bigg[1-\frac{\omega^2_n + k^2}{\omega_n^2} {\cal T}_\rmi{K}(i\omega_n,k) \bigg] \, , \\
\Pi_\rmi{L}(i\omega_n,k) &\equiv& m_\rmi{D}^2 \Big[1-{\cal T}_\rmi{K}(i\omega_n,k)\Big] \, , \\
A_\rmi{0}(i\widetilde{\omega}_n+\mu_f,k) &\equiv& i\widetilde{\omega}_n+\mu_f - \frac{m_\rmi{q$_f$}^2}{i\widetilde{\omega}_n+\mu_f}\  \widetilde{{\cal T}}_\rmi{K}(i\widetilde{\omega}_n+\mu_f,k) \, , \\
A_\rmi{S}(i\widetilde{\omega}_n+\mu_f,k) &\equiv& k+\frac{m_\rmi{q$_f$}^2}{k} \Big[1-\widetilde{{\cal T}}_\rmi{K}(i\widetilde{\omega}_n+\mu_f,k)\Big] \, .
\eeqa}
\hspace{-0.12cm}In $d=3-2\epsilon$ spatial dimensions, the HTLpt functions ${\cal T}_{\rm K}$ and $\widetilde{{\cal T}}_{\rm K}$ can finally be written in terms of hypergeometric functions,
{\allowdisplaybreaks
\beqa
{\cal T}_\rmi{K}(i\omega_n,k) &\equiv& \frac{\Gamma\left(\frac{3}{2}-\epsilon\right)}{\Gamma\left(\frac{3}{2}\right)\Gamma\left(1-\epsilon\right)}\int^{1}_{0}\kern-0.5em\mathop{{\rm d}\!}\nolimits c \left(1-c^2\right)^{-\epsilon} \frac{(i\omega_n)^2}{(i\omega_n)^2-k^2 c^2} \nonumber \\
&=& {}_2F_1\left(\frac{1}{2},1;\frac{3}{2}-\epsilon;\frac{k^2}{(i\omega_n)^2}\right) \label{HTL_function_B} \, , \\
\widetilde{{\cal T}}_\rmi{K}(i\widetilde{\omega}_n+\mu_f,k)&\equiv& \frac{\Gamma\left(\frac{3}{2}-\epsilon\right)}{\Gamma\left(\frac{3}{2}\right)\Gamma\left(1-\epsilon\right)}\int^{1}_{0}\kern-0.5em\mathop{{\rm d}\!}\nolimits c \left(1-c^2\right)^{-\epsilon} \frac{(i\widetilde{\omega}_n+\mu_f)^2}{(i\widetilde{\omega}_n+\mu_f)^2-k^2 c^2} \nonumber \\
&=& {}_2F_1\left(\frac{1}{2},1;\frac{3}{2}-\epsilon;\frac{k^2}{(i\widetilde{\omega}_n+\mu_f)^2}\right) 
 \, . \label{HTL_function_F}
\eeqa}

In appendix~\ref{Computing_Full_HTLpt}, we will present a detailed evaluation of the exact one-loop HTLpt pressure using the above expressions. There, we will in particular show that after cancelling all the $1/\epsilon$ divergences upon renormalization, the remaining finite result reads
{\allowdisplaybreaks
\beqa\label{Full_HTLpt_pressure}
p_\rmi{HTLpt}\left(T,\bm{\mu}\right)&=& d_\rmi{A}\Bigg\{\frac{m_\rmi{D}^4}{64\pi^2}\left(\log\frac{\lmsb}{m_\rmi{D}}+C_\rmi{g}\right)+\frac{1}{2\pi^3}\inteddw\frac{1}{e^{\beta\omega}-1}\inteddk k^2\bigg(2\phi_\rmi{T}-\phi_\rmi{L}\bigg) \, \nonumber \\
&-&\frac{T}{2\pi^2}\intedk k^2\bigg[2\log\bigg(1-e^{-\beta\omega_\rmi{T}}\bigg)+\log\bigg(1-e^{-\beta\omega_\rmi{L}}\bigg)\bigg]-\frac{\pi^2\,T^4}{90}\Bigg\} \ \ \ \ \ \ \ \nonumber \\
&+&N_\rmi{c}\sum_{f,\,s=\pm 1}\Bigg\{\frac{C_\rmi{q}}{2}\ m_\rmi{q$_f$}^4+\frac{2\ T^4}{\pi^2}\mbox{Li}_4\bigg(-e^{s\,\beta\mu_f}\bigg)-\frac{1}{\pi^3}\intdwdk\frac{k^2\,\theta_\rmi{q$_f$}}{e^{\beta\left(\omega+s\,\mu_f\right)}+1} \, \nonumber \\
&+&\frac{T}{\pi^2}\intedk k^2\ \bigg[\log\bigg(1+e^{-\beta\left(\omega_{f_+}+s\,\mu_f\right)}\bigg)+\log\bigg(1+e^{-\beta\left(\omega_{f_-}+s\,\mu_f\right)}\bigg)\bigg]\Bigg\} \, , 
\nonumber \\
\eeqa}
\hspace{-0.12cm}where the angles $\phi_\rmi{T,L}\equiv\phi_\rmi{T,L}\left(T,\bm{\mu}\right)$ and $\theta_\rmi{q$_f$}\equiv\theta_\rmi{q$_f$}\left(T,\bm{\mu}\right)$, as well as the dispersion relations $\omega_\rmi{T,L,{$f_\pm$}} \equiv\omega_\rmi{T,L,{$f_\pm$}}\left(T,\bm{\mu}\right)$ and the constants $C_\rmi{g}\approx 1.17201$ and $C_\rmi{q}\approx -0.03653$ are all defined in section~\ref{sec:full_one_loop_HTLpt_pressure}. The mass parameters $m_\rmi{D}\equiv m_\rmi{D}\left(T,\bm{\mu}\right)$ and $m_\rmi{q$_f$}\equiv m_\rmi{q$_f$}\left(T,\bm{\mu}\right)$ are in turn given by eq.~(\ref{HTLpt_mD_mq_parameters}). It should be noted that when expanded in powers of the gauge coupling, this expression will differ from the correct weak coupling expansion of the QCD pressure already at order $g^2$, even though it does reproduce the correct plasmon term. This issue is automatically taken care of at NLO in the HTLpt expansion. Notice also that unlike in DR, the fundamental expansion parameter of HTLpt is $\delta$ (instead of the coupling $g$), which makes the two schemes different by construction.

\subsection{Dimensional reduction \label{sec:DR}}

In addition to the HTLpt reorganization of perturbative finite-temperature QCD, there exists another natural framework for including physically important higher order corrections to thermodynamic quantities. It is based on the fact that at high temperatures, the compact temporal direction of the imaginary time formalism shrinks as $1/T$, rendering the system effectively three-dimensional. Taking advantage of this observation, dubbed \textit{dimensional reduction}, it can be shown that the dynamics of length scales of order $1/(gT)$ and larger can be described using a three-dimensional effective theory for the static bosonic field modes, Electrostatic QCD~\cite{Kajantie1,braaten2}. This becomes particularly relevant for the perturbative determination of various thermodynamic quantities, as it is exactly these field modes that are responsible for the IR problems (and the associated poor convergence) of unresummed perturbation theory.

The EQCD Lagrangian can be formally obtained by integrating out the hard degrees of freedom from the full theory, exhibiting thermal masses of order $T$. This leads to a three-dimensional SU($N_\rmi{c}$) Yang-Mills theory  coupled to an adjoint Higgs field $A_0$ that corresponds (at leading order) to the zero Matsubara mode of the four-dimensional temporal gauge field. Up to higher order operators that enter the weak coupling expansion of the pressure beyond ${\cal O}(g^6)$ (cf.~the $\delta{\mathcal L}_\rmi{E}$ term below), the Lagrangian of EQCD reads\footnote{This applies only for $N_\rmi{c}\leq 3$; beyond this, there are two independent operators quartic in $A_0$.}
\beqa
{\mathcal L}_{\rmi{EQCD}}&\equiv&\frac{1}{2}\,{\rm Tr}\left[G_{ij}^2\right]+{\rm Tr}\!\left[(D_i\,A_0)^2\right]+m_\rmi{E}^2\,{\rm Tr}\left[A_0^2\right] \nonumber \\
&+& i\zeta\,{\rm Tr}\left[A_0^3\right]+\lambda_\rmi{E}\,{\rm Tr}\left[A_0^4\right]+\delta{\mathcal L}_\rmi{E} \, ,
\eeqa
where $D_i$ now denotes the covariant derivative in the adjoint representation and the different parameters can be determined via matching computations in the full theory. Of these constants, the parameter $\zeta$ is somewhat special, as it is nonvanishing only in the presence of nonzero quark chemical potentials and thus contributes to the finite-density equation of state and the quark number susceptibilities, but not the $\mu_f=0$ EoS.

EQCD turns out to be an extremely efficient tool not only in organizing high-order perturbative calculations, but also in performing nonperturbative studies of the IR sector of the full theory. In the latter case, it is nevertheless good to recall that this construction explicitly breaks the Z($N_\rmi{c}$) center symmetry of four-dimensional Yang-Mills theory, which can only be remedied by generalizing the degrees of freedom of the effective theory to coarse grained Wilson loop operators~\cite{aleksi4,deforcrand1,zhang1}. For our purposes, this however plays no role, as we are only interested in weak coupling expansions, which in any case involve expanding the functional integral corresponding to the partition function around the trivial Z($N_\rmi{c}$) vacuum (and moreover, the Z($N_\rmi{c}$) symmetry is explicitly broken by quarks). In our calculation, EQCD is merely used to remove the IR divergences encoutered in the evaluation of the pressure, as well as to resum an important class of higher order contributions to this quantity. As we will see in the following sections, this has a remarkable effect on the convergence properties of the corresponding weak coupling expansion.

\subsubsection{QCD pressure via DR motivated resummation}

Using EQCD to account for the contributions of the soft IR sector of QCD, the pressure of the theory obtains the simple form (for details, see e.g.~\cite{braaten2,kajantie3,aleksi3})
\beq
p_\rmi{QCD}\left(T,\bm{\mu}\right) \equiv p_\rmi{HARD}\left(T,\bm{\mu}\right) + T\, p_\rmi{SOFT}\left(T,\bm{\mu}\right)\, ,
\label{drpressure}
\eeq
where $p_\rmi{HARD}$ is available through a strict loop expansion in the four-dimensional theory, while $p_\rmi{SOFT}$ denotes the (nonperturbative) pressure obtained from the partition function of EQCD. The first of these terms has a simple physical interpretation as the contribution of the hard scale $T$ to the pressure and is organized in a power series in even powers of $g$, while the second term contains all contributions from the soft and ultrasoft scales $gT$ and $g^2T$. Due to the nonperturbative nature of the ultrasoft sector of the theory, the function $p_\rmi{SOFT}$ does not allow for a diagrammatic evaluation to all loop orders in $g$, but contains fundamentally nonperturbative terms. They enter the quantity at order $g^6$ and must be determined via three-dimensional lattice simulations and a conversion of the results to continuum regularization (see e.g.~\cite{direnzo1} for details).

At nonzero chemical potentials, the functions $p_\rmi{HARD}$ and $p_\rmi{SOFT}$ have been worked out up to and partially including order $g^6$, with the only contribution missing from the full ${\mathcal O}(g^6)$ term originating from the four-loop full theory diagrams needed in $p_\rmi{HARD}$ (see~\cite{gynther1} for the evaluation of some of these integrals and~\cite{York} for some recent progress in the general evaluation of sum-integrals). Following the procedure of~\cite{mikkoyork}, we write these two functions in the forms
\beqa
\frac{p_\rmi{HARD}\left(T,\bm{\mu}\right)}{T^4} &=& \aE{1} + \hat g_3^2 \ \aE{2} + \frac{\hat g_3^4}{(4\pi)^2} \bigg(\aE{3}-\aE{2}\ \aE{7}-\frac{1}{4} d_\rmi{A} C_\rmi{A} \ \aE{5}\bigg) \, \label{pHARD} \\
&+& \frac{\hat g_3^6}{(4\pi)^4}\left[d_\rmi{A} C_\rmi{A}\bigg(\aE{6}-\aE{4}\,\aE{7}\bigg)-d_\rmi{A} C_\rmi{A}^3\left(\frac{43}{3}-\frac{27}{32}\pi^2\right)\right]\log\frac{\lmsb}{4\pi T} +{\mathcal O}(g^6)\, , \ \ \ \ \ \ \nonumber \\
\frac{p_\rmi{SOFT}\left(T,\bm{\mu}\right)}{T^3} &=& \frac{\hat m_\rmi{E}^3}{12\pi}\ d_\rmi{A} - \frac{\hat g_3^2 \ \hat m_\rmi{E}^2}{(4\pi)^2}\ d_\rmi{A} C_\rmi{A} \bigg( \log\frac{\lmsb}{2T \hat m_\rmi{E}} + \frac{3}{4}\bigg) \, \nonumber \\
&-& \frac{\hat g_3^4 \ \hat m_\rmi{E}}{(4\pi)^3}\,d_\rmi{A} C_\rmi{A}^2 \bigg(\frac{89}{24} + \frac{\pi^2}{6} - \frac{11}{6} \log 2\bigg) \, \nonumber \\
&+& \frac{\hat g_3^6}{(4\pi)^4}\,d_\rmi{A} \Bigg[C_\rmi{A}^3\left(\frac{43}{4}-\frac{491}{768}\pi^2\right)\log\frac{\lmsb}{2T\hat m_\rmi{E}}+C_\rmi{A}^3\left(\frac{43}{12}-\frac{157}{768}\pi^2\right)\log\frac{\lmsb}{2C_\rmi{A} T\hat g_3^2} \ \, \nonumber \\
& & \ \ \ \ \ \ \ \ \ \ \ \ - \frac{4}{3}\frac{N_\rmi{c}^2-4}{N_\rmi{c}}\Bigg(\,\sum_f\hat\mu_f\,\Bigg)^2\,\log\frac{\lmsb}{2T\hat m_\rmi{E}}\,\Bigg] +{\mathcal O}(g^6)\, ,
\eeqa
where $C_\rmi{A}\equiv N_\rmi{c}$. In this expression, we have also rescaled the electric screening mass as well as the three-dimensional gauge coupling of EQCD to be dimensionless via (note the factor $2\pi$ difference to our HTLpt notation)
\beq
\hat m_\rmi{E}\equiv \frac{m_\rmi{E}}{T} \, ,\quad \hat g_3^2 \equiv \frac{g_3^2}{T} \, ,
\eeq
and have in addition used the result~\cite{Hart}
\beqa
\zeta&=&\frac{\hat g_3^3}{3\pi^2}\sum_f \mu_f + {\mathcal O}(\hat g_3^5)\, .
\eeqa
The matching coefficients $\alpha_\rmi{E1}-\alpha_\rmi{E7}$ appearing in $p_\rmi{HARD}$ depend on $\mu_f/T$, $\bar{\Lambda}/T$, as well as various group theory invariants, but are by definition independent of $g$. Of them, the three first ones are defined via the strict weak coupling expansion of the full theory pressure, while the others originate from the relations
\beqa
\hat m_\rmi{E}^2 & = & g^2
 \Bigl( \aE{4} +
 \aE{5} \epsilon + {\cal O}(\epsilon^2) \Bigr)
 + \frac{g^4}{(4\pi)^2}
 \Bigl( \aE{6}  +
 {\cal O}(\epsilon) \Bigr) + {\cal O}(g^6) , \hspace*{0.5cm} \\
\hat g_\rmi{3}^2 & = &  g^2 + \frac{g^4}{(4\pi)^2}
 \Bigl( \aE{7} + {\cal O}(\epsilon) \Bigr)
 + {\cal O}(g^6)  \, .
\eeqa
At $\mu_f=0$, these constants are available e.g.~from~\cite{kajantie3} (see also~\cite{Moeller}), while their finite density counterparts can be found from ref.~\cite{aleksi3}.\footnote{During the preparation of v2 of our manuscript, we were informed by the authors of~\cite{haque} that they had discovered an error in one of the coefficients of $\alpha_\rmi{E3}$, quoted in eq.~(3.16) of~\cite{aleksi3} (in the $C_\rmi{F} T_\rmi{F}$ part, one should replace ``$-24(1-4\overline{\mu}^2)\times (...)$'' $\rightarrow$ ``$-24(1-12\overline{\mu}^2)\times (...)$''). This change has been reflected in eq.~(\ref{AE3_Corrected}) as well as in all numerical results displayed in the paper at hand.} The latter are for convenience listed in appendix~\ref{sec:DR_matching_para}.

It should be noted that in eq.~(\ref{pHARD}) we have on purpose written both the hard and soft contributions in terms of the EQCD parameters $\hat g_3$ and $\hat m_\rmi{E}$, anticipating a resummation of higher order contributions; due to this, some terms have been added to $p_\rmi{HARD}$ to ensure that when expanded in powers of the four-dimensional coupling, the correct weak coupling expansion will be recovered. The explicit logarithms of the renormalization scale shown in eq.~(\ref{pHARD}) have been chosen (somewhat arbitrarily) so that they cancel the scale dependence of $p_\rmi{SOFT}$ at order $g^6$, cf.~\cite{mikkoyork}. This is in accordance with the fact that our result in any case misses non-logarithmic contributions of order $g^6$.

The above results provide a straightforward recipe to evaluate the QCD pressure to order $g^6\ln\,g$, but at the same time leave some freedom to deal with the higher order terms. If the final outcome is expanded in powers of the four-dimensional gauge coupling and all contributions of order $g^6$ and higher are thrown away, the resulting expression corresponds to the unresummed weak coupling expansion of the pressure. However, another natural alternative is to consider both $p_\rmi{HARD}$ and $p_\rmi{SOFT}$ functions of the three-dimensional gauge coupling $g_3$ and the electric screening mass $m_\rmi{E}$, and to keep these parameters unexpanded in $g$. This leads to a resummed expression for the pressure, which was first suggested in the $\mu_f=0$ case in ref.~\cite{blaizot4} and later demonstrated to lead to a dramatic decrease in the renormalization scale dependence (and thus an improvement in the convergence properties) of the quantity in ref.~\cite{mikkoyork}. Just like in ref.~\cite{sylvain1}, in the present paper we will implement this procedure at nonzero quark chemical potentials --- only now for the full finite-density pressure. The resulting expression will be refered to as the `DR pressure' from now on.

\section{Mass expansion in HTLpt}\label{sec:Section4_for_Intro}

In this section, we study the convergence of the $m/T$-\,expansion of the one-loop HTLpt pressure and quark number susceptibilities, first calculated utilizing the expansion in ref.~\cite{sylvain1} and generalized to exact HTLpt in the present paper.

\subsection{Expressions for the $m/T$-\,expansion}

As will be shown in the appendices of this paper, both the pressure and the QNS can be calculated exactly to one-loop order in HTLpt by combining analytical and numerical techniques. However, at higher-loop orders it turns out that the sum-integrals encountered are very difficult to handle, and hence one must in practice resort to various approximations~\cite{htlpt2loop}. One such approximation, which greatly simplifies the evaluation of the HTLpt sum-integrals, involves expanding the integrands in powers of the ratios of the (Debye and quark thermal) masses and the hard scale $T$. This is expected to be a reasonable approximation at least at high temperatures, as the weak coupling expansions of the thermal masses start at order $g\,T$.

%%%%%%%%%%%%%%%%%%%%%%%%%%%%%%%%%%%%%%%%%%%%%%%%%%%%%%%%%%%%%%%%%%%%%%%%
\begin{figure}[!t]
\centering\includegraphics[scale=0.38]{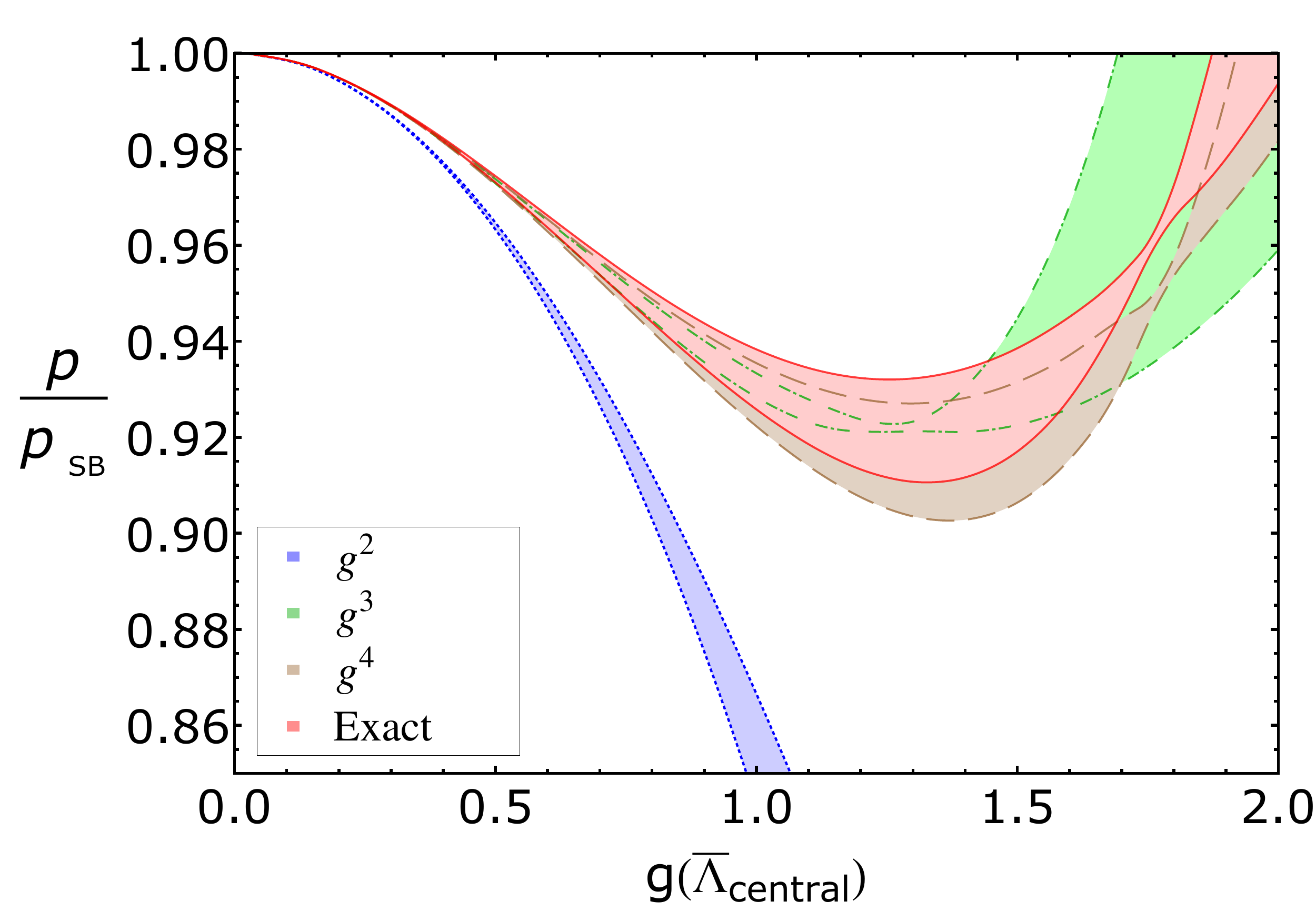}
\caption{The $N_\rmi{f}=3$ zero density normalized pressure plotted as a function of the coupling $g(\bar{\Lambda}_{\rm central})$, evaluated to one-loop order in HTLpt and truncated at orders $g^2$ (blue, dotted lines), $g^3$ (green, dot-dashed), $g^4$ (brown, dashed), and not at all (red, solid).\label{presconv}}
\end{figure}
%%%%%%%%%%%%%%%%%%%%%%%%%%%%%%%%%%%%%%%%%%%%%%%%%%%%%%%%%%%%%%%%%%%%%%%%

In the $m/T$-\,expansion, the one-loop expressions for the pressure as well as the second and fourth-order QNS were first obtained in ref.~\cite{sylvain1}. These purely analytical results read
{\allowdisplaybreaks
\beqa
p_\rmi{HTLpt}^\rmi{\tiny High-T}\left(T,\bm{\mu}\right)&=&\frac{d_\rmi{A}\pi^2 T^4}{45}\Bigg\{1+\frac{N_\rmi{c}}{d_\rmi{A}}\sum_f\bigg(\frac{7}{4}+30\ \hat \mu_f^2 +60\ \hat \mu_f^4\bigg) -\frac{15}{2}\hat m_\rmi{D}^2  \, \nonumber \\
& & \ \ \ \ \ \ \ \ \ \ \ \ - \frac{30\ N_\rmi{c}}{d_\rmi{A}}\sum_f\bigg( 1+12\ \hat \mu_f^2\bigg) \hat m_\rmi{q$_f$}^2+30\ \hat m_\rmi{D}^3  \, \nonumber \\
& &  \ \ \ \ \ \ \ \ \ \ \ \ +\frac{45}{4}\bigg(\gamma_\rmi{\tiny E}-\frac{7}{2}+\frac{\pi^2}{3}+\log\frac{\lmsb}{4\,\pi\,T}\bigg)\ \hat m_\rmi{D}^4 \, \nonumber \\
& & \ \ \ \ \ \ \ \ \ \ \ \ +\frac{60\ N_\rmi{c}}{d_\rmi{A}} \bigg(6-\pi^2\bigg) \sum_f \hat m_\rmi{q$_f$}^4+{\cal O}\left(\hat m_\rmi{D}^6,\hat m_\rmi{q$_f$}^6\right)\Bigg\} \, ,\\ \nonumber
\frac{\chi_{\rm u2}}{\chi_{\rm u2,SB}}&=&1-\frac{3\,d_\rmi{A}}{8\,N_\rmi{c}\,\pi^2}\,\,g^2+d_\rmi{A}\left(1+\frac{N_\rmi{f}}{2\,N_\rmi{c}}\right)^{1/2}\frac{\sqrt{3/N_\rmi{c}}}{8\,\pi^3}\,\,\,g^{3}+\frac{d_\rmi{A}}{32\,\pi^4}\left(1+\frac{N_\rmi{f}}{2\,N_\rmi{c}}\right) \nonumber \\
&&\ \times\Bigg[\frac{\pi^2}{3}-\frac{7}{2}+\gamma_\rmi{\tiny E}+\log\frac{\lmsb}{4\,\pi\,T}+\frac{d_\rmi{A}\,\left(6-\pi^2\right)}{4\,N_\rmi{c}\,\left(2\,N_\rmi{c}+N_\rmi{f}\right)}\,\Bigg]\,g^{4}+{\cal O}(g^6)\, , \\
\frac{\chi_{\rm u4}}{\chi_{\rm u4,SB}}&=&1-\frac{3\,d_\rmi{A}}{8\,N_\rmi{c}\,\pi^2}\,\,g^2+3\,d_\rmi{A}\,\frac{N_\rmi{f}}{N_\rmi{c}}\left(1+\frac{N_\rmi{f}}{2\,N_\rmi{c}}\right)^{-1/2}\frac{\sqrt{3/\,N_\rmi{c}}}{32\,\pi^3}\,\,g^{3}+\frac{3\,d_\rmi{A}}{32\,\pi^4} \nonumber \\
&&\ \times\left(\frac{N_\rmi{f}}{2\,N_\rmi{c}}\right)\Bigg[\frac{\pi^2}{3}-\frac{7}{2}+\gamma_\rmi{\tiny E}+\log\frac{\lmsb}{4\,\pi\,T}+\frac{d_\rmi{A}\,\left(6-\pi^2\right)}{12\,N_\rmi{c}\,N_\rmi{f}}\,\Bigg]\,g^{4}+{\cal O}(g^6)\, , \ \ \ \ \ \ \ \ 
\eeqa}
\hspace{-0.12cm}the detailed derivation of which we present in appendix~\ref{sec:Mass_expansion}. The mass parameters $m_\rmi{D}$ and $m_\rmi{q$_f$}$ are again given by eq.~(\ref{HTLpt_mD_mq_parameters}).

\subsection{Convergence of the $m/T$-\,expansion}

To study the convergence of the mass expansion, we specialize to the $N_\rmi{f}=3$ case and inspect the $\mu_f=0$ pressure as a function of $g(\bar{\Lambda}_{\rm central})$, normalized to the corresponding noninteracting Stefan-Boltzmann result. We note that $\bar{\Lambda}_{\rm central}$ corresponds to the central renormalization scale, as defined in section~(\ref{sec:Fixing_Para}). The different bands shown in figure~\ref{presconv} correspond to truncations at different orders in $m/T\sim g$, and are generated by varying the renormalization scale $\lmsb$ by a factor of two around its central value. We see from here that with the exception of the most naive truncation at order $g^2$, the truncated results are not only in good agreement with each other, but even with the exact untruncated band. For values of $g$ relevant for our calculation, the differences between the various bands are of the order of one per cent, which can be considered an extremely positive outcome. Note also that for values of $g$ below 1.5, the widths of the bands somewhat surprisingly become larger as we go to higher orders in the $m/T$-\,expansion.

\begin{figure}[!t]
\centering\includegraphics[scale=0.31]{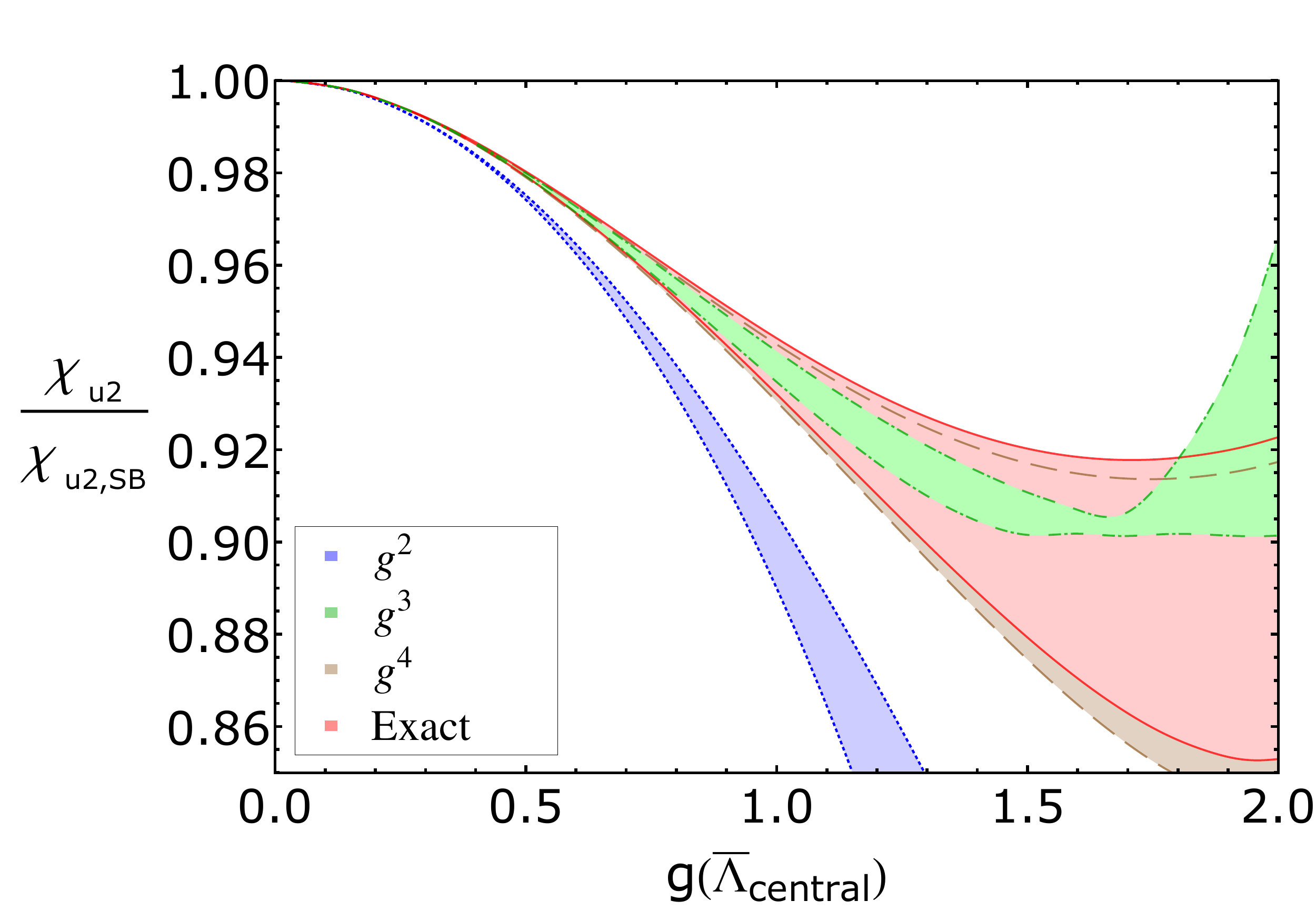}\;\includegraphics[scale=0.31]{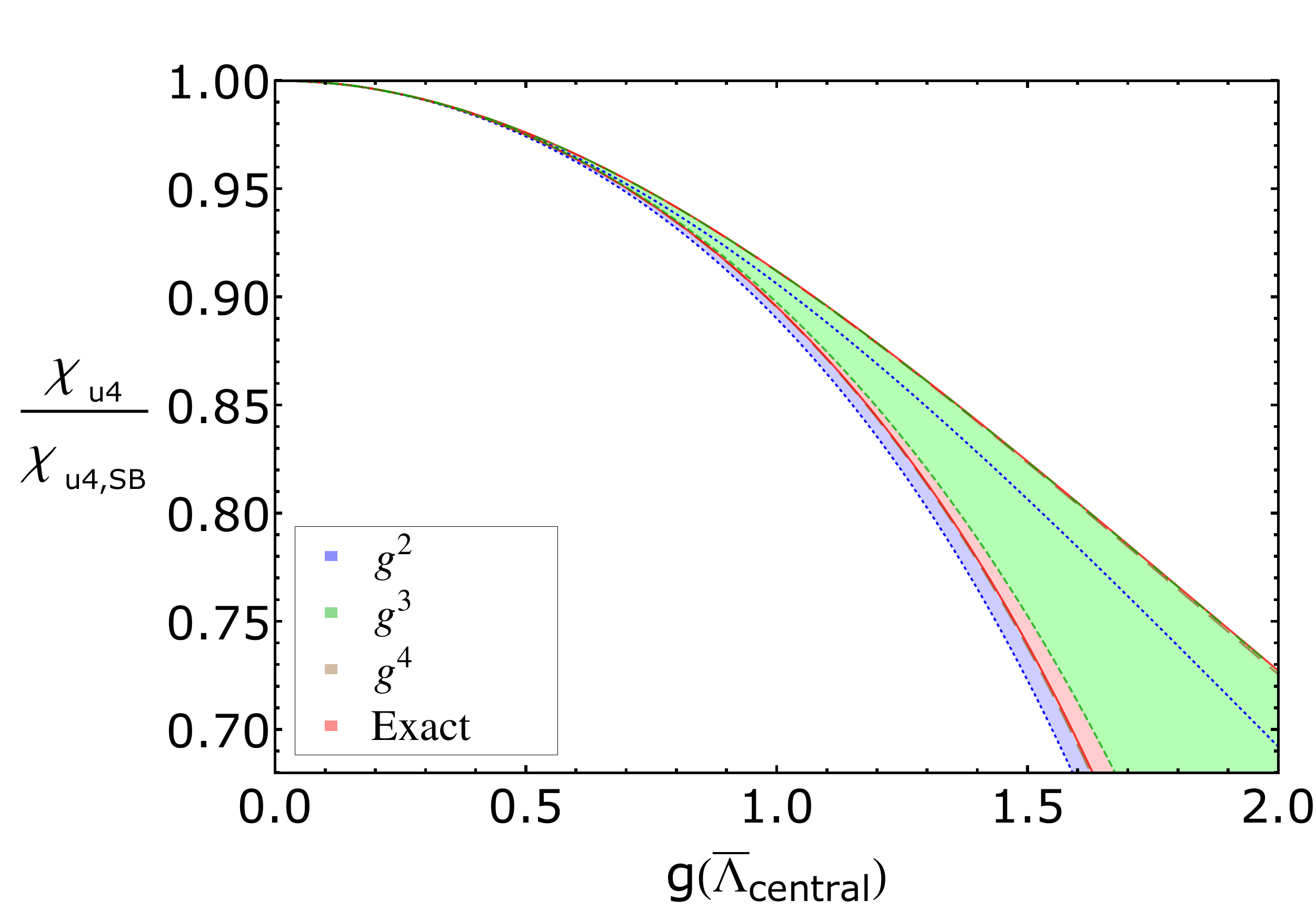}
\caption{The properly normalized $N_\rmi{f}=3$ second (left) and fourth (right) order QNS plotted as functions of the coupling $g(\bar{\Lambda}_{\rm central})$ within one-loop HTLpt. The results are also truncated at orders $g^2$ (blue, dotted lines), $g^3$ (green, dot-dashed), $g^4$ (brown, dashed), and not at all (red, solid).\label{chiu2u4conv}}
\end{figure}

Finally, in figure~\ref{chiu2u4conv} we perform a similar analysis of the second (left) and fourth (right) order QNS, also normalized to their Stefan-Boltzmann limits. We again see that the $g^3$ and $g^4$ bands are in good agreement with each other as well as with the exact results. We also observe that the agreement between the $g^4$ results and the exact ones are especially good and much faster than for the $\mu_f=0$ pressure; in particular for the fourth-order QNS, the $g^4$ truncation is seen to almost coincide with the exact band. This excellent convergence of the $m/T$-\,expansion is in agreement with observations made for the three-loop scalar $\phi^4$ theory in ref.~\cite{motspt}. Although it was noted in this work that the convergence slightly worsens as one proceeds to higher loop orders, this gives us some confidence that the $m/T$-expansion is in general a good approximation in HTLpt calculations.

%%%%%%%%%%%%%%%%%%%%%%%%%%%%%%%%%%%%%%%%%%%%%%%%%%%%%%%%%%%%%%%%%%%%%%%%%%%%%
%%%%%%%%%%%%%%%%%%%%%%%%%%%%%%%%%%%%%%%%%%%%%%%%%%%%%%%%%%%%%%%%%%%%%%%%%%%%%

\section{Results and comparison with lattice data}\label{sec:Section3_for_Intro}

In this section, we present our results for the $\mu_f\neq 0$ pressure as well as the associated quark number susceptibilities obtained both from our exact one-loop HTLpt calculation as well as the DR motivated resummation introduced in section \ref{sec:DR}. We then compare these results with state-of-the-art lattice data~\cite{wuppertalcharge,bieleHighT,biele1,biele2,Borsanyi_delta_P,handsu6u2,gupta} as well as a very recent three-loop HTLpt calculation~\cite{haque}. We begin by explaining how various parameters in the DR and HTLpt results are fixed,  and then move on to consider first the three- and later the two-flavor case.

\subsection{Fixing the parameters}\label{sec:Fixing_Para}

In perturbative calculations at high temperature and zero quark chemical potentials, the renormalization scale $\bar{\Lambda}$ is typically chosen to be of order $\bar{\Lambda}\approx 2\pi T$, dictated by the thermal mass of the lowest nonvanishing (bosonic) Matsubara mode. With high order perturbative results available for a number of quantities, different refinements of this choice are however also possible, based on schemes such as the Fastest Apparent Convergence (FAC) or the Principle of Minimal Sensitivity (PMS). As is customary in recent literature (see e.g.~\cite{kajantie4}), we apply in all of our results the FAC scheme to the NLO gauge coupling of EQCD, providing the values $\bar{\Lambda}_{\rm central}=1.445\times 2\pi T$ for three quark flavors as well as $\bar{\Lambda}_{\rm central}=1.291\times 2\pi T$ for two flavors.

The most straightforward way to generalize the above setup to nonzero density is to perform the exact same procedure, i.e.~solve $\bar{\Lambda}$ from the equation $\alpha_\rmi{E7}=0$, at nonvanishing $\mu_f$. This leads to the results
\beqa
\bar{\Lambda}_{\rm central}^{\rm N_\rmi{f}=3}&=&\frac{0.9344\times 2\pi T}{\exp\left(\frac{1}{27}\sum_f\left[\Psi(\frac{1}{2}+i\hat\mu_f)+\Psi(\frac{1}{2}-i\hat\mu_f)\right]\right)} \, , \\
\bar{\Lambda}_{\rm central}^{\rm N_\rmi{f}=2}&=&\frac{0.9847\times 2\pi T}{\exp\left(\frac{1}{29}\sum_f\left[\Psi(\frac{1}{2}+i\hat\mu_f)+\Psi(\frac{1}{2}-i\hat\mu_f)\right]\right)}  \, ,
\eeqa
where $\Psi$ denotes the digamma function. To assess the sensitivity of our results to the choice of the renormalization scale, we will in addition always vary $\bar{\Lambda}$ by a factor of two around these central values.

With the running gauge coupling $g$, we follow the standard procedure of working with a two-loop expression in the DR case and a one-loop coupling in LO HTLpt. To determine the value of $\Lambda_{\rm\overline{MS}}$, we on the other hand take the recent lattice result $\alpha_\rmi{s}(1.5\;{\rm GeV})=0.326$~\cite{runningalpha}, and demand that our running coupling agrees with it for $\bar{\Lambda}_{\rm central}=1.5$ GeV. For  $N_\rmi{f}=3$, this yields the values $\Lambda_{\rm\overline{MS}}=176$ MeV and 283 MeV for the one- and two-loop couplings,  while for $N_\rmi{f}=2$, the corresponding results read $\Lambda_{\rm\overline{MS}}=204$ MeV and 324 MeV. These parameters are also varied by 30 MeV around the central values quoted here.

In the recent three-loop HTLpt results of~\cite{haque}, to which we will compare our calculations below, the authors used two separate renormalization scales $\Lambda_\rmi{g}$ and $\Lambda_\rmi{q}$ for purely gluonic and fermionic graphs, respectively. They took the central values $\Lambda_\rmi{g}=2\pi T$ and $\Lambda_\rmi{q}=2\pi\sqrt{T^2+\mu^2/\pi^2}$ and varied both scales by a factor of two in order to estimate the renormalization scale sensitivity of their results. For the gauge coupling, they used a one-loop running with $\Lambda_{\rm \overline{MS}}=176$ MeV, which for $N_\rmi{f}=3$ gives $\alpha_\rmi{s}(1.5\;{\rm GeV})=0.326$ as well.

\subsection{Results for three flavors}

\begin{figure}[!t]\centering\includegraphics[scale=0.3]{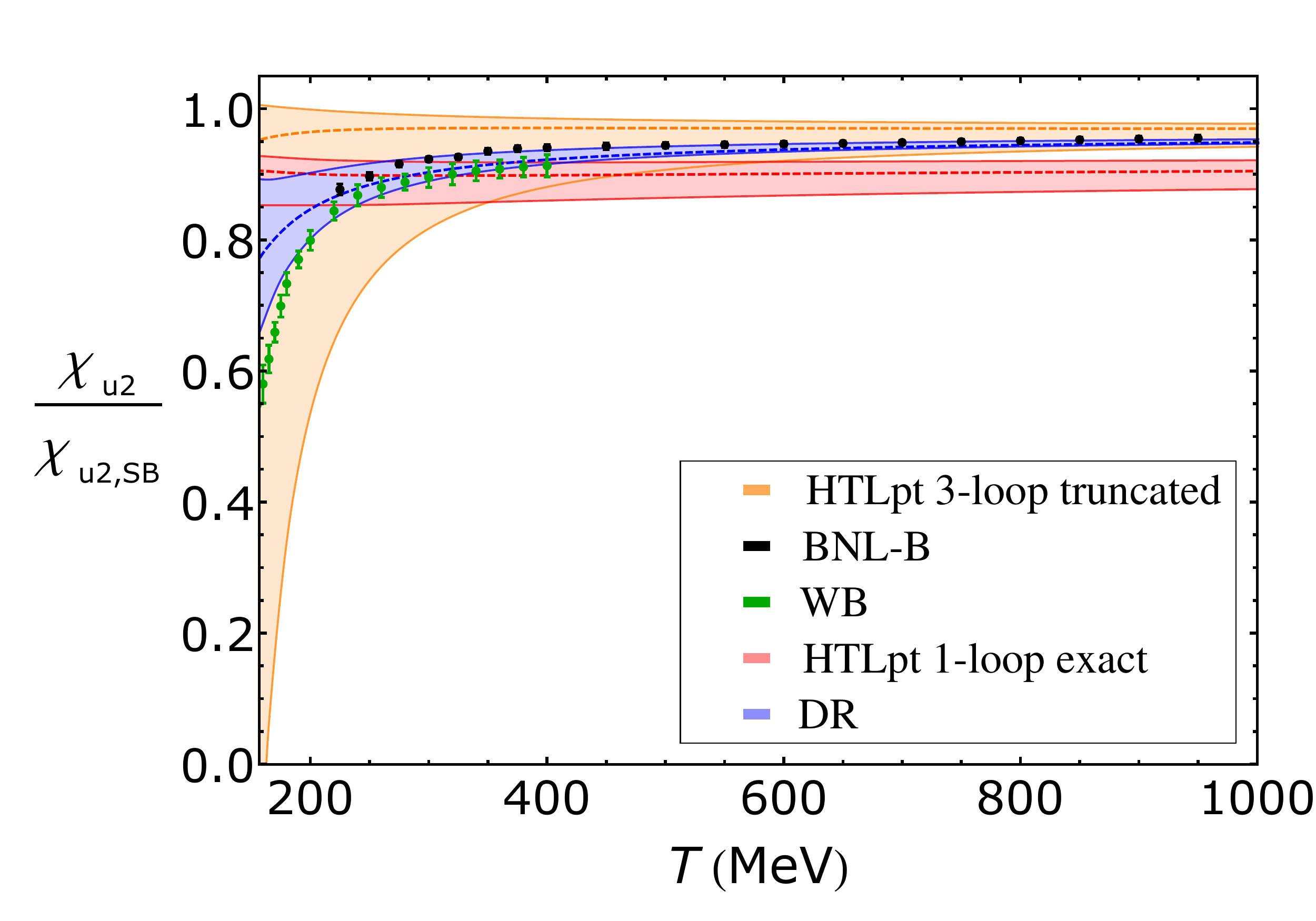}\!\!\!\includegraphics[scale=0.3]{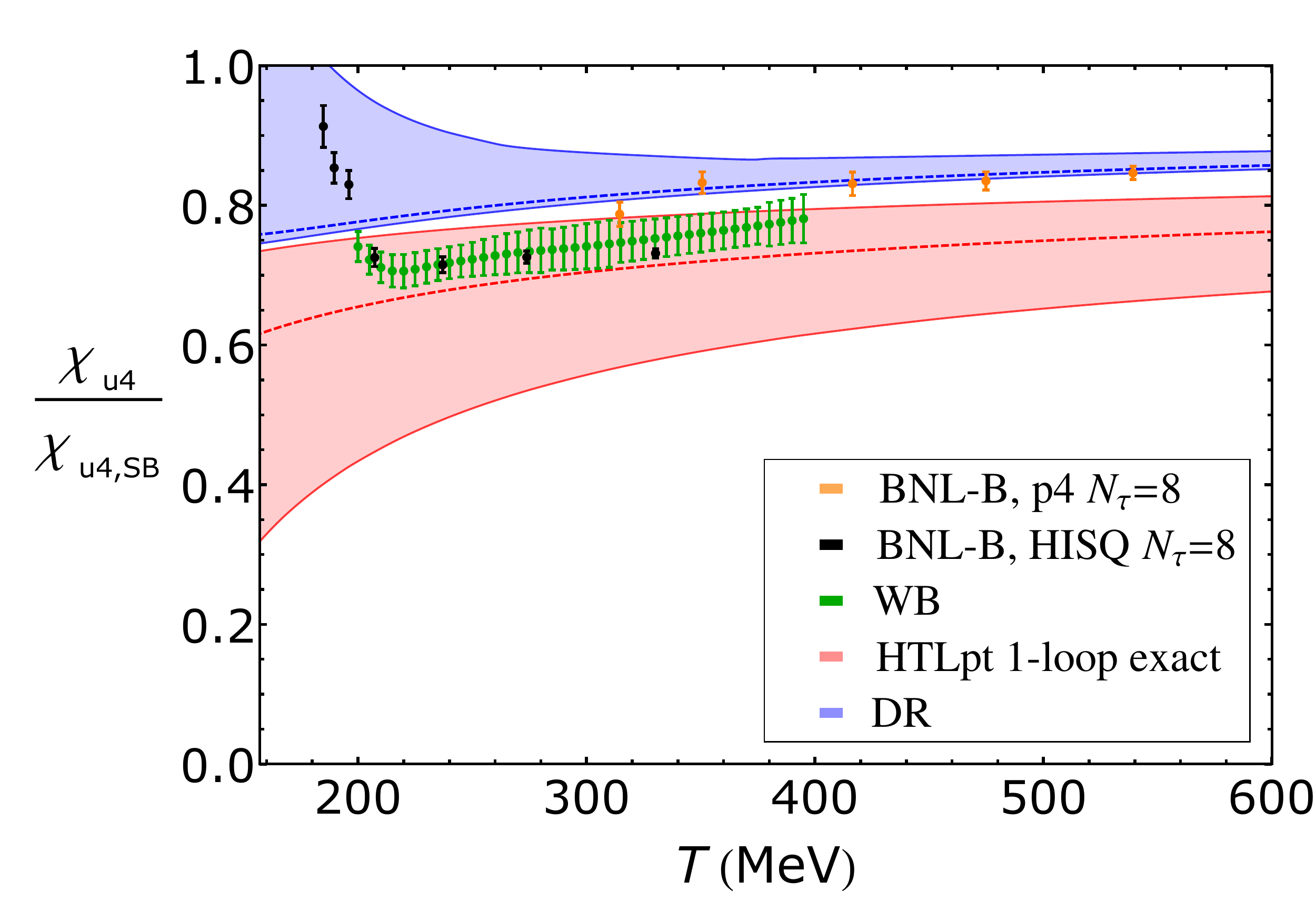}
\caption{\label{fig:CHIu2u4Nf3} The $N_\rmi{f}=3$ second (left) and fourth (right) order diagonal QNS normalized to their respective Stefan-Boltzmann values. The truncated three-loop HTLpt results are from~\cite{haque} and the lattice data are from BNL-Bielefeld (BNL-B)~\cite{bieleHighT,biele1,biele2} and Wuppertal-Budapest (WB)~\cite{wuppertalcharge,chi4}.}
\end{figure}

Let us begin the analysis of our results from the quark number susceptibilities in the physically most interesting case of $N_\rmi{f}=3$. In figure~\ref{fig:CHIu2u4Nf3} (left), we display the second order diagonal susceptibility $\chi_{\rm u2}$ normalized to its Stefan-Boltzmann limit $\chi_{\rm u2,SB}=T^2$. The blue band in the figure corresponds to the DR result, obtained by varying the values of both $\bar{\Lambda}$ and $\Lambda_{\rm\overline{MS}}$ in the ranges explained above, while the red and orange bands are the exact one-loop and truncated three-loop HTLpt results. The thick dashed lines inside the bands correspond to the central values of the renormalization and QCD scales. Finally, we note that the three-loop HTLpt band in fact corresponds to the baryon (and not quark) number susceptibility~\cite{haque}; however, for the second order susceptibilities the difference between these two quantities should be hardly visible~\cite{bieleHighT}.

%%%%%%%%%%%%%%%%%%%%%%%%%%%%%%%%%%%%%%%%%%%%%%%%%%%%%
%%%%%%%%%%%%%%%%%%%%%%%%%%%%%%%%%%%%%%%%%%%%%%%%%%%%%

The widths of the bands shown indicate that the scale dependence of the DR result is extremely weak, except for the very lowest temperatures. At the same time, the one- and three-loop HTLpt results are also quite close to one another for temperatures above 500 MeV, indicating that the quantity under consideration nicely converges at these temperatures.\footnote{Although we do not show it in figure~\ref{fig:CHIu2u4Nf3} (left), the two-loop HTLpt result for the second-order susceptibility is also quite close to the three-loop HTLpt result for temperatures above 500 MeV~\cite{mikehaque}.} In figure~\ref{fig:CHIu2u4Nf3} (left) we also display lattice results from both the BNL-Bielefeld (BNL-B, black dots)~\cite{bieleHighT} and Wuppertal-Budapest (WB, green dots)~\cite{wuppertalcharge} collaborations. Both sets of data have been continuum extrapolated. We observe that the DR and three-loop HTLpt results are all in good agreement with the two lattice results for temperatures of roughly 500 MeV and higher; at even lower $T$, some differences do, however, occur and it is the resummed DR result that seems to agree better with the lattice data points.

%%%%%%%%%%%%%%%%%%%%%%%%%%%%%%%%%%%%%%%%%%%%%%%%%%%%%
\begin{figure}[!t]\centering\includegraphics[scale=0.42]{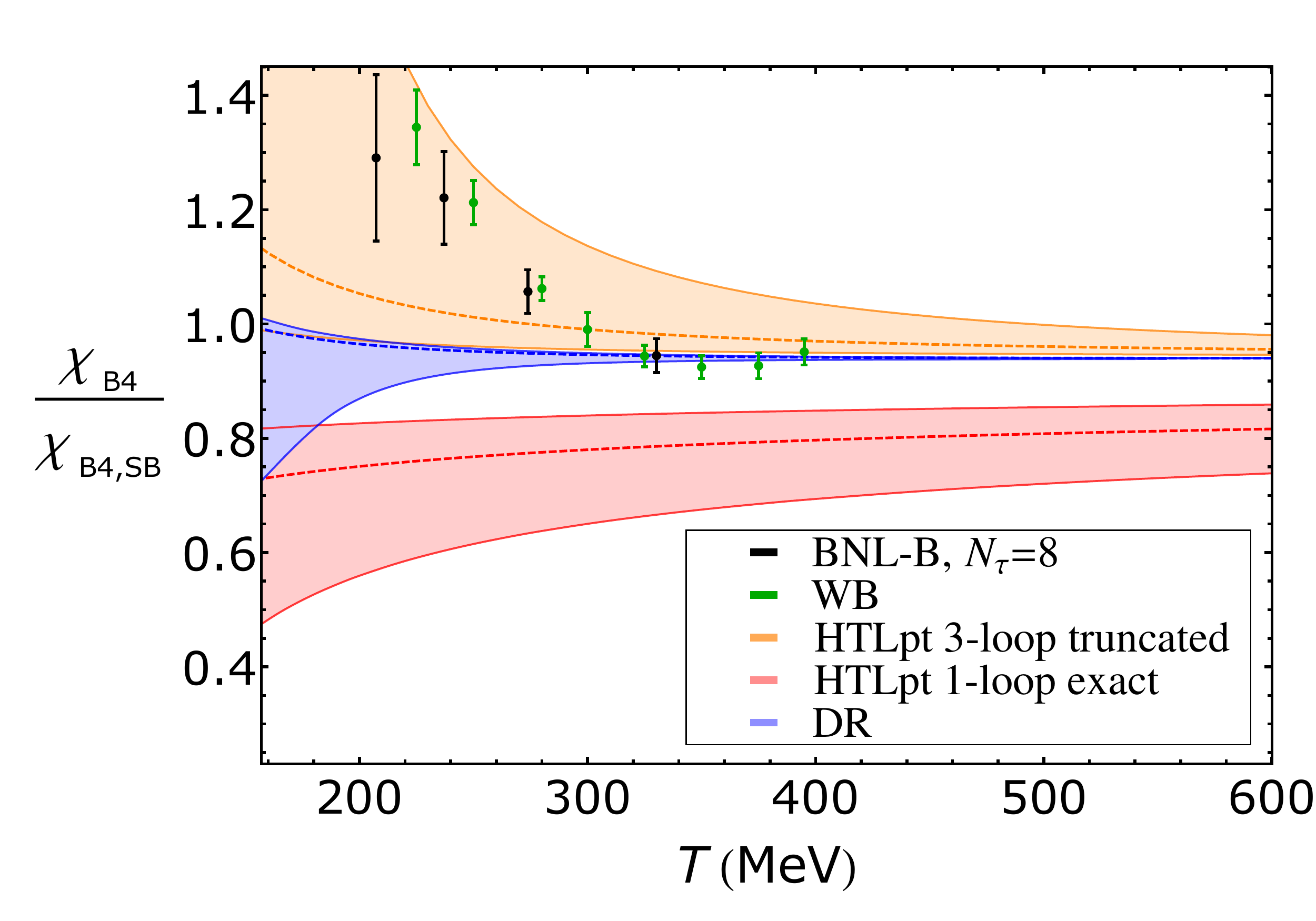}
\caption{\label{fig:CHIb4Nf3} The $N_\rmi{f}=3$ fourth order diagonal baryon number susceptibility normalized to its Stefan-Boltzmann value. The truncated three-loop HTLpt results are from~\cite{haque}, and the lattice data from Wuppertal-Budapest (WB)~\cite{chi4}
and BNL-Bielefeld (BNL-B)~\cite{Bazavov13}.}
\end{figure}
%%%%%%%%%%%%%%%%%%%%%%%%%%%%%%%%%%%%%%%%%%%%%%%%%%%%%

In figure~\ref{fig:CHIu2u4Nf3} (right), we next show our results for the fourth order diagonal QNS $\chi_{\rm u4}$ normalized to the corresponding Stefan-Boltzmann limit $\chi_{\rm u4,SB}=6/\pi^2$. Once again, the thick dashed line in each band corresponds to the central value of the renormalization and QCD scales, while the width of the band indicates the sensitivity of the corresponding result with respect to these parameters. The continuum extrapolated WB lattice data are this time taken from ref.~\cite{chi4}, while the HISQ $N_\tau = 8$ BNL-B results are from refs.~\cite{biele1,biele2}. Our resummed DR and exact one-loop HTLpt results appear to reproduce the qualitative trend of the lattice results for most temperatures, but this time there is a more sizable difference between these two theoretical predictions over the entire temperature range. While for the lowest temperatures the lattice data seem to favor our one-loop HTLpt result, the difference between the lattice points and the DR result is seen to decrease with increasing temperature. It will thus be very interesting to see, how future high-precision lattice data at $T= 500$ MeV and higher will affect these conclusions.

Note that figure~\ref{fig:CHIu2u4Nf3} (right) lacks a band corresponding to the three-loop HTLpt calculation of~\cite{haque}. This is due to the fact that, as pointed out above, at present the three-loop HTLpt results are only available for baryon number susceptibilities, which differ from the quark number ones by off-diagonal contributions, estimated to be non-negligible in this case~\cite{bieleHighT}. To this end, in figure~\ref{fig:CHIb4Nf3} we compare our results for the fourth order baryon number susceptibility $\chi_\rmi{B4}$ normalized to the corresponding Stefan-Boltzmann limit $\chi_{\rm B4,SB}=2/(9\pi^2)$, obtained using the resummed DR approach as well as exact one-loop HTLpt, to the three-loop HTLpt result of ref.~\cite{haque} and lattice data from  both the Wuppertal-Budapest (WB)~\cite{chi4} and BNL-Bielefeld (BNL-B)~\cite{Bazavov13} collaborations. Our notations for the bands and the dashed curves therein are defined as before.

%%%%%%%%%%%%%%%%%%%%%%%%%%%%%%%%%%%%%%%%%%%%%%%%%%%%%
\begin{figure}[!t]\centering\includegraphics[scale=0.42]{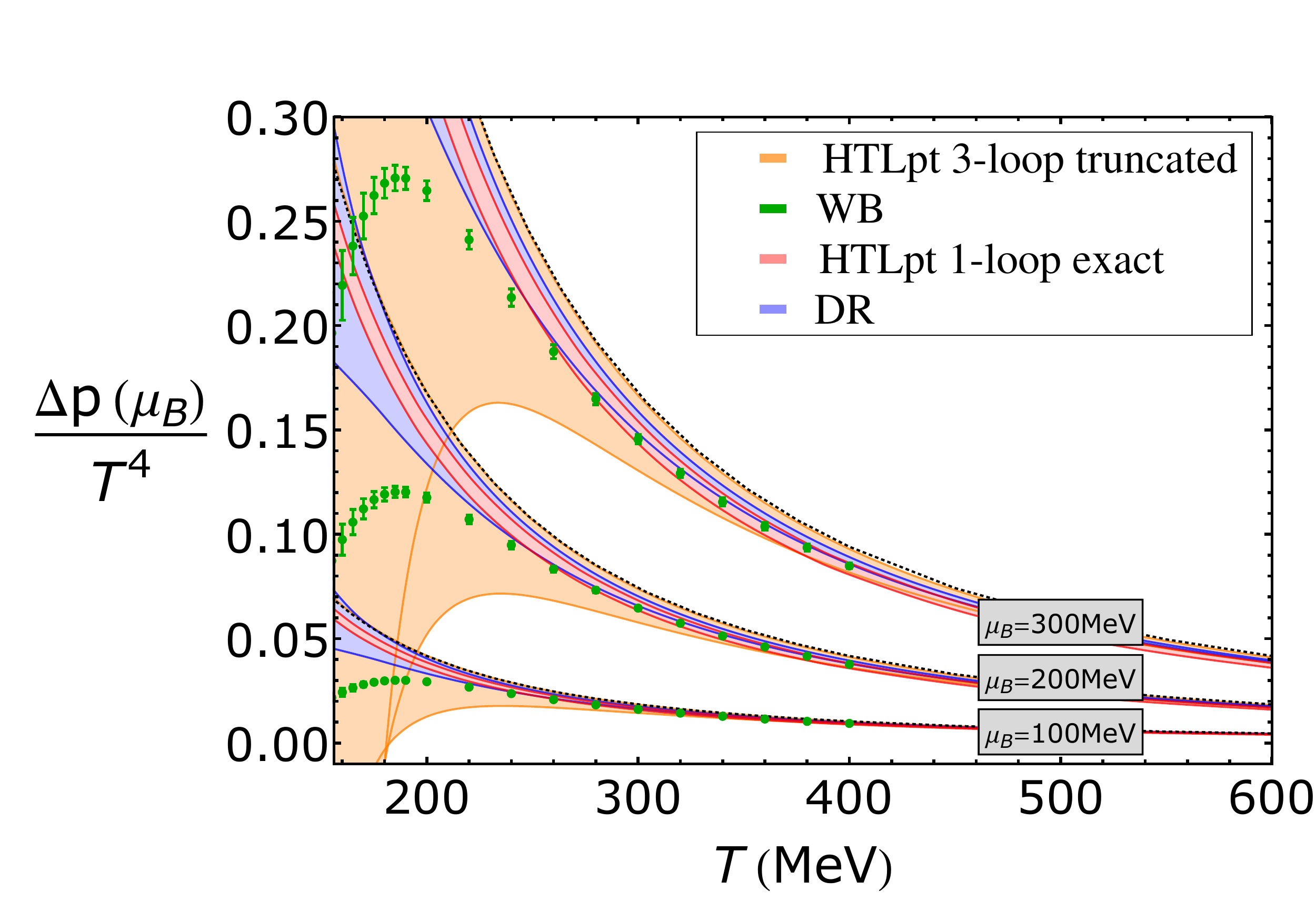}
\caption{The difference between the $N_\rmi{f}=3$ pressure evaluated at nonvanishing chemical potentials and at $\mu_f=0$. The dashed lines indicate the respective Stefan-Boltzmann limits. The three-loop HTLpt results are again from~\cite{haque}, and the lattice data from Wuppertal-Budapest (WB)~\cite{Borsanyi_delta_P}.\label{fig:delta_pressure}}
\end{figure}
%%%%%%%%%%%%%%%%%%%%%%%%%%%%%%%%%%%%%%%%%%%%%%%%%%%%%

From the figure, we see that the central lines from the resummed DR and three-loop HTLpt results for $\chi_{\rm B4}$ are in good agreement in the entire temperature range shown, and moreover overlap with the lattice data from 350 MeV onwards. The resummed DR result, however, has a smaller band than the three-loop HTLpt result. We also note that the convergence of the successive HTLpt loop approximations for the fourth-order susceptibility is not as good as the convergence found for the second-order susceptibility.  This seems to be due to the fact that the fourth-order susceptibility is more sensitive to over-counting which occurs in low loop-order HTLpt.  At three-loop order, this over-counting is fixed through order $g^5$, if the result is perturbatively expanded.

Inspecting next the equation of state itself, in figure~\ref{fig:delta_pressure} we show the difference of the pressure evaluated with nonzero and vanishing quark chemical potentials, which are chosen to be the same for each flavor with $\mu_\rmi{B}=100$, 200, and 300 MeV. The `continuum estimated' lattice data quoted in this figure are from~\cite{Borsanyi_delta_P}, and are based on an expansion of the pressure up to and including ${\cal O}(\mu_f^2)$. The HTLpt and DR results are on the other hand accurate to all orders in $\mu_f$, being only restricted by assumptions inherent in the HTLpt and DR resummation schemes; in the latter case, the results have been shown to be valid for all $T$ and $\mu_f$ satisfying $\pi T \gtrsim g\mu_f$~\cite{ipp3}, which clearly is the case here. As expected based on the earlier analysis of~\cite{aleksi3}, we observe a good agreement between our results and the lattice data down to temperatures of the order of 250 MeV. Also, it is important to note that both the DR and LO HTLpt results are clearly distinct from the corresponding Stefan-Boltzmann limits, indicated by the dashed black lines in the figure.

\subsection{Results for two flavors}

\begin{figure}[!t]\centering\includegraphics[scale=0.29]{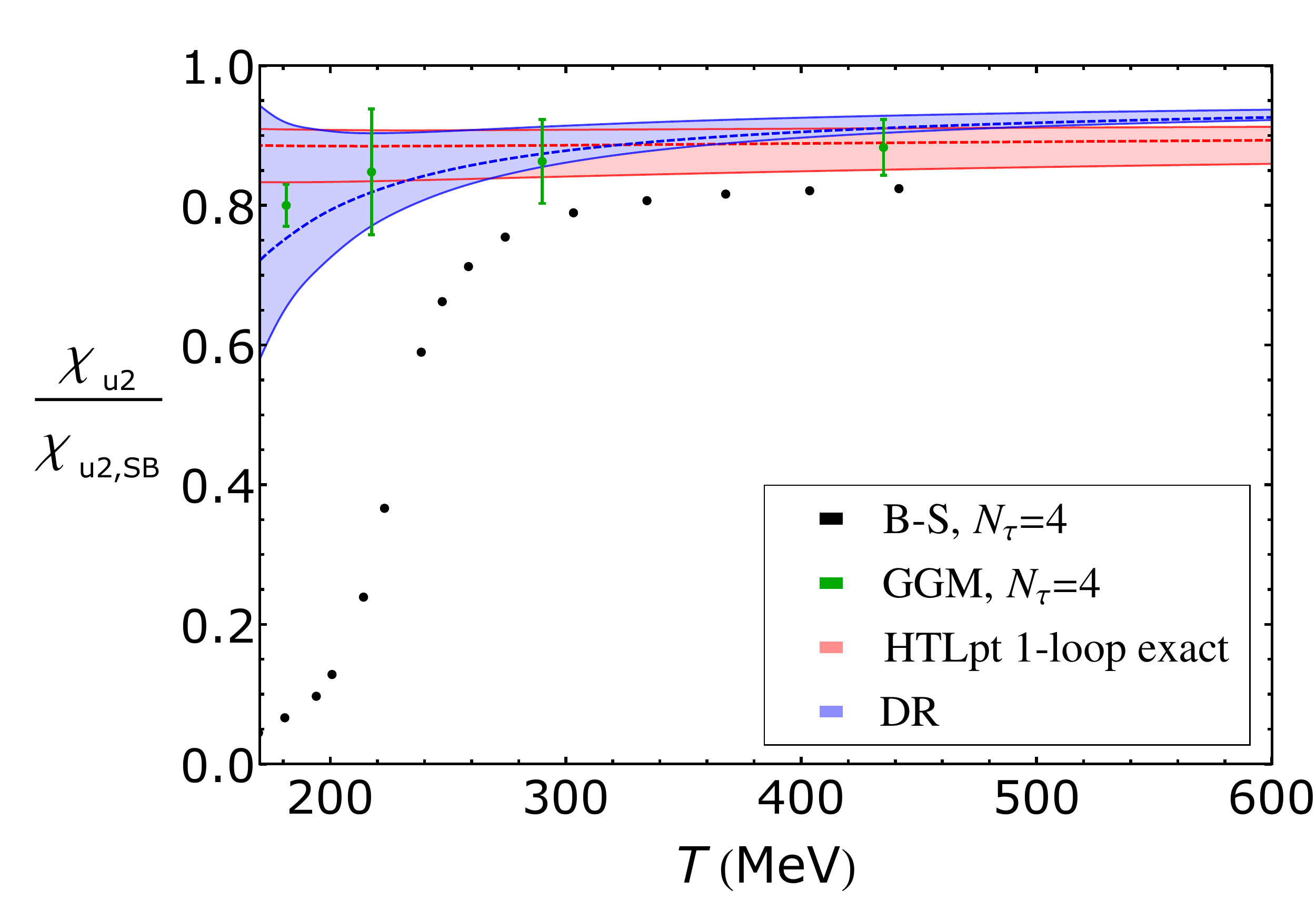}\!\!\!\!\includegraphics[scale=0.31]{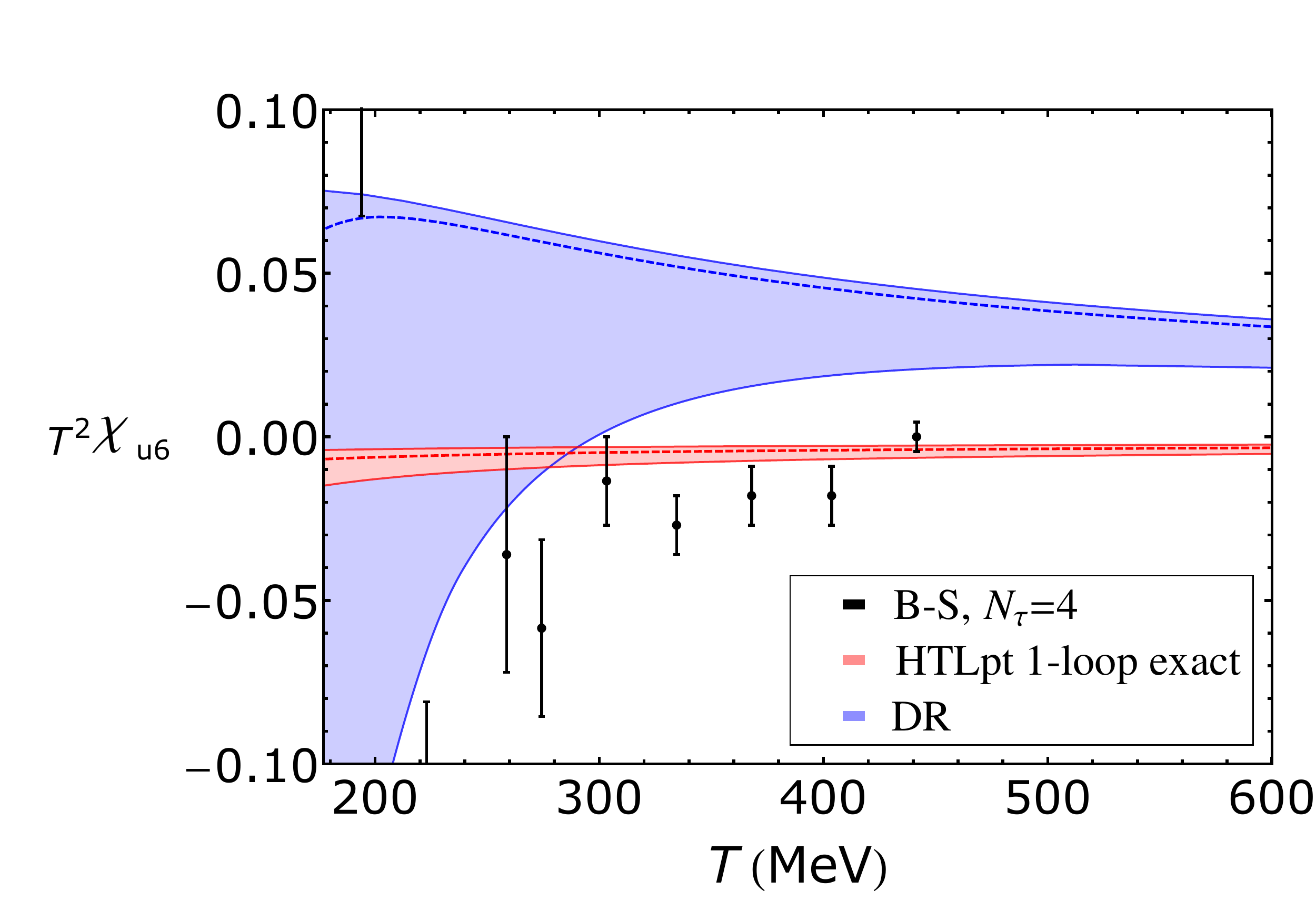}
\caption{The second-order diagonal QNS normalized to its Stefan-Boltzmann value (left), and the sixth-order diagonal QNS multiplied by $T^2$ (right), both evaluated for $N_\rmi{f}=2$. The lattice data sets have been taken from Bielefeld-Swansea (B-S)~\cite{handsu6u2} as well as from~\cite{gupta} (GGM). Note that both data sets in the left figure in fact do include error bars, even though for the B-S data they are hardly visible. \label{fig:CHIu2u6_Nf2}}
\end{figure}

Moving on to the two flavor case, widely studied with lattice methods, we first examine the behavior of the second-order diagonal QNS normalized to its Stefan-Boltzmann limit, displayed in figure~\ref{fig:CHIu2u6_Nf2} (left). As can be seen from here, both of our perturbative bands are now somewhat wider (and reside lower) than in the three flavor case, but their general trends are very similar.\footnote{Note that at the moment three-loop HTLpt results are not yet available in the case of two quark flavors.} Also shown in this figure are lattice results from the Bielefeld-Swansea collaboration (B-S, black dots)~\cite{handsu6u2} as well as those by Gavai, Gupta and Majumdar (GGM, green dots)~\cite{gupta}, both of which were computed on $N_\tau=4$ lattices. One issue in transfering the lattice results to physical units is the determination of the critical temperature appropriate for the lattice spacings used. For the results of~\cite{gupta}, the problem is easily resolved, as we may simply use the value $T_\rmi{c}=145$ MeV, quoted as the pseudo-critical temperature in the same reference. In the case of the B-S results, obtained using p4-improved staggered fermions, more care must, however, be exercised; here, we have chosen to follow the approach of~\cite{Peikert,Peikert2}, which leads to the value $T_\rmi{c}=223$ MeV. As neither of these values correspond to the correct pseudo-critical temperature (in the chiral limit, $T_\rmi{c}=173$ MeV~\cite{AbsoluteScale}), and they moreover deviate from it in opposite directions, we urge the reader to take the lattice data points quoted in the figure with some reservation.

%%%%%%%%%%%%%%%%%%%%%%%%%%%%%%%%%%%%%%%%%%%%%%%%%%%%%
%%%%%%%%%%%%%%%%%%%%%%%%%%%%%%%%%%%%%%%%%%%%%%%%%%%%%

From figure~\ref{fig:CHIu2u6_Nf2} (left), we see that the agreement between our HTLpt and DR bands is very good over the entire range of temperatures displayed. However, unlike in the three flavor case, the agreement between the perturbative bands and the lattice data of~\cite{handsu6u2} is clearly not optimal. We suspect that this may well be related to the discrepancy between the results of the two different lattice groups, which itself can be traced back to the difference between the values of $T_\rmi{c}$ used in converting the lattice data to physical units. Naively, the perturbative results seem to be in better agreement with~\cite{gupta}, but in the absence of continuum extrapolated lattice results, firm conclusions are difficult to draw.

%%%%%%%%%%%%%%%%%%%%%%%%%%%%%%%%%%%%%%%%%%%%%%%%%%%%%
\begin{figure}[!t]\centering\includegraphics[scale=0.42]{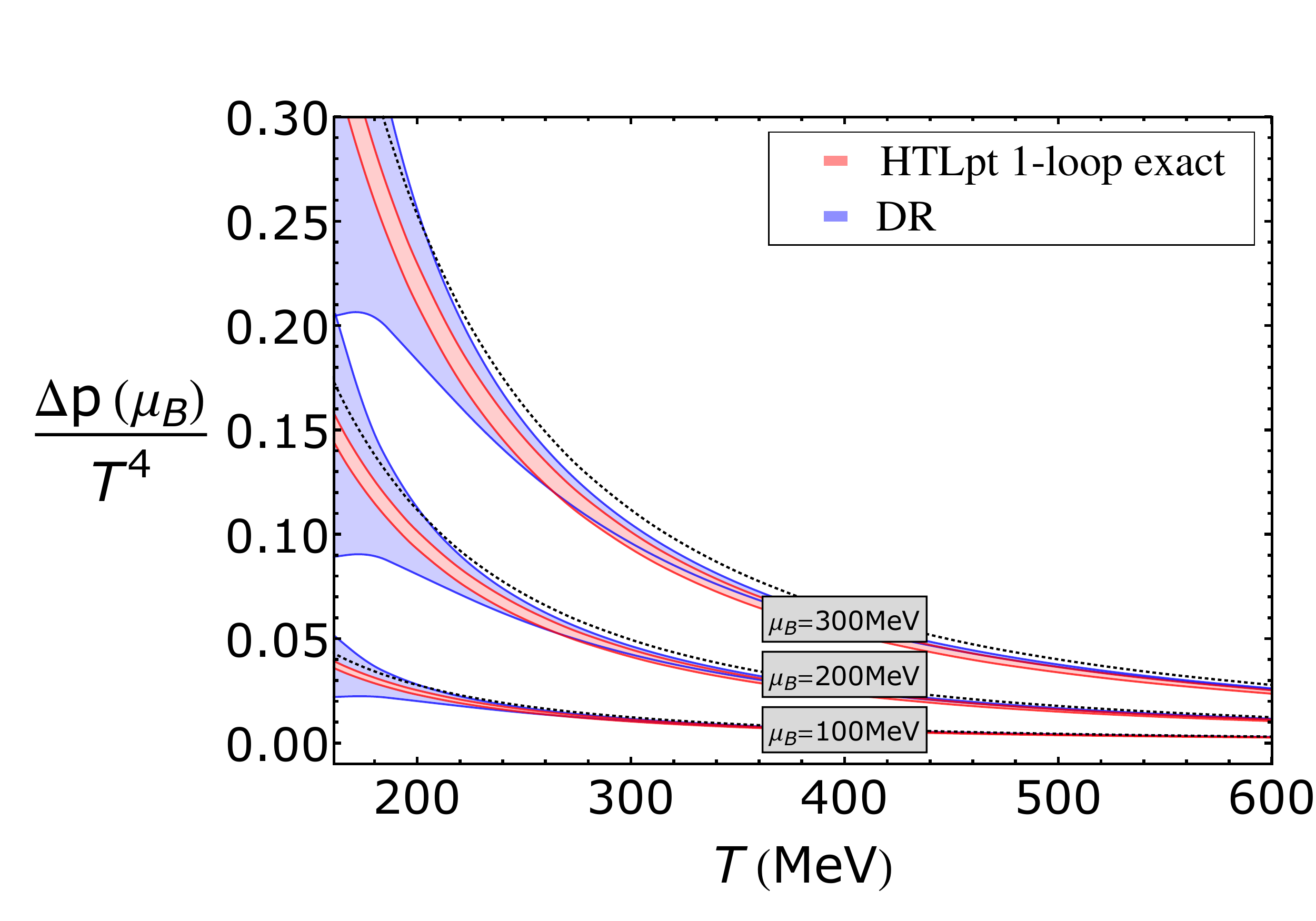}
\caption{The difference between the pressure evaluated at finite and vanishing chemical potentials, for $N_\rmi{f}=2$. The dashed lines denote the Stefan-Boltzmann limits in the three cases considered. \label{fig:delta_pressure_Nf2}}
\end{figure}
%%%%%%%%%%%%%%%%%%%%%%%%%%%%%%%%%%%%%%%%%%%%%%%%%%%%%

Next, consider the sixth order diagonal QNS $\chi_{\rm u6}$, which we display in figure~\ref{fig:CHIu2u6_Nf2} (right) together with lattice data again from Bielefeld-Swansea (B-S)~\cite{handsu6u2}. As we now encounter the first case, where the weak coupling expansion only begins at order $g^3$, it is not surprising that our exact one-loop HTLpt and DR results are seen to disagree: While the DR band is above zero for most of the relevant temperatures, the one-loop HTLpt result is consistently negative. As noted above, at the moment there are unfortunately no published three-loop HTLpt results available for this quantity; given the good agreement of our resummed DR results with the three-loop HTLpt ones for all other susceptibilities, we however suspect that this will be the case also for the sixth order QNS. This despite the fact that according to figure~\ref{fig:CHIu2u6_Nf2} (right) the non-continuum extrapolated lattice data of~\cite{handsu6u2} appear to favor our exact one-loop HTLpt results over the DR ones at least at low temperatures.

Finally, in addition to the susceptibilities, we can inspect the behavior of the pressure at nonzero chemical potentials, also for the case of $N_\rmi{f}=2$. This is done in figure~\ref{fig:delta_pressure_Nf2}, where in the absence of lattice and three-loop HTLpt results we only display two perturbative bands corresponding to our one-loop HTLpt and four-loop DR calculations. Not surprisingly, the HTLpt and DR results are again in good agreement with each other, though the DR bands widen at low temperatures in a more pronounced way than in the three flavor case. Finally, we notice that both sets of bands again differ from their Stefan-Boltzmann limits in a noticeable way.

\section{Conclusions and outlook}\label{sec:Section5_for_Intro}

In this paper, we have computed the pressure of deconfined quark-gluon plasma at high temperature and nonzero quark number density using two variations of resummed perturbation theory. First, we employed hard-thermal-loop perturbation theory (HTLpt) to derive an exact one-loop result, after which we performed a resummation inspired by dimensional reduction (DR) to its existing unresummed four-loop weak coupling expansion. The computation within HTLpt also included some new technical improvements, as we were able to implement the renormalization procedure of HTLpt in a significantly more straightforward and elegant manner than was done in the original works of~\cite{Jens2,Jens3}, as well as to refine the calculation of a couple of the renormalization constants originally obtained in these references. Finally, we also studied the convergence of the $m/T$-expansion within one-loop HTLpt, observing that the truncated results rapidly converge towards the exact unexpanded ones. This gives us confidence in higher-order calculations within HTLpt~\cite{mikehaque,Haque1,haque}, where the $m/T$-expansion has been used.

After obtaining the pressure as a function of temperature and quark chemical potentials, we derived predictions for the second, fourth and sixth order diagonal quark number susceptibilities (QNS) as well as for the chemical potential dependence of the pressure itself. Agreement between the HTLpt and resummed DR results, as well as between them and lattice data, was consistently observed to be good starting at temperatures below 500 MeV with the sole exception of the sixth order QNS, which we studied for two quark flavors. For this quantity, the weak coupling expansion of which starts only at order $g^3$, the one-loop HTLpt and resummed DR predictions were seen to be in clear disagreement. At the moment, no continuum extrapolated lattice data exist for this observable, but the available $N_\tau=4$ data of the Bielefeld-Swansea collaboration~\cite{handsu6u2} appear to favor the one-loop HTLpt prediction for most temperatures. However, given the good agreement between our DR resummation and three-loop HTLpt for the other susceptibilities, we have reasons to believe that the forthcoming three-loop HTLpt~\cite{htlptforth} and our present DR results will be in agreement for this quantity as well.

In summary, our results can be taken as an indication that for temperatures above 250-500 MeV (somewhat dependent on the quantity under study), the behavior of quark number susceptibilities and the chemical potential dependence of the pressure can be accurately described via resummed perturbation theory. This is an important observation for two separate reasons. First, it indicates that at least the fermionic sector of QCD appears to allow a description in terms of weakly interacting quasiparticles not only at asymptotically high energy density, but already in a regime that can be reached in modern colliders. Secondly and perhaps even more importantly, one should recall that the methods employed in our calculation work well even at very large values of the ratio $\mu_f/T$, implying that our results extend to a region of the QCD phase diagram, where no other first principles method is available. We are indeed hopeful that the perturbative results discussed in this paper --- in particular those based on our DR resummation and the recent three-loop HTLpt calculation of~\cite{haque} --- will turn out to be of practical phenomenological use in the eventual analysis of heavy ion data from the FAIR facility of GSI, whose goal is to probe the finite-density regime of the QCD phase diagram.

%%% ACKNOWLEDGMENTS
\acknowledgments
We are grateful to Szabolcs Bors\'{a}nyi, Gergely Endr\H{o}di, Pietro Giudice, Simon Hands, Najmul Haque, Frithjof Karsch, Igor Kondrashuk, Edwin Laermann, Mikko Laine, Munshi Mustafa, P\'eter Petreczky, Anton Rebhan, Kari Rummukainen, Christian Schmidt, Sayantan Sharma, and Mathias Wagner for useful discussions. We would also like to thank Szabolcs Bors\'{a}nyi and Frithjof Karsch for providing us with their latest lattice data. S.M.~and A.V.~were supported by the Sofja Kovalevskaja program and  N.S.~by the Postdoctoral Research Fellowship of the Alexander von Humboldt Foundation. M.S.~was supported in part by the DOE Grant No.~DE-SC0004104.

%%% APPENDICES
\appendix

\section{Notation}\label{sec:Notations}

In the imaginary time formalism of thermal field theory, the four-momentum $K=(K_0,{\bf k})$ is Euclidean, $K^2=K_0^2+{\bf k}^2$. The norm of the spatial component of this momentum is denoted by $k \equiv \left|{\bf k}\right|$, while its temporal component is discrete,
\beqa
K_0&=&\omega_n\;,\hspace{2cm} ({\rm bosons})\;, \\
K_0&=&\tilde{\omega}_n-i\mu_f\;,\hspace{1cm} ({\rm fermions})\;,
\eeqa
where $\omega_n=2n\pi T$ and $\tilde{\omega}_n=(2n+1)\pi T$, with $n$ an integer. Here $\mu_f$ is the quark chemical potential of the flavor $f$. We define the dimensionally regularized sum-integrals by
\beqa
\sumint_K&=&\left(\frac{\lmsb^2\,e^{\gamma_\rmi{\tiny E}}}{4\pi}\right)^{\epsilon}T\sum_{\omega_n} \int\kern-0.5em\frac{\mathop{{\rm d}^{3-2\epsilon}\!}\nolimits {\bf k}}{(2\pi)^{3-2\epsilon}}\;,\\
\sumint_{\{K\}}&=&\left(\frac{\lmsb^2\,e^{\gamma_\rmi{\tiny E}}}{4\pi}\right)^{\epsilon}T\sum_{\tilde{\omega}_n} \int\kern-0.5em\frac{\mathop{{\rm d}^{3-2\epsilon}\!}\nolimits {\bf k}}{(2\pi)^{3-2\epsilon}}\;,
\eeqa
and from here onwards will abbreviate the integral over the three-momenta by
\beqa
\int_{\bf k}&\equiv&\left(\frac{\lmsb^2\,e^{\gamma_\rmi{\tiny E}}}{4\pi}\right)^{\epsilon}\int\kern-0.5em\frac{\mathop{{\rm d}^{3-2\epsilon}\!}\nolimits {\bf k}}{(2\pi)^{3-2\epsilon}}\;.
\eeqa

Throughout our work, we use the modified minimal subtraction scheme ${\overline{\mbox{\rmi{MS}}}}$ of dimensional regularization in $d=3-2\epsilon$ dimensions, where the renormalization scale $\lmsb$ is related to the minimal subtraction ($\mbox{\rmi{MS}}$) one via $\lmsb^2\equiv 4 \pi\Lambda^2 \exp\left(-\gamma_\rmi{\tiny E}\right)$, with $\gamma_\rmi{\tiny E}\approx 0.577216$ the Euler-Mascheroni constant. Finally, the dimensionalities of the fermion and gluon representations of the SU($N_\rmi{c}$) group are denoted by $d_\rmi{F}=N_\rmi{c} N_\rmi{f}$ and $d_\rmi{A}=N_\rmi{c}^2-1$, while some further group theory constants, needed in appendix \ref{sec:DR_matching_para}, read $C_\rmi{A}=N_\rmi{c}$, $C_\rmi{F}=d_\rmi{A}/(2N_\rmi{c})$ as well as $T_\rmi{F}=N_\rmi{f}/2$.

\section{Matching coefficients of EQCD}\label{sec:DR_matching_para}

The matching constants of EQCD, defined in section \ref{sec:DR}, all have analytical expressions as functions of the ratios $\hat \mu_f \equiv \mu_f/(2\pi T)$, derived in ref.~\cite{aleksi1}. They are expressed in terms of derivatives of the generalized Riemann Zeta function, $\zeta'(n,z)\equiv \partial_n \zeta(n,z)$ and the digamma function $\Psi(z)=\Gamma'(z)/\Gamma(z)$, appearing in the combinations
{\allowdisplaybreaks
\beqa
\aleph(n,\mu_1,\mu_2)&\equiv&\aleph(n,\mu_1+\mu_2) \, , \label{Def_aleh_functions_1} \\
\aleph(n,\mu)&\equiv&\zeta'(-n,1/2-i\ \hat \mu)+(-1)^{n+1}\zeta'(-n,1/2+i\ \hat \mu) \, , \label{Def_aleh_functions_2} \\
\aleph(\mu)&\equiv&\Psi(1/2-i\ \hat \mu)+\Psi(1/2+i\ \hat \mu) \; , \label{Def_aleh_functions_3}
\eeqa}
\hspace{-0.12cm}where $n$ is a non-negative integer and $\hat \mu$ is real. The explicit expressions read
{\allowdisplaybreaks
\beqa
\aE{1}&=&\frac{\pi^2}{45\,N_\rmi{f}}\ \sum_f\bigg\{d_\rmi{A}+d_\rmi{F}\bigg(\frac{7}{4}+30\ \hat \mu_f^2 + 60\ \hat \mu_f^4\bigg)\bigg\} \, ,\\
\aE{2}&=&-\ \frac{d_\rmi{A}}{144\,N_\rmi{f}}\ \sum_f\bigg\{C_\rmi{A}+\frac{T_\rmi{F}}{2}\bigg(1+12\ \hat \mu_f^2\bigg)\bigg(5+12\ \hat \mu_f^2\bigg)\bigg\} \, ,\\
\aE{3}&=&\frac{d_\rmi{A}}{144\,N_\rmi{f}}\sum_f\Bigg\{C_\rmi{A}^2\bigg(\frac{194}{3}\log\frac{\bar{\Lambda}}{4\pi T}+\frac{116}{5}+4\gamma_\rmi{\tiny E}-\frac{38}{3}\frac{\zeta'(-3)}{\zeta(-3)}+\frac{220}{3}\frac{\zeta'(-1)}{\zeta(-1)}\bigg) \, \nonumber \\
& & \ \ \ \ \ \ \ \ \ \ \ +C_\rmi{A}\ T_\rmi{F} \bigg[\bigg(\frac{169}{3}+600\ \hat \mu_f^2-528\ \hat \mu_f^4\bigg)\log\frac{\bar{\Lambda}}{4\pi T}+\frac{1121}{60}+8\gamma_\rmi{\tiny E} \, \nonumber \\
& & \ \ \ \ \ \ \ \ \ \ \ +2\,\bigg(127+48\gamma_\rmi{\tiny E}\bigg)\ \hat \mu_f^2+ \frac{4}{3}\bigg(11+156\ \hat \mu_f^2\bigg)\frac{\zeta'(-1)}{\zeta(-1)}-644\ \hat \mu_f^4 \, \nonumber \\
& & \ \ \ \ \ \ \ \ \ \ \ +\frac{268}{15}\frac{\zeta'(-3)}{\zeta(-3)}+24\bigg(52\ \aleph(3,\mu_f)+4\ i\ \hat \mu_f\ \aleph(0,\mu_f) \, \nonumber \\
& & \ \ \ \ \ \ \ \ \ \ \ +144\ i\ \hat \mu_f\ \aleph(2,\mu_f)+\bigg(17-92\ \hat \mu_f^2\bigg)\ \aleph(1,\mu_f)\bigg)\bigg] \ \ \ \ \ \ \ \ \ \ \, \nonumber \\
& & \ \ \ \ \ \ \ \ \ \ \ +C_\rmi{F}\,T_\rmi{F} \bigg[\frac{3}{4}\bigg(1+4\ \hat \mu_f^2\bigg)\bigg(35+332\ \hat \mu_f^2\bigg)-24\bigg(1-12\ \hat \mu_f^2\bigg)\frac{\zeta'(-1)}{\zeta(-1)} \, \nonumber \\
& & \ \ \ \ \ \ \ \ \ \ \ -144\,\bigg(12\ i\ \hat \mu_f\ \aleph(2,\mu_f)-2\bigg(1+8\ \hat \mu_f^2\bigg)\ \aleph(1,\mu_f) \, \nonumber \\
& & \ \ \ \ \ \ \ \ \ \ \ -\ i\ \hat \mu_f\bigg(1+4\ \hat \mu_f^2\bigg)\ \aleph(0,\mu_f)\bigg)\bigg] \, \nonumber \\
& & \ \ \ \ \ \ \ \ \ \ \ + T_\rmi{F}^2 \bigg[\frac{4}{3}\bigg(1+12\ \hat \mu_f^2\bigg)\bigg(5+12\ \hat \mu_f^2\bigg) \log\frac{\bar{\Lambda}}{4\pi T}+\frac{1}{3}+4\,\gamma_\rmi{\tiny E} \, \nonumber \\
& & \ \ \ \ \ \ \ \ \ \ \ +8\,\bigg(7+12\gamma_\rmi{\tiny E}\bigg)\ \hat \mu_f^2+112\ \hat \mu_f^4-\frac{32}{3}\bigg(1+12\ \hat \mu_f^2\bigg)\frac{\zeta'(-1)}{\zeta(-1)} \, \nonumber \\
& & \ \ \ \ \ \ \ \ \ \ \ -\frac{64}{15}\,\frac{\zeta'(-3)}{\zeta(-3)}-96\bigg(8\ \aleph(3,\mu_f) + 12\ i\ \hat \mu_f\ \aleph(2,\mu_f) \, \nonumber \\
& & \ \ \ \ \ \ \ \ \ \ \ - 2\,(1+2\ \hat \mu_f^2)\,\aleph(1,\mu_f)-\ i\ \hat \mu_f\ \aleph(0,\mu_f)\bigg)\bigg] \, \nonumber \\
& & \ \ \ \ \ \ \ \ \ \ \ +\frac{288\,T_\rmi{F}^2}{N_\rmi{f}}\, \sum_g\bigg\{2\ \bigg(1+\gamma_\rmi{\tiny E}\bigg)\ \hat \mu_f^2\ \hat \mu_g^2-\,\aleph(3,\mu_f,\mu_g)-\,\aleph(3,\mu_f,-\mu_g) \, \nonumber \\
& & \ \ \ \ \ \ \ \ \ \ \ \ \ \ \ \ \ \ \ \ \ \ +4\ \hat \mu_g^2\ \aleph(1,\mu_f)-4\ i\ \hat \mu_f\bigg(\aleph(2,\mu_f,\mu_g) + \aleph(2,\mu_f,-\mu_g)\bigg) \, \nonumber \\
& & \ \ \ \ \ \ \ \ \ \ \ \ \ \ \ \ \ \ \ \ \ \ +\bigg(\ \hat \mu_f+\ \hat \mu_g\bigg)^2\aleph(1,\mu_f,\mu_g)+4\ i\ \hat \mu_f\ \hat \mu_g^2\ \aleph(0,\mu_f) \, \nonumber \\
& & \ \ \ \ \ \ \ \ \ \ \ \ \ \ \ \ \ \ \ \ \ \ + \bigg(\ \hat \mu_f-\ \hat \mu_g\bigg)^2\aleph(1,\mu_f,-\mu_g)\bigg\}\Bigg\} \, , \label{AE3_Corrected}\\
\aE{4}&=&\frac{1}{3\,N_\rmi{f}}\ \sum_f\bigg\{C_\rmi{A}+T_\rmi{F}\bigg(1+12\ \hat \mu_f^2\bigg)\bigg\} \, ,\\
\aE{5}&=&\frac{1}{3\,N_\rmi{f}}\ \sum_f\bigg\{2\ C_\rmi{A}\bigg(\log\frac{\bar{\Lambda}}{4\pi T}+\frac{\zeta'(-1)}{\zeta(-1)}\bigg) \, \nonumber \\
& & \ \ \ \ \ \ \ \ \ +T_\rmi{F} \bigg[\bigg(1+12\ \hat \mu_f^2\bigg)\bigg(2\ \log\frac{\bar{\Lambda}}{4\pi T}+1\bigg)+ 24\ \aleph(1,\mu_f)\bigg]\bigg\} \, ,\\
\aE{6}&=&\frac{1}{9\,N_\rmi{f}}\ \sum_f\bigg\{C_\rmi{A}^2\bigg(22\ \log\frac{e^{\gamma_\rmi{\tiny E}}\bar{\Lambda}}{4\pi T}+5\bigg)-18\,C_\rmi{F}\,T_\rmi{F} \bigg(1+12\ \hat \mu_f^2\bigg) \, \nonumber \\
& & \ \ \ \ \ \ \ \ \ + C_\rmi{A}\,T_\rmi{F} \bigg[2\,\bigg(7+132\ \hat \mu_f^2\bigg)\log\frac{e^{\gamma_\rmi{\tiny E}}\bar{\Lambda}}{4\pi T}+9+132\ \hat \mu_f^2+8\gamma_\rmi{\tiny E} + 4\ \aleph(\mu_f)\bigg] \, \nonumber \\
& & \ \ \ \ \ \ \ \ \ -4\,T_\rmi{F}^2\bigg(1+12\ \hat \mu_f^2\bigg)\bigg(2 \log\frac{\bar{\Lambda}}{4\pi T}-1-\aleph(\mu_f)\bigg)\bigg\} \, ,\\
\aE{7}&=&\frac{1}{3\,N_\rmi{f}}\sum_f\bigg\{C_\rmi{A}\bigg(22 \log\frac{e^{\gamma_\rmi{\tiny E}}\bar{\Lambda}}{4\pi T}+1\bigg)-4\ T_\rmi{F}\bigg(2 \log\frac{\bar{\Lambda}}{4\pi T}-\aleph(\mu_f)\bigg)\bigg\}\;.
\eeqa}

To accurately evaluate the pressure for arbitrary values of the chemical potentials $\mu_f$, one needs to use the full expressions of the $\aleph$ functions. If one is interested in calculating the QNS at vanishing chemical potential, it is however more practical to work with expressions already expanded in powers of $\hat \mu_f$. These expressions are given by
{\allowdisplaybreaks
\beqa
\aleph(\mu_i)&=&-2 \bigg(\gamma_\rmi{\tiny E}+\log 4\bigg)+14\ \zeta(3) \ \hat \mu_i^2-62\ \zeta(5) \ \hat \mu_i^4+254\ \zeta(7) \ \hat \mu_i^6 \, \nonumber \\
&-& 1022\ \zeta(9) \ \hat \mu_i^8+4094\ \zeta(11) \ \hat \mu_i^{10}+ {\cal O}\bigg(\hat \mu_i^{12}\bigg) \, ,\\
-i\ \aleph(0,\mu_i)&=&2\ \hat \mu_i \Big(\gamma_\rmi{\tiny E}+\log 4\Big)- \frac{14}{3}\ \zeta(3)\ \hat \mu_i^3+\frac{62}{5}\ \zeta(5)\ \hat \mu_i^5-\frac{254}{7}\ \zeta(7)\ \hat \mu_i^7 \, \nonumber \\
&+& \frac{1022}{9}\ \zeta(9)\ \hat \mu_i^9+ {\cal O}\bigg(\hat \mu_i^{11}\bigg) \, ,\\
\aleph(1,\mu_i)&=&-\zeta'(-1)-\frac{\log 2}{12}+\bigg(\log 4-1+\gamma_\rmi{\tiny E}\bigg)\ \hat \mu_i^2-\frac{7}{6}\ \zeta(3)\ \hat \mu_i^4 \, \nonumber \\
&+& \frac{31}{15}\ \zeta(5)\ \hat \mu_i^6-\frac{127}{28}\ \zeta(7)\ \hat \mu_i^8+\frac{511}{45}\ \zeta(9)\ \hat \mu_i^{10}+ {\cal O}\bigg(\hat \mu_i^{12}\bigg) \, ,\\
-i\ \aleph(2,\mu_i)&=&\frac{1}{12}\Big(1+\log 4+24\ \zeta'(-1)\Big)\ \hat \mu_i-\frac{1}{3}\Big(2\ \gamma_\rmi{\tiny E}-3+\log 16\Big)\ \hat \mu_i^3 \, \nonumber \\
&+& \frac{7}{15}\ \zeta(3)\ \hat \mu_i^5-\frac{62}{105}\ \zeta(5)\ \hat \mu_i^7+\frac{127}{126}\ \zeta(7)\ \hat \mu_i^9+ {\cal O}\bigg(\hat \mu_i^{11}\bigg) \, ,\\
\aleph(3,\mu_i)&=&\frac{1}{480}\Big(\log 2-840\ \zeta'(-3)\Big)+\frac{1}{24}\Big(5+\log 64+72\ \zeta'(-1)\Big)\ \hat \mu_i^2 \, \nonumber \\
&-& \frac{1}{12}\Big(6\ \gamma_\rmi{\tiny E}-11+\log 4096\Big)\ \hat \mu_i^4+\frac{7}{30}\ \zeta(3)\ \hat \mu_i^6-\frac{31}{140}\ \zeta(5)\ \hat \mu_i^8 \, \nonumber \\
&+& \frac{127}{420}\ \zeta(7)\ \hat \mu_i^{10}+{\cal O}\bigg(\hat \mu_i^{12}\bigg) \, ,\\
\aleph(1,\mu_i,\mu_j)&=&2\ \zeta'(-1)+\bigg(\gamma_\rmi{\tiny E}-1\bigg)\ \left(\hat \mu_i+\hat \mu_j\right)^2-\frac{\zeta(3)}{6}\ \left(\hat \mu_i+\hat \mu_j\right)^4 \, \nonumber \\
&+& \frac{\zeta(5)}{15}\ \left(\hat \mu_i+\hat \mu_j\right)^6-\frac{\zeta(7)}{28}\ \left(\hat \mu_i+\hat \mu_j\right)^8 \, \nonumber \\
&+& \frac{\zeta(9)}{45}\ \left(\hat \mu_i+\hat \mu_j\right)^{10}+ {\cal O}\bigg(\hat \mu_i^{12},\hat \mu_j^{12}\bigg) \, ,\\
-i\ \aleph(2,\mu_i,\mu_j)&=&-\bigg(4\ \zeta'(-1)+\frac{1}{6}\bigg)\left(\hat \mu_i+\hat \mu_j\right)+\bigg(1-\frac{2}{3}\gamma_\rmi{\tiny E}\bigg)\left(\hat \mu_i+\hat \mu_j\right)^3 \, \nonumber \\
&+& \frac{\zeta(3)}{15}\ \left(\hat \mu_i+\hat \mu_j\right)^5-\frac{2\ \zeta(5)}{105}\left(\hat \mu_i+\hat \mu_j\right)^7 \, \nonumber \\
&+& \frac{\zeta(7)}{126}\left(\hat \mu_i+\hat \mu_j\right)^9+{\cal O}\bigg(\hat \mu_i^{11},\hat \mu_j^{11}\bigg) \, ,\\
\aleph(3,\mu_i,\mu_j)&=&2\ \zeta'(-3)-\bigg(6\ \zeta'(-1)+\frac{5}{12}\bigg)\left(\hat \mu_i+\hat \mu_j\right)^2 \, \nonumber \\
&+& \bigg(\frac{11}{12}-\frac{\gamma_\rmi{\tiny E}}{2}\bigg)\left(\hat \mu_i+\hat \mu_j\right)^4+\frac{\zeta(3)}{30}\left(\hat \mu_i+\hat \mu_j\right)^6 \, \nonumber \\
&-& \frac{\zeta(5)}{140}\left(\hat \mu_i+\hat \mu_j\right)^8+\frac{\zeta(7)}{420}\left(\hat \mu_i+\hat \mu_j\right)^{10} +{\cal O}\bigg(\hat \mu_i^{12},\hat \mu_j^{12}\bigg) \, . \ \ \ \ \ \ \ 
\eeqa}
\hspace{-0.12cm}Note that all of these results can be found in a Mathematica file named DREoS.nb, which is available at~\cite{mathfile}.

\section{Computing the exact one-loop HTLpt pressure}\label{Computing_Full_HTLpt}

In this section, we will provide details concerning the evaluation of the exact one-loop HTLpt pressure $p_\rmi{HTLpt}$, defined in eq.~(\ref{oneloophtl}). This includes a detailed treatment of the contributions $p_\rmi{T}$, $p_\rmi{L}$, and $p_\rmi{q$_f$}$, respectively given by eqs.~(\ref{Full_Transverse_Gluon}),~(\ref{Full_Longitudinal_Gluon}), and~(\ref{Full_Quarks}).

\subsection{Transverse gluon contribution}

\begin{figure}[!t]
\begin{center}
\includegraphics[scale=0.16]{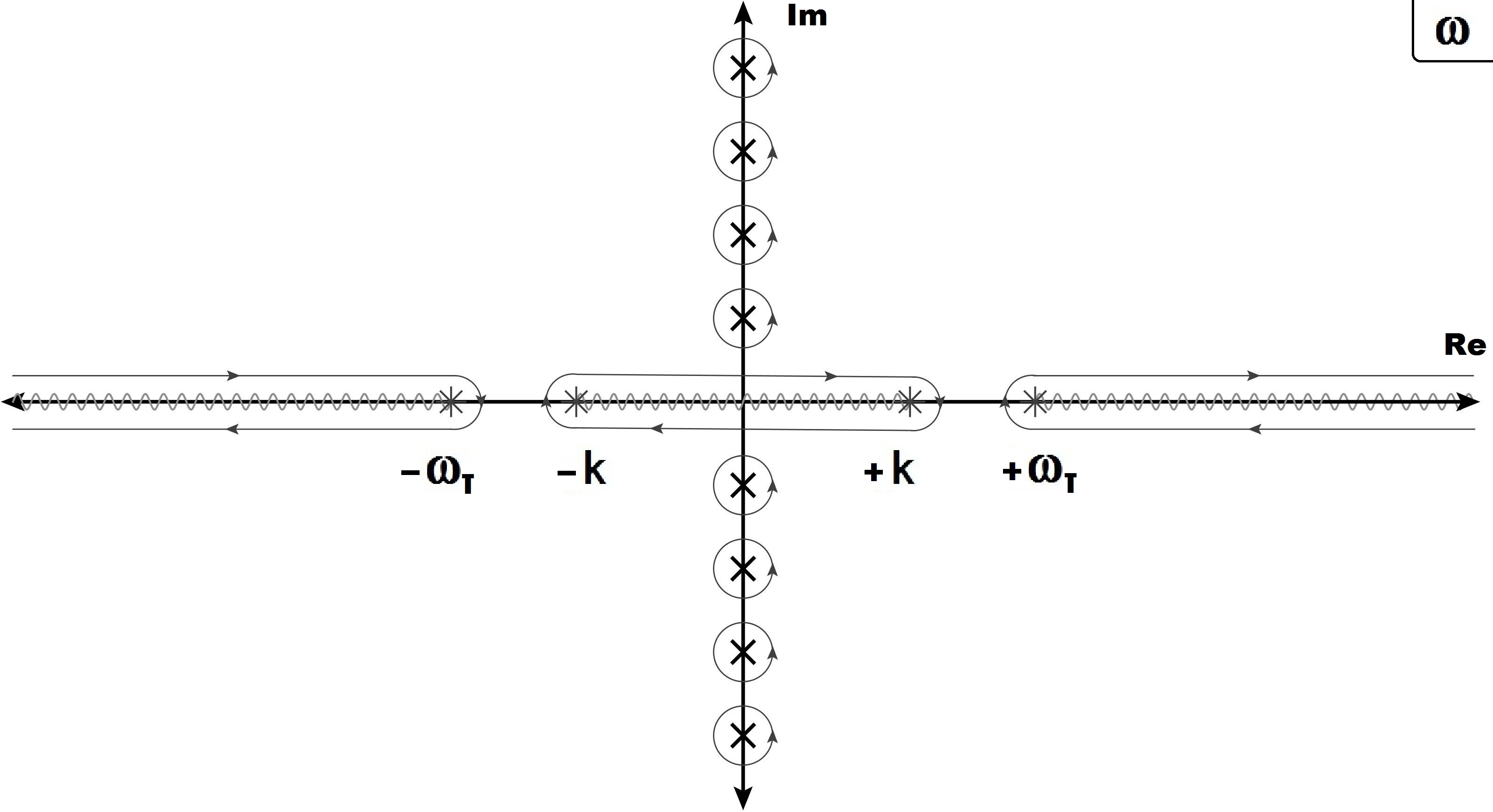}
\caption{The contours $C$ and $C_\rmi{T}$ used to evaluate the transverse gluon contribution of the one-loop HTLpt pressure, as well as the corresponding branch cuts. See main text for details.}
\label{contour}
\end{center}
\end{figure}

Let us begin from the transverse gluon contribution $p_\rmi{T}$, which is given by eq.~(\ref{Full_Transverse_Gluon}). We first rewrite the sum in this expression as
\beqa\label{Each_term_in_sum}
p_\rmi{T}&=&-\frac{1}{2}\sumint_{K}\log(k^2)-\frac{1}{2}T\int_{\bf k}\sum_{n\neq 0}\log\left[\frac{k^2+\omega_n^2+\Pi_\rmi{T}(i\omega_n,k)}{k^2}\right]\;,
\eeqa
where we have used the fact that $\Pi_\rmi{T}(0,k)=0$ to drop the $n=0$ contribution in the second term. The first term on the right-hand side of eq.~(\ref{Each_term_in_sum}) also vanishes as a scale free integral in dimensional regularization. Next, we then use the familiar contour trick to write the sum over Matsubara frequencies as a contour integral in the complex energy plane. This is shown in figure~\ref{contour}, where the contour $C$ encloses the points $\omega=i\omega_n$ with $n\neq0$, and leads to the result
\beqa
p_\rmi{T}&=&-\frac{1}{2}T\,\int_{\bf k}\sum_{n\neq 0}\log\left[\frac{k^2+\omega_n^2+\Pi_\rmi{T}(i\omega_n,k)}{k^2}\right] \nonumber \\
&=&-\frac{1}{4}\,\int_{\bf k}\intec{C}{0.25em}{0em}\log\left[\frac{k^2-\omega^2+\Pi_\rmi{T}(\omega,k)}{k^2}\right]\,\coth\left(\frac{\beta\,\omega}{2}\right) \, .
\label{Sum_to_C}
\eeqa
The integrand in eq.~(\ref{Sum_to_C}) has branch cuts that we choose to run from $-\infty$ to $-\omega_\rmi{T}(k)$ and from $+\omega_\rmi{T}(k)$ to $+\infty$, where $\omega_\rmi{T}(k)$ is the quasiparticle dispersion relation for the transverse gluons. This dispersion relation satisfies
\beq\label{General_Tgluon_dispersion_relation}
k^2-\omega_\rmi{T}^2+\Pi_\rmi{T}(\omega_\rmi{T},k)=0 \;.
\eeq

The integrand also has a remaining branch cut that we choose to run from $-k$ to $+k$, due to the function $\Pi_\rmi{T}(\omega,k)$. The contour $C$ can then be deformed into a contour $C_\rmi{T}$ that wraps around the branch cuts, as shown in figure~\ref{contour}. Thus we can write
\beqa\label{Sum_to_Inte_GluonT}
p_\rmi{T}&=&-\frac{1}{4}\,\int_{\bf k}\intec{C_\rmi{T}}{0.25em}{0em}\log\left[\frac{k^2-\omega^2+\Pi_\rmi{T}(\omega,k)}{k^2}\right]\,\coth\left(\frac{\beta\,\omega}{2}\right) \;.
\eeqa
The contribution from the branch cut that runs from $\omega=-k$ to $\omega=+k$ is identified with the Landau damping part of $p_\rmi{T}$, and reads after the collapse of $C_\rmi{T}$
\beqa
p_{\rmi{T}_\rmi{Ld}}&\equiv&-\int_{\bf k} \! \inteddwbis \! \Bigg\{\!\disc\!\!\arctan\!\!\left[\frac{\frac{m_\rmi{D}^2}{2-2\epsilon}\frac{k^2-\omega^2}{k^2} \ \ImPart{\left\{_2F_1\left(\mbox{$\frac{1}{2},1;\frac{3}{2}-\epsilon;\frac{k^2}{\omega^2}$}\right)\right\}}}{k^2-\omega^2+\frac{m_\rmi{D}^2}{2-2\epsilon} \frac{\omega^2}{k^2} \left[ 1 + \frac{k^2-\omega^2}{\omega^2} \ \RePart{\left\{_2F_1\left(\mbox{$\frac{1}{2},1;\frac{3}{2}-\epsilon;\frac{k^2}{\omega^2}$}\right)\right\}}\right]}\right]\nonumber\\
&& \hspace*{8truecm} \times\frac{1}{2\,\pi}\,\bigg[\frac{1}{e^{\beta\omega}-1}+
\frac{1}{2}\bigg]\Bigg\} \label{General_Tgluon_contribution_after_collapse_Ld}\; ,
\eeqa
where the symbol $\disc$ stands for the discontinuity of the $\arctan$ function accross the positive part of the cut. More precisely, regarding the negative part, we have performed the change of variable $\omega\rightarrow -\omega$, treating the corresponding discontinuity in the same fashion as for the positive part, and used the relation
\beq
\frac{1}{2}\left(\frac{1}{e^{\beta\omega}-1}-\frac{1}{e^{-\beta\omega}-1}\right)=\frac{1}{e^{\beta\omega}-1}+\frac{1}{2} \, ,
\eeq
in such a way that the sum of both positive and negative parts factorizes to eq.~(\ref{General_Tgluon_contribution_after_collapse_Ld}). We give more details on this discontinuity in appendix~\ref{sec:BC_Disc}.

The contribution from the cuts running from $-\infty$ to $-\omega_\rmi{T}(k)$ and from $+\omega_\rmi{T}(k)$ to $+\infty$ is identified with the quasiparticle contribution to $p_\rmi{T}$, and reads  after the collapse of $C_\rmi{T}$
\beqa
p_{\rmi{T}_\rmi{qp}}&\equiv&-\int_{\bf k}\,\,\bigg\{\frac{1}{2}\omega_\rmi{T}(k)+T\,\log\left(1-e^{-\beta\omega_\rmi{T}}\right)\bigg\} \, ,\label{General_Tgluon_contribution_after_collapse_qp} 
\eeqa
where for the time being, we consider the dispersion relation $\omega_\rmi{T}(k)$ in $d=3-2\epsilon$ dimensions.\\*
In order to compute the corresponding discontinuities, we proceeded in a similar way as for the Landau damping contribution. We first made the change of variable $\omega\rightarrow -\omega$, which allowed to treat only the positive part of the cuts. Then, we collapsed the contour onto the branch cut, noticing that the $\log$ function has this time a discontinuity which reduces to the rather simple form of $-2\pi i$ across the branch cut. Finally, we analytically performed the $\omega$-integration, using whenever it is needed a convergence factor for which we refer to ref.~\cite{Jens2}. As a matter of fact, this integration gives a factor $-1$ and therefore compensates the sign of the discontinuity itself.

Finally, the transverse gluon contribution obtains the form
\beqa
\label{General_Tgluon_contribution_after_collapse}
p_\rmi{T}=p_{\rmi{T}_\rmi{qp}}+p_{\rmi{T}_\rmi{Ld}}\;.
\eeqa

\subsection{Longitudinal gluon contribution}

\begin{figure}[!t]
\centering\includegraphics[scale=0.17]{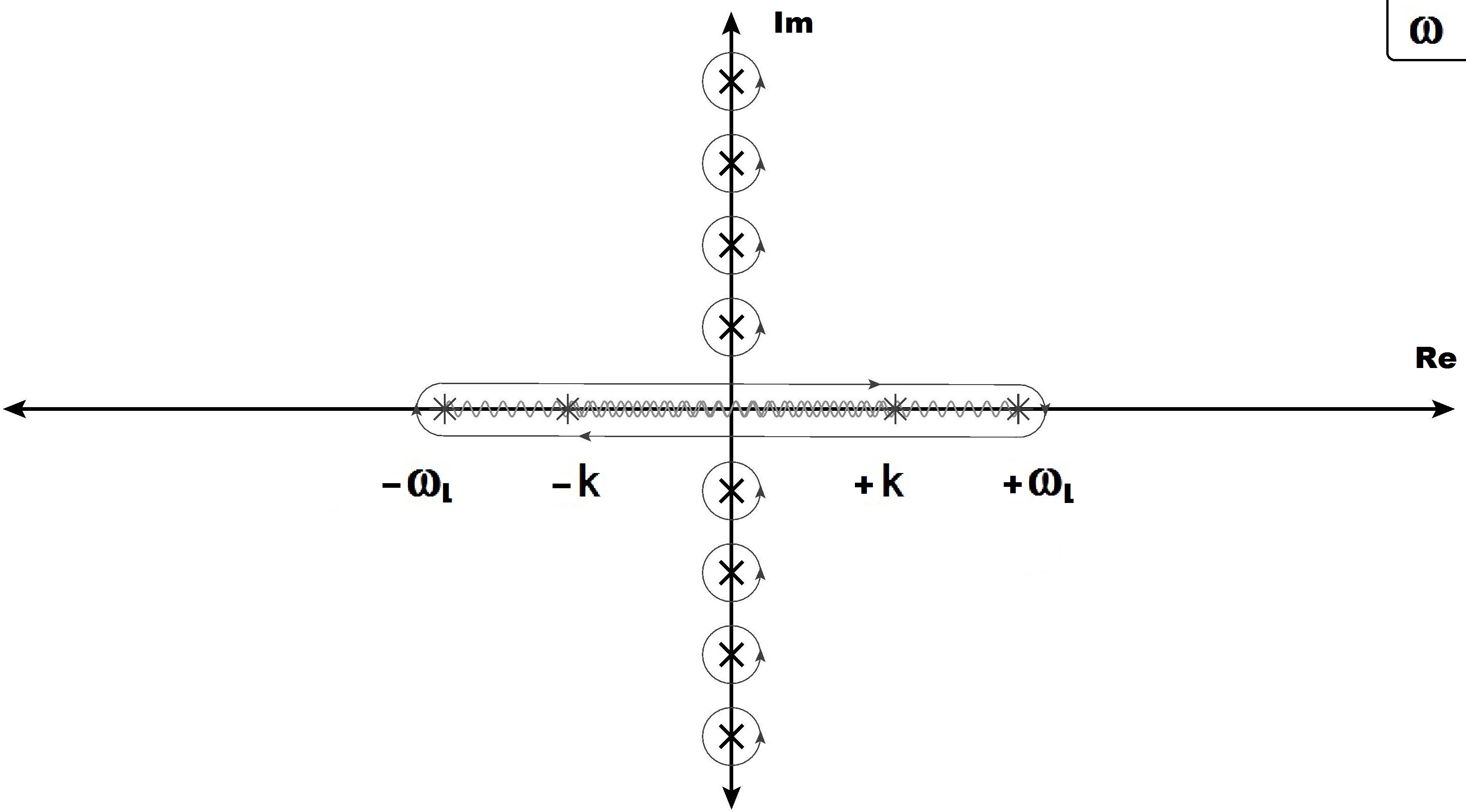}
\caption{\label{fig:wL_Contour}
Contours $C,\,C_\rmi{L}$ and branch cuts of the longitudinal gluon contribution. See main text for details.}
\end{figure}

We next consider the longitudinal part $p_\rmi{L}$ given by eq.~(\ref{Full_Longitudinal_Gluon}). Isolating the contribution from the $n=0$ mode, we obtain
\beqa\label{Each_term_in_sum_2}
p_\rmi{L}&=&-\frac{1}{2}T\int_{\bf k}\,\sum_{n\neq 0}\log\left[k^2+\Pi_\rmi{L}(i\omega_n,k)\right] - \frac{1}{2}T\int_{\bf k}\,\log\left[k^2+m_\rmi{D}^2\right]\;.
\eeqa
Using the residue theorem, we rewrite these two terms as a contour integral that encircles the points $\omega=i\omega_n$ with $n\neq0$. This is shown in figure~\ref{fig:wL_Contour}, and the result is
\beqa\label{Contour_Inte_GluonL}
p_\rmi{L}&=&-\frac{1}{4}\int_{\bf k}\,\intec{C}{0.25em}{0em}\log\left[\frac{k^2+\Pi_\rmi{L}(\omega,k)}{k^2+m_\rmi{D}^2}\right]\,\coth\left(\frac{\beta\,\omega}{2}\right)\;.
\eeqa
The integrand has branch cuts that we choose to run from $-\omega_\rmi{L}(k)$ to $+\omega_\rmi{L}(k)$, where $\omega_\rmi{L}(k)$ is the quasiparticle dispersion relation for the longitudinal gluons. This dispersion relation satisfies the equation
\beq\label{General_Lgluon_dispersion_relation}
0\equiv k^2+\Pi_\rmi{L}(\omega_\rmi{L},k)\;.
\eeq
The integrand also has another cut chosen to run from $-k$ to $+k$, due to the function $\Pi_\rmi{L}(\omega,k)$. The contour $C$ can then be deformed into a contour $C_\rmi{L}$ that wraps around the branch cuts, as shown in figure~\ref{fig:wL_Contour}.

The contributions from the cuts are identified with the Landau damping part and the quasiparticle part, respectively. In analogy with the transverse gluon contribution, one finds
\beq\label{General_Lgluon_contribution_after_collapse}
p_\rmi{L}\equiv p_{\rmi{L}_\rmi{qp}}+p_{\rmi{L}_\rmi{Ld}} \, ,
\eeq
with
\beqa
p_{\rmi{L}_\rmi{qp}}&=&-\int_{\bf k}\,\,\bigg\{\frac{1}{2}\left(\omega_\rmi{L}(k)-k\right)+T\,\log\left(\frac{1-e^{-\beta\omega_\rmi{L}}}{1-e^{-\beta k}}\right)\bigg\} \, ,\label{General_Lgluon_contribution_after_collapse_qp}
\eeqa
{\allowdisplaybreaks
and
\beqa
p_{\rmi{L}_\rmi{Ld}}&=&\int_{\bf k}\,\inteddwbis\left\{\disc\arctan\left[\frac{m_\rmi{D}^2 \ \ImPart{\left\{_2F_1\left(\mbox{$\frac{1}{2},1;\frac{3}{2}-\epsilon;\frac{k^2}{\omega^2}$}\right)\right\}}}{k^2 + m_\rmi{D}^2 - m_\rmi{D}^2 \ \RePart{\left\{_2F_1\left(\mbox{$\frac{1}{2},1;\frac{3}{2}-\epsilon;\frac{k^2}{\omega^2}$}\right)\right\}}}\right]\right.\nonumber\\
&& \hspace*{8truecm} \,\times\left.\frac{1}{2\,\pi}\bigg[\frac{1}{e^{\beta\omega}-1}+\frac{1}{2}\bigg]\Bigg.\Bigg.\setlength{\delimitershortfall}{-5pt}\right\} \label{General_Lgluon_contribution_after_collapse_Ld}\, , \ \ \ \ \ \ \ \ \ \ 
\eeqa}
\hspace{-0.12cm}where both discontinuities are actually of the same type as for $p_{\rmi{T}_\rmi{Ld}}$, and are therefore handled in a similar manner. Considering the Landau damping discontinuity, we also give details in appendix~\ref{sec:BC_Disc}. The dispersion relation is again considered in $d=3-2\epsilon$ dimensions.

\subsection{Quark contribution}

\begin{figure}[!t]\centering\includegraphics[scale=0.17]{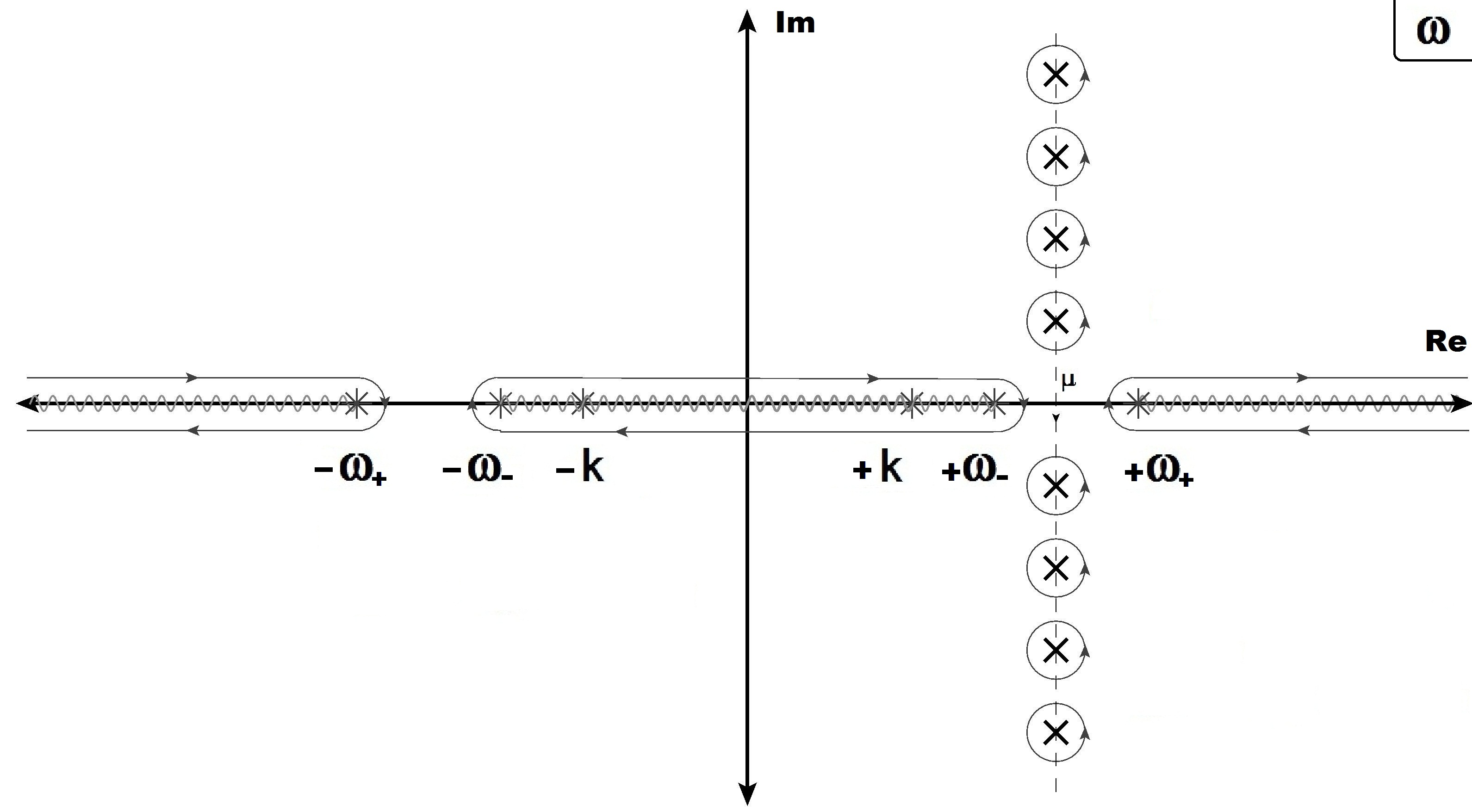}
\caption{\label{fig:wq_Contour}
Contours $C,\,C_\rmi{q}$ and branch cuts of the quark contribution. See main text for details.}
\end{figure}

Finally, let us consider the quark contribution which is given by eq.~(\ref{Full_Quarks}). As usual, by means of the contour trick, adding and subtracting the contribution to the pressure from an ideal gas of massless quarks leads to
{\allowdisplaybreaks
\beqa
p_\rmi{q$_f$}&=&2\,\sumint_{\{K\}}\log\Big[k^2+(\widetilde{\omega}_n-i\mu_f)^2\Big] \nonumber \\
&& \ \ \ \ \ +2\int_{\bf k}\,T\,\sum_{n}\log\left[\frac{A_\rmi{S}^2(i\widetilde{\omega}_n+\mu_f,k)-A_\rmi{0}^2(i\widetilde{\omega}_n+\mu_f,k)}{k^2-(i\widetilde{\omega}_n+\mu_f)^2}\right]\; \nonumber \\
&=&\frac{\pi^2 T^4}{45}\left(\frac{7}{4}+30\hat \mu_f^2+60\hat \mu_f^4\right) \nonumber \\
&& \ \ \ \ \ +\int_{\bf k}\,\intec{C}{0.25em}{0em}\log\left[\frac{A_\rmi{S}^2(\omega,k)-A_\rmi{0}^2(\omega,k)}{k^2-\omega^2}\right]\,\tanh\left(\frac{\beta\,(\omega-\mu_f)}{2}\right) \, , \label{Contour_Inte_Quark}
\eeqa}
\hspace{-0.12cm}where the contour $C$ is shown in figure~\ref{fig:wq_Contour}, and the flavor index f runs from 1 to $N_\rmi{f}$. The integrand has cuts starting at $\pm\omega_\rmi{\tiny $f_\pm$}(k)$, where $\omega_\rmi{\tiny $f_\pm$}(k)$ are the quasiparticle dispersion relations for quarks and plasminos, satisfying
\beq\label{General_Qarks_dispersion_relation}
0\equiv A_\rmi{0}(\omega_\rmi{\tiny $f_\pm$},k)\mp A_\rmi{S}(\omega_\rmi{\tiny $f_\pm$},k)\;.
\eeq
The integrand has another set of cuts starting at $\pm k$. We choose the cuts to run from the branch points as shown in figure~\ref{fig:wq_Contour}.

The contour is finally deformed to wrap around the cuts and gives $C_\rmi{q}$. After collapsing it onto the branch cuts, the quark contribution to the pressure can be written as 
\beq\label{General_Quark_contribution_after_collapse}
p_\rmi{q$_f$}\equiv 
p_{\rmi{q}_{\rmi{qp}_f}}+p_{\rmi{q}_{\rmi{Ld}_f}} \, ,
\eeq
with
{\allowdisplaybreaks
\beqa
p_{\rmi{q}_{\rmi{qp}_f}}&=&2\int_{\bf k}\,\,\bigg\{T\,\log\left(1+e^{-\beta\left(\omega_{f_+}+\mu_f\right)}\right)+T\,\log\left(1+e^{-\beta\left(\omega_{f_+}-\mu_f\right)}\right) \nonumber \\
& & \ \ \ \ \ \ \ \ \ +T\,\log\left(\frac{1+e^{-\beta\left(\omega_{f_-}+\mu_f\right)}}{1+e^{-\beta (k+\mu_f)}}\right)+T\,\log\left(\frac{1+e^{-\beta\left(\omega_{f_-}-\mu_f\right)}}{1+e^{-\beta (k-\mu_f)}}\right)\bigg\} \nonumber \\
&+&2\int_{\bf k}\,\,\bigg\{\omega_{f_+}(k)+\left(\omega_{f_-}(k)-k\right)\bigg\} \, ,
\label{General_Quark_contribution_after_collapse_qp}
\eeqa}
\hspace{-0.12cm}and
\beqa
p_{\rmi{q}_{\rmi{Ld}_f}}&=&\frac{1}{\pi}\int_{\bf k}\,\inteddwbis\left\{\disc\theta_\rmi{q$_f$,$\epsilon$}\,\bigg[\frac{1}{e^{\beta(\omega+\mu_f)}+1}+\frac{1}{e^{\beta(\omega-\mu_f)}+1}-1\bigg]\right\}\label{General_Quark_contribution_after_collapse_Ld}\;,
\eeqa
where $\theta_\rmi{q$_f$,$\epsilon$}\equiv\arctan\left[\Xi_\rmi{$f$,$\epsilon$}\right]$, with its argument defined as
\beqa
&&\Xi_\rmi{$f$,$\epsilon$}\equiv \\
&&\frac{\frac{m_{q_{\mbox{\tiny$f$}}}^4}{k^2} \left[2 \ \ImPart{\left\{_2F_1\left(\mbox{$\frac{1}{2},1;\frac{3}{2}-\epsilon;\frac{k^2}{\omega^2}$}\right)\right\}}+\frac{k^2-\omega^2}{\omega^2} \ \ImPart{\left\{_2F_1\left(\mbox{$\frac{1}{2},1;\frac{3}{2}-\epsilon;\frac{k^2}{\omega^2}$}\right)^2\right\}}\right]}{k^2-\omega^2+2 m_{q_{\mbox{\tiny$f$}}}^2+\frac{m_{q_{\mbox{\tiny$f$}}}^4}{k^2}\,\left[1-2\,\RePart{\left\{_2F_1\left(\mbox{$\frac{1}{2},1;\frac{3}{2}-\epsilon;\frac{k^2}{\omega^2}$}\right)\right\}}-\frac{k^2-\omega^2}{\omega^2}\,\RePart{\left\{_2F_1\left(\mbox{$\frac{1}{2},1;\frac{3}{2}-\epsilon;\frac{k^2}{\omega^2}$}\right)^2\right\}}\right]}\;. \nonumber\ \ \ \ \ 
\label{Xi}
\eeqa
The subscript $\epsilon$ stands to remind that these quantities are considered in $d=3-2\epsilon$ dimensions, as for the dispersion relations. Again, both discontinuities are of the same type as for $p_{\rmi{T}_\rmi{Ld}}$ and $p_{\rmi{L}_\rmi{Ld}}$, and have to be handled in a similar manner. We refer to appendix~\ref{sec:BC_Disc} for more details about the Landau damping discontinuity.

\subsection{Renormalization}

The expressions for $p_\rmi{T}$, $p_\rmi{L}$, and $p_\rmi{q$_f$}$ given by eqs.~(\ref{General_Tgluon_contribution_after_collapse_Ld}), (\ref{General_Tgluon_contribution_after_collapse_qp}), (\ref{General_Lgluon_contribution_after_collapse_qp}), (\ref{General_Lgluon_contribution_after_collapse_Ld}), (\ref{General_Quark_contribution_after_collapse_qp}), and~(\ref{General_Quark_contribution_after_collapse_Ld}) are UV divergent and are regulated by using dimensional regularization in $d=3-2\epsilon$ dimensions. Some of the divergences are explicitly independent of temperature and chemical potentials, while others depend on the temperature and the chemical potentials via the Bose-Einstein and Fermi-Dirac distribution functions. The strategy will be to first calculate the divergences of the temperature and chemical potential dependent terms, and then calculate the terms that are explicitly independent of $(T,\,{\boldsymbol \mu})$, by taking the limits $T\rightarrow0$ and ${\boldsymbol \mu}\rightarrow 0$.

We first consider the second term in eqs.~(\ref{General_Lgluon_contribution_after_collapse_qp}). This term is finite and we therefore just keep it as it stands. The next term is the temperature dependent piece of the Landau damping contribution in eq.~(\ref{General_Lgluon_contribution_after_collapse_Ld}), and reads
\beqa
&&\frac{1}{2\,\pi}\int_{\bf k}\,\inteddwbis\left\{\disc\arctan\left[\frac{m_\rmi{D}^2 \ \ImPart{\left\{_2F_1\left(\mbox{$\frac{1}{2},1;\frac{3}{2}-\epsilon;\frac{k^2}{\omega^2}$}\right)\right\}}}{k^2 + m_\rmi{D}^2 - m_\rmi{D}^2 \ \RePart{\left\{_2F_1\left(\mbox{$\frac{1}{2},1;\frac{3}{2}-\epsilon;\frac{k^2}{\omega^2}$}\right)\right\}}}\right]\,\,\frac{1}{e^{\beta\omega}-1}\Bigg.\Bigg.\setlength{\delimitershortfall}{-5pt}\right\}\;.\label{ld222}\ \ \ \ \ \ \ \ \ \ 
\eeqa
In ref.~\cite{Jens2}, this term was analyzed with the integrand directly in $d=3$ dimensions, and it was shown that UV divergences appear for $k\rightarrow\infty$ with $\omega$ fixed. A similar analysis can be carried out here (i.e. in $d=3-2\epsilon$ dimensions), by noticing that
\beqa
\label{expf2}
\disc \ImPart{\left\{_2F_1\left(\mbox{$\frac{1}{2},1;\frac{3}{2}-\epsilon;\frac{k^2}{\omega^2}$}\right)\right\}} &\underset{\underset{\mbox{\tiny $\omega$ fixed}}{k\longrightarrow \infty}}{=}& -\,4\,\, \frac{\Gamma\left(\frac{3}{2}-\epsilon\right)\,\Gamma\left(\frac{3}{2}\right)}{\Gamma\left(1-\epsilon\right)}\,\frac{\omega}{k}+{\cal O}\left(\frac{\omega^2}{k^2}\right) \; .
\eeqa
%{\stackrel{k\longrightarrow \infty}{\stackrel{\mbox{$=$}}{\mbox{\tiny $\omega$ fixed}}}}
%\underset{\underset{\mbox{\tiny $\omega$ fixed}}{k\longrightarrow \infty}}{=}
We refer to appendix~\ref{sec:BC_Disc} for details on the computation of the branch cut discontinuity of the imaginary part of this hypergeometric function, in its assymptotic expansion.

By Taylor expanding the integrand in eq.~(\ref{ld222}) around $k=+\infty$, using eq.~(\ref{expf2}), we obtain an integral which contains the relevant UV divergences in three dimensions, and reads
\beqa
\mathcal{D}_{\rmi{L}_\rmi{Ld}}&\equiv& -\frac{2\,m_\rmi{D}^2}{\pi}\,\,\frac{\Gamma\left(\frac{3}{2}-\epsilon\right)\Gamma\left(\frac{3}{2}\right)}{\Gamma\left(1-\epsilon\right)}\int_{\bf k}\,\inteddwbis\left\{\frac{\omega}{k^3}\,\,\frac{1}{e^{\beta\omega}-1}\Bigg.\Bigg.\setlength{\delimitershortfall}{-5pt}\right\}\;.
\eeqa
This integral can be easily evaluated, and the result through ${\cal O}\left(\epsilon^{0}\right)$ is
\beqa
\label{ldiv}
\mathcal{D}_{\rmi{L}_\rmi{Ld}}&=& \frac{m_\rmi{D}^2\,T^2}{48}\,\left[\frac{1}{\epsilon}+2\log\left(\frac{\Lambda}{2\pi T}\right)+2\,\,\frac{\zeta^{\prime}\left(-1\right)}{\zeta\left(-1\right)}+{\cal O}\left(\epsilon\right)\right]\;.
\eeqa
Similarly, the temperature dependent term from the transverse gluon contribution~(\ref{General_Tgluon_contribution_after_collapse_Ld}) is
\beqa
&&-\frac{1}{2\,\pi}\int_{\bf k}\,\inteddwbis\left\{\disc\arctan\left[\frac{\frac{m_\rmi{D}^2}{2-2\epsilon}\frac{k^2-\omega^2}{k^2} \ \ImPart{\left\{_2F_1\left(\mbox{$\frac{1}{2},1;\frac{3}{2}-\epsilon;\frac{k^2}{\omega^2}$}\right)\right\}}}{k^2-\omega^2+\frac{m_\rmi{D}^2}{2-2\epsilon} \frac{\omega^2}{k^2}\left[ 1 + \frac{k^2-\omega^2}{\omega^2} \ \RePart{\left\{_2F_1
\left(\mbox{$\frac{1}{2},1;\frac{3}{2}-\epsilon;\frac{k^2}{\omega^2}$}\right)\right\}}\right]}\right]\right.\nonumber\\
&& \hspace*{8truecm} \,\left.\times\bigg[\frac{1}{e^{\beta\omega}-1}\bigg]\Bigg.\Bigg.\setlength{\delimitershortfall}{-5pt}\right\}\;.
\eeqa
This leads to the integral
\beqa
\mathcal{D}_{\rmi{T}_\rmi{Ld}}&\equiv& \frac{m_\rmi{D}^2}{\pi}\,\,\frac{\Gamma\left(\frac{3}{2}-\epsilon\right)\Gamma\left(\frac{3}{2}\right)}{(1-\epsilon)\,\Gamma\left(1-\epsilon\right)}\int_{\bf k}\,\inteddwbis\left\{\frac{\omega}{k^3}\,\,\frac{1}{e^{\beta\omega}-1}\Bigg.\Bigg.\setlength{\delimitershortfall}{-5pt}\right\}\,,
\eeqa
which through ${\cal O}\left(\epsilon^{0}\right)$ reads
\beqa
\label{tdiv}
\mathcal{D}_{\rmi{T}_\rmi{Ld}}&=& -\frac{m_\rmi{D}^2\,T^2}{96}\,\left[\frac{1}{\epsilon}+1+2\log\left(\frac{\Lambda}{2\pi\,T}\right)+2\,\,\frac{\zeta^{\prime}\left(-1\right)}{\zeta\left(-1\right)}+{\cal O}\left(\epsilon\right)\right] \;.
\eeqa
We notice that the divergences in eqs.~(\ref{ldiv}) and~(\ref{tdiv}) cancel in the combination
\beq
d_\rmi{A} \Big[(2-2\epsilon)\ \mathcal{D}_{\rmi{T}_\rmi{Ld}} + \mathcal{D}_{\rmi{L}_\rmi{Ld}} \Big] ={\cal O}\left(\epsilon\right) \;,
\eeq
which is actually the one given by eq.~(\ref{oneloophtl}). Hence, there is no need for adding and subtracting an integral for the Landau damping divergent pieces in $p_\rmi{T}$ and $p_\rmi{L}$, in order to renormalize the result for the gluon Landau damping contribution.

Let us finally discuss the quark terms given by eqs.~(\ref{General_Quark_contribution_after_collapse_qp}) and~(\ref{General_Quark_contribution_after_collapse_Ld}). In eq.~(\ref{General_Quark_contribution_after_collapse_qp}), the only divergent term involves the dispersion relation $\omega_{f_+}(k)$ and is independent of $T$ and $\bm{\mu}$. Thus, we will deal with it later. In eq.~(\ref{General_Quark_contribution_after_collapse_Ld}), there are no divergent terms, even when $k\rightarrow\infty$ with fixed $\omega$, since the expansion of eq.~(\ref{Xi}) starts to contribute at ${\cal O}({m_{q_f}^2{\omega}/k^3})$ instead of ${\cal O}({\omega/k})$.

The fact that all the temperature dependent divergences cancel, and that there is no divergence dependent on the quark chemical potentials, allows us to write the one-loop pressure as follows
{\allowdisplaybreaks
\beqa\label{Pre_result_full_HTLpt}
p_\rmi{HTLpt}\left(T,\bm{\mu}\right)&=& d_\rmi{A}\Bigg\{-(2-2\epsilon)\,T\int_{\bf k}\log\bigg(1-e^{-\beta\omega_\rmi{T}}\bigg)-\,T\int_{\bf k}\log\bigg(\frac{1-e^{-\beta\omega_\rmi{L}}}{1-e^{-\beta\,k}}\bigg) + \ p^{\star}_\rmi{L}\nonumber\\
&&\ \ \ \ +\ (2-2\epsilon)\, p^{\star}_\rmi{T} -\frac{1}{2\,\pi}\int_{\bf k}\,\inteddwbis \bigg[(2-2\epsilon)\,\disc\phi_{\rmi{T},\epsilon}-\disc\phi_{\rmi{L},\epsilon}\bigg]\,\,\frac{1}{e^{\beta\omega}-1}\Bigg\} \nonumber\\
&+&N_\rmi{c}\sum_{f,\,s=\pm 1}\Bigg\{2\,T\int_{\bf k}\log\bigg[1+e^{-\beta\left(\omega_{f_+}+s\,\mu_f\right)}\bigg]+2\,T\int_{\bf k}\log\left[\frac{1+e^{-\beta\left(\omega_{f_-}+s\,\mu_f\right)}}{1-e^{-\beta\left(k+s\,\mu_f\right)}}\right] \nonumber\\
&&\ \ \ \ \ \ \ \ \ + \ \frac{p^{\star}_\rmi{q$_f$}}{2} + \frac{1}{\pi}\,\int_{\bf k}\,\inteddwbis\,\,\disc\theta_\rmi{q$_f$,$\epsilon$}\left[\frac{1}{{e^{\beta\left(\omega+s\,\mu_f\right)}+1}}\right]\Bigg\} + \Delta p\;,
\eeqa}
\hspace{-0.12cm}where the subscript $\epsilon$ in $\phi_{\rmi{T},\epsilon}$ and $\phi_{\rmi{L},\epsilon}$ serves to remind us that these angles, respectively corresponding to the $\arctan$ functions in eqs.~(\ref{General_Tgluon_contribution_after_collapse_Ld}) and (\ref{General_Lgluon_contribution_after_collapse_Ld}), are considered in $d=3-2\epsilon$ dimensions. Note also that the dispersion relations are still in $d=3-2\epsilon$ dimensions.

The terms $p^{\star}_\rmi{T}$, $p^{\star}_\rmi{L}$, and $p^{\star}_\rmi{q$_f$}$ are the terms from $p_\rmi{T}$, $p_\rmi{L}$, and $p_\rmi{q$_f$}$ that are explicitly independent of $T$ and ${\boldsymbol \mu}$. They contain all the UV divergences. Moreover, $p^{\star}_\rmi{L}$ also contains the finite contribution coming from the integral over $(\omega_\rmi{L}(k)-k)/2$, and $p^{\star}_\rmi{q$_f$}$ contains the finite contribution coming from the integral over $(\omega_{f_-}(k)-k)$, since both are explicitly temperature and chemical potential independent. We need therefore to compute them.

In the limit $T\rightarrow0$, the transverse gluon term $p_\rmi{T}$ approaches $p^{\star}_\rmi{T}$ and can be expressed as an integral over continuous Euclidean energy, which we denote $\omega_\rmi{E}$. We find
\beq
p^{\star}_\rmi{T}=-\frac{1}{4\,\pi}\inteddwters{-\infty}{+\infty}\int_{\bf k}\,\log\Big[k^2+\omega_\rmi{E}^2+\Pi_\rmi{T}(i\omega_\rmi{E},k)\Big] \, .
\eeq
It is then convenient to rescale the Euclidean energy as $\omega_\rmi{E}\rightarrow k\,\omega_\rmi{E}$. This yields
\beq
p^{\star}_\rmi{T}=-\frac{1}{2\,\pi}\inteddwters{0}{\infty}\int_{\bf k}\,
k\log\Big[(1+\omega_\rmi{E}^2)\,k^2+\Pi_\rmi{T}(i\omega_\rmi{E},1)\Big] \, .
\eeq
Similarly, the rescaled expressions for the longitudinal gluon contribution~(\ref{Full_Longitudinal_Gluon}) and the quark contribution~(\ref{Full_Quarks}), taking the limit ${\boldsymbol \mu}\rightarrow0$, are
{\allowdisplaybreaks
\beqa
p^{\star}_\rmi{L}&=&-\frac{1}{2\,\pi}\inteddwters{0}{\infty}\int_{\bf k}\,k\,\log\Big[k^2+\Pi_\rmi{L}(i\omega_\rmi{E},1)\Big] \, , \\
p^{\star}_\rmi{q$_f$}&=&\frac{2}{\pi}\inteddwters{0}{\infty}\int_{\bf k}\,k\,\log\left[\left(1+\frac{m_\rmi{q$_f$}^2}{k^2}\,\left\{\frac{1}{1+i\,\omega_\rmi{E}}-\frac{\widetilde{{\cal T}}_\rmi{K}(i\omega_\rmi{E},1)}{i\,\omega_\rmi{E}}\right\}\right)\right. \nonumber\\
&& \ \ \ \ \ \ \ \ \ \ \ \ \ \ \ \ \ \ \ \ \ \ \ \ \ \ \ \times\left.\left(1+\frac{m_\rmi{q$_f$}^2}{k^2}\,\left\{\frac{1}{1-i\,\omega_\rmi{E}}+\frac{\widetilde{{\cal T}}_\rmi{K}(i\omega_\rmi{E},1)}{i\,\omega_\rmi{E}}\right\}\right)\right] \, . \label{quark_zero_T_and_mu}
\eeqa}

For the latter, notice that before rescaling, we added and subtracted a $\log \left(k^2+\omega_\rmi{E}^2\right)$ term to the main logarithm in eq.~(\ref{quark_zero_T_and_mu}). The added piece is computed directly and vanishes thanks to dimensional regularization. The subtracted one is combined with the main $\log$ for convenience during manipulations of its argument. Integrating over $k$, we obtain
{\allowdisplaybreaks
\beqa
p^{\star}_\rmi{T}&=&\frac{e^{\gamma_\rmi{\tiny E}\epsilon}\,\lmsb^{2\epsilon}}{16\,\pi^{5/2}}\,\,\frac{\Gamma\left(2-\epsilon\right)\Gamma\left(\epsilon-2\right)}{\Gamma\left(\frac{3}{2}-\epsilon\right)}\inteddwters{0}{\infty}\,\left(\frac{\Pi_\rmi{T}(i\omega_\rmi{E},1)}{1+\omega_\rmi{E}^2}\right)^{2-\epsilon} \, , \\
p^{\star}_\rmi{L}&=&\frac{e^{\gamma_\rmi{\tiny E}\epsilon}\,\lmsb^{2\epsilon}}{16\,\pi^{5/2}}\,\,\frac{\Gamma\left(2-\epsilon\right)\Gamma\left(\epsilon-2\right)}{\Gamma\left(\frac{3}{2}-\epsilon\right)}\inteddwters{0}{\infty}\,\bigg(\Pi_\rmi{L}(i\omega_\rmi{E},1)\bigg)^{2-\epsilon} \, , \\
p^{\star}_\rmi{q$_f$}&=&-\,m_\rmi{q$_f$}^{4-2\epsilon}\,\frac{e^{\gamma_\rmi{\tiny E}\epsilon}\,\lmsb^{2\epsilon}}{4\,\pi^{5/2}}\,\,\frac{\Gamma\left(2-\epsilon\right)\Gamma\left(\epsilon-2\right)}{\Gamma\left(\frac{3}{2}-\epsilon\right)} \nonumber \\
&\times&\ \inteddwters{0}{\infty}\,\Bigg[\left(\frac{1}{1+i\,\omega_\rmi{E}}-\frac{\widetilde{{\cal T}}_\rmi{K}(i\omega_\rmi{E},1)}{i\,\omega_\rmi{E}}\right)^{2-\epsilon}+\,\left(\frac{1}{1-i\,\omega_\rmi{E}}+\frac{\widetilde{{\cal T}}_\rmi{K}(i\omega_\rmi{E},1)}{i\,\omega_\rmi{E}}\right)^{2-\epsilon}\Bigg]\; . \ \ \ \ 
\eeqa}

At this point, we recall that we are only interested in calculating the poles in $\epsilon$ analytically. Each integral above is multiplied by a factor $\Gamma(\epsilon-2)$, which has a simple pole in $\epsilon$, and since the integrals are finite in $d=3$, it is sufficient to expand the integrand to order $\epsilon$ only. This yields
{\allowdisplaybreaks
\beqa
&&\left(\frac{\Pi_\rmi{T}(i\omega_\rmi{E},1)}{1+\omega_\rmi{E}^2}\right)^{2-\epsilon}=\,\left(\frac{\Pi_\rmi{T}^{{}^{\mbox{\tiny $(0)$}}}(i\omega_\rmi{E},1)}{1+\omega_\rmi{E}^2}\right)^{2}-\left(\frac{\Pi_\rmi{T}^{{}^{\mbox{\tiny $(0)$}}}(i\omega_\rmi{E},1)}{1+\omega_\rmi{E}^2}\right)^{2} \\
&&\hspace*{6.5truecm}\times\left\{\log\bigg(\frac{\Pi_\rmi{T}^{{}^{\mbox{\tiny $(0)$}}}(i\omega_\rmi{E},1)}{1+\omega_\rmi{E}^2}\bigg)-\frac{2\,\Pi_\rmi{T}^{{}^{\mbox{\tiny $(1)$}}}(i\omega_\rmi{E},1)}{\Pi_\rmi{T}^{{}^{\mbox{\tiny $(0)$}}}(i\omega_\rmi{E},1)}\right\}\,\epsilon \, , \nonumber \\
&&\bigg(\Pi_\rmi{L}(i\omega_\rmi{E},1)\bigg)^{2-\epsilon}=\,\bigg(\Pi_\rmi{L}^{{}^{\mbox{\tiny $(0)$}}}(i\omega_\rmi{E},1)\bigg)^{2}-\bigg(\Pi_\rmi{L}^{{}^{\mbox{\tiny $(0)$}}}(i\omega_\rmi{E},1)\bigg)^{2} \\
&&\hspace*{6.25truecm}\times\left\{\log\bigg(\Pi_\rmi{L}^{{}^{\mbox{\tiny $(0)$}}}(i\omega_\rmi{E},1)\bigg)-\frac{2\,\Pi_\rmi{L}^{{}^{\mbox{\tiny $(1)$}}}(i\omega_\rmi{E},1)}{\Pi_\rmi{L}^{{}^{\mbox{\tiny $(0)$}}}(i\omega_\rmi{E},1)}\right\}\,\epsilon \, , \nonumber \\
&&\left(\frac{1}{1\pm i\,\omega_\rmi{E}}\mp\frac{\widetilde{{\cal T}}_\rmi{K}(i\omega_\rmi{E},1)}{i\,\omega_\rmi{E}}\right)^{2-\epsilon}=\left(\frac{1}{1\pm i\,\omega_\rmi{E}}\pm\frac{i\,\widetilde{{\cal T}}^{{}^{\mbox{\tiny $(0)$}}}_\rmi{K}(i\omega_\rmi{E},1)}{\omega_\rmi{E}}\right)^{2}-\left(\frac{1}{1\pm i\,\omega_\rmi{E}}\pm\frac{i\,\widetilde{{\cal T}}^{{}^{\mbox{\tiny $(0)$}}}_\rmi{K}(i\omega_\rmi{E},1)}{\omega_\rmi{E}}\right)^{2} \hspace*{1.45truecm} \\
&&\hspace*{5.25truecm}\times\left\{\log\left(\frac{1}{1\pm i\,\omega_\rmi{E}}\pm\frac{i\,\widetilde{{\cal T}}^{{}^{\mbox{\tiny $(0)$}}}_\rmi{K}(i\omega_\rmi{E},1)}{\omega_\rmi{E}}\right)\mp\frac{2\,i\,\widetilde{{\cal T}}^{{}^{\mbox{\tiny $(1)$}}}_\rmi{K}(i\omega_\rmi{E},1)}{\frac{\omega_\rmi{E}}{1\pm\,i\,\omega_\rmi{E}}\pm\,i\,\widetilde{{\cal T}}^{{}^{\mbox{\tiny $(0)$}}}_\rmi{K}(i\omega_\rmi{E},1)}\right\}\,\epsilon\, \,\nonumber ,
\eeqa}
\hspace{-0.12cm}where the superscripts in $\Pi_\rmi{T,L}^{{}^{\mbox{\tiny $(0)$}}},\,\Pi_\rmi{T,L}^{{}^{\mbox{\tiny $(1)$}}}$ or $\widetilde{{\cal T}}^{{}^{\mbox{\tiny $(0)$}}}_\rmi{K},\,\widetilde{{\cal T}}^{{}^{\mbox{\tiny $(1)$}}}_\rmi{K}$ denote the order of derivative with respect to $\epsilon$, before setting $\epsilon$ to zero at the end. In the last equation, we notice that the sum of the first term with its complex conjugate vanishes, so that there will be no pole coming from the quark contribution.

Accounting for the proper degrees of freedom, one gets after integration over $\omega_\rmi{E}$
{\allowdisplaybreaks
\beqa
(2-2\epsilon)\,p^{\star}_\rmi{T}&=&\frac{m_\rmi{D}^4}{64\,\pi^2}\,\bigg\{\frac{\log 256-5}{6}\,\left(\frac{1}{\epsilon}+\frac{7}{2}-\log 2+\log\frac{\lmsb^2}{m_\rmi{D}^2}\right)+\frac{2\,\kappa_\rmi{T}}{\pi}\bigg\} \label{p_star_T}\, , \\
p^{\star}_\rmi{L}&=&\frac{m_\rmi{D}^4}{64\,\pi^2}\,\bigg\{\frac{4-\log 16}{3}\,\left(\frac{1}{\epsilon}+\frac{5}{2}-2\,\log 2+\log\frac{\lmsb^2}{m_\rmi{D}^2}\right)+\frac{4\,\kappa_\rmi{L}}{\pi}\bigg\} \label{p_star_L}\, , \\
p^{\star}_\rmi{q$_f$}&=&m_\rmi{q$_f$}^4\,\bigg\{\frac{\kappa_\rmi{q}+\kappa_\rmi{q}^{\star}}{4\,\pi^{3}}\bigg\} 
\label{p_star_q} \, ,
\eeqa}
\hspace{-0.12cm}where $\kappa_\rmi{T}$, $\kappa_\rmi{L}$, and $\kappa_\rmi{q}$ are defined in section~\ref{sec:full_one_loop_HTLpt_pressure}. According to eqs.~(\ref{p_star_T})--(\ref{p_star_q}), we see that the counterterm needed to cancel the UV divergences of the pressure is
\beq\label{Counter_term}
\Delta p\equiv -\,d_\rmi{A}\,\frac{m_\rmi{D}^4}{128\,\pi^2\,\epsilon}\;.
\eeq

\subsection{Renormalized one-loop HTLpt pressure}\label{sec:full_one_loop_HTLpt_pressure}

Writing eq.~(\ref{Pre_result_full_HTLpt}) using eqs.~(\ref{p_star_T})--(\ref{Counter_term}), and taking the limit $\epsilon \rightarrow 0$, we obtain the renormalized result for the one-loop HTLpt pressure, which reads
{\allowdisplaybreaks
\beqa\label{Full_HTLpt_pressure2}
p_\rmi{HTLpt}\left(T,\bm{\mu}\right)&=& d_\rmi{A}\Bigg\{\frac{m_\rmi{D}^4}{64\pi^2}\left(\log\frac{\lmsb}{m_\rmi{D}}+C_\rmi{g}\right)+\frac{1}{2\pi^3}\inteddw\frac{1}{e^{\beta\omega}-1}\inteddk k^2\bigg(2\phi_\rmi{T}-\phi_\rmi{L}\bigg) \, \nonumber \\
& & \ \ \ \ -\frac{T}{2\pi^2}\intedk k^2\bigg[2\log\bigg(1-e^{-\beta\omega_\rmi{T}}\bigg)+\log\bigg(1-e^{-\beta\omega_\rmi{L}}\bigg)\bigg]-\frac{\pi^2\,T^4}{90}\Bigg\} \ \ \ \ \ \nonumber \\
&+&N_\rmi{c}\sum_{f,\,s=\pm 1}\Bigg\{\frac{C_\rmi{q}}{2}\ m_\rmi{q$_f$}^4+\frac{2\ T^4}{\pi^2}\mbox{Li}_4\bigg(-e^{s\,\beta\mu_f}\bigg)-\frac{1}{\pi^3}\intdwdk\frac{k^2\,\theta_\rmi{q$_f$}}{e^{\beta\left(\omega+s\,\mu_f\right)}+1} \, \nonumber \\
& & \ \ \ \ \ \ \ +\frac{T}{\pi^2}\intedk k^2\ \bigg[\log\bigg(1+e^{-\beta\left(\omega_{f_+}+s\,\mu_f\right)}\bigg)+\log\bigg(1+e^{-\beta\left(\omega_{f_-}+s\,\mu_f\right)}\bigg)\bigg]\Bigg\} \, , \nonumber \\
\eeqa}
\hspace{-0.12cm}for which the angles $\phi_\rmi{T,L}$ and $\theta_\rmi{q$_f$}$, the dispersion relations $\omega_\rmi{T,L,{$f_\pm$}}$, and the constants $C_\rmi{g}\approx 1.17201$ and $C_\rmi{q}\approx -0.03653$ are listed below. The mass parameters $m_\rmi{D}\left(T,\bm{\mu}\right)$ and $m_\rmi{q$_f$}\left(T,\bm{\mu}\right)$ are given by the prescription in eq.~(\ref{HTLpt_mD_mq_parameters}), and we refer to appendix~\ref{sec:BC_Disc} for details about the $\epsilon \rightarrow 0$ limit on the branch cut discontinuities of the Landau damping angles.

In order to numerically evaluate the exact one-loop HTLpt pressure at finite temperature and quark chemical potentials, one needs the angles $\phi_\rmi{T}$, $\phi_\rmi{L}$ and $\theta_\rmi{q$_f$}$ which are given by the following expressions
{\allowdisplaybreaks
\beqa
\phi_\rmi{T}&=&\arctan\left[\frac{\frac{\pi}{4}m_\rmi{D}^2\frac{\omega}{k^3}(k^2-\omega^2)}{k^2-\omega^2+\frac{m_\rmi{D}^2}{2}\frac{\omega^2}{k^2}\Big[1+\frac{k^2-\omega^2}{2k\omega}\log\left(\frac{k+\omega}{k-\omega}\right)\Big]}\right] \, , \\
\phi_\rmi{L}&=&\arctan\left[\frac{\frac{\pi}{2}m_\rmi{D}^2\frac{\omega}{k}}{k^2+m_\rmi{D}^2\Big[1-\frac{\omega}{2k}\log\left(\frac{k+\omega}{k-\omega}\right)\Big]}\right]\, , \\
\theta_\rmi{q$_f$}&=&\arctan\left[\frac{\frac{\pi m_\rmi{q$_f$}^4}{k^2}\Big[\frac{\omega}{k}+\frac{k^2-\omega^2}{2k^2}\log\left(\frac{k+\omega}{k-\omega}\right)\Big]}{k^2-\omega^2+2m_\rmi{q$_f$}^2+\frac{m_\rmi{q$_f$}^4}{k^2}\bigg[1-\frac{\omega}{k}\log\Big(\frac{k+\omega}{k-\omega}\Big)-\frac{k^2-\omega^2}{4k^2}\bigg[\log\Big(\frac{k+\omega}{k-\omega}\Big)^2-\pi^2\bigg]\bigg]}\right]\,.
\hspace*{2.75em}
\eeqa}
\hspace{-0.12cm}Furthermore, one also needs the dispersion relations for $\omega_\rmi{T}$, $\omega_\rmi{L}$, and $\omega_{f_\pm}$ in $d=3$ dimensions, which are the solutions to the following transcendental equations
{\allowdisplaybreaks
\beqa\label{Dispersion_relations}
\omega_\rmi{T}^2&=&k^2+\frac{1}{2}m_\rmi{D}^2\frac{\omega_\rmi{T}^2}{k^2} \left[1-\frac{\omega_\rmi{T}^2-k^2}{2\,\omega_\rmi{T} k}\log\left(\frac{\omega_\rmi{T}+k}{\omega_\rmi{T}-k}\right)\right] \, , \\
0&=&k^2+m_\rmi{D}^2\left[1-\frac{\omega_\rmi{L}}{2k}\log\left(\frac{\omega_\rmi{L}+k}{\omega_\rmi{L}-k}\right)\right] \, , \\
0&=&\omega_{f_\pm}\mp k-\frac{m_\rmi{q$_f$}^2}{2k}\left[\left(1\mp\frac{\omega_{f_\pm}}{k}\right)\log\left(\frac{\omega_{f_\pm}+k}{\omega_{f_\pm}-k}\right)\pm 2\right] \, .
\eeqa}
\hspace{-0.12cm}Finally, the constants $C_\rmi{g}$ and $C_\rmi{q}$ are given by
\beqa\label{KappaT}
C_\rmi{g}&=& \frac{2\ \kappa_\rmi{T}}{\pi}+\frac{4\ \kappa_\rmi{L}}{\pi}+\frac{1}{12}\bigg[5+\Big(\log 256-3\Big)\log 4\bigg]\ \approx 1.17201 \, , \\
C_\rmi{q}&=& \frac{\kappa_\rmi{q}+\kappa_\rmi{q}^{\star}}{4\pi^3}\approx -0.03653\, ,
\eeqa
\hspace{-0.12cm}where $\kappa_\rmi{T}\approx 0.082875$, $\kappa_\rmi{L}\approx 0.320878$, and $\kappa_\rmi{q}+\kappa_\rmi{q}^{\star}\approx -4.53025$ are defined as
{\allowdisplaybreaks
\beqa
\kappa_\rmi{T}&=& -\inteddwters{0}{\infty}\bigg[\omega_\rmi{E}\Big(\frac{\pi}{2}-\arctan\left(\omega_\rmi{E}\right)\Big)-\frac{\omega_\rmi{E}^2}{1+\omega_\rmi{E}^2}\bigg]^2 \\
& & \ \ \ \ \ \ \ \ \ \ \ \times\left[\frac{2\ \ {}_2F_1^{{}^{\mbox{\tiny $(0,0;1;0)$}}}\left(\frac{1}{2},1;\frac{3}{2};-\frac{1}{\omega_\rmi{E}^2}\right)}{\omega_\rmi{E}\Big(\frac{\pi}{2}-\arctan\left(\omega_\rmi{E}\right)\Big)-\frac{\omega_\rmi{E}^2}{1+\omega_\rmi{E}^2}}+\log\left[\omega_\rmi{E}\Big(\frac{\pi}{2}-\arctan\left(\omega_\rmi{E}\right)\Big)-
\frac{\omega_\rmi{E}^2}{1+\omega_\rmi{E}^2}\right]\right] \, , \nonumber\\
\kappa_\rmi{L}&=& \inteddwters{0}{\infty}\bigg[1+\omega_\rmi{E}\Big(\arctan\left(\omega_\rmi{E}\right)-\frac{\pi}{2}\Big)\bigg]^2 \\
& & \ \ \ \ \ \ \ \ \, \times\left[\frac{2\ \ {}_2F_1^{{}^{\mbox{\tiny $(0,0;1;0)$}}}\left(\frac{1}{2},1;\frac{3}{2};-\frac{1}{\omega_\rmi{E}^2}\right)}{1+\omega_\rmi{E}\Big(\arctan\left(\omega_\rmi{E}\right)-\frac{\pi}{2}\Big)}-\log\left[1+\omega_\rmi{E}\Big(\arctan\left(\omega_\rmi{E}\right)-\frac{\pi}{2}\Big)\right]\right] \, , \nonumber \\
\kappa_\rmi{q}&=& \inteddwters{0}{\infty}\left[i\,\bigg(\arctan\left(\omega_\rmi{E}\right)-\frac{\pi}{2}\bigg)-\frac{1}{1+i\,\omega_\rmi{E}}\right]^2 \\
& & \ \ \ \ \ \ \ \ \ \times\left[\log\left[i\,\bigg(\frac{\pi}{2}-\arctan\left(\omega_\rmi{E}\right)\bigg)+\frac{1}{1+i\,\omega_\rmi{E}}\right]+\frac{2\,i\,\,\,{}_2F_1^{{}^{\mbox{\tiny $(0,0;1;0)$}}}\left(\frac{1}{2},1;\frac{3}{2};-\frac{1}{\omega_\rmi{E}^2}\right)}{\frac{\omega_\rmi{E}}{1+i\,\omega_\rmi{E}}-i\,\omega_\rmi{E}\,\bigg(\arctan\left(\omega_\rmi{E}\right)-\frac{\pi}{2}\bigg)}\right]\;.\nonumber
\eeqa}
\hspace{-0.12cm}Notice that the present derivative of the hypergeometric function has the following real-valued representation
{\allowdisplaybreaks
\beqa
{}_2F_1^{{}^{\mbox{\tiny $(0,0;1;0)$}}}\left(\mbox{\small $\frac{1}{2},1;\frac{3}{2};-\frac{1}{\omega_\rmi{E}^2}$}\right)&=&\omega_\rmi{E}\bigg(\frac{\pi}{2}-\arctan(\omega_\rmi{E})\bigg)\bigg(2-\log 4\bigg)-\omega_\rmi{E}\,\log 2\,\arg\left(\frac{i\omega_\rmi{E}+1}{i\omega_\rmi{E}-1}\right) \nonumber \\
&+& \frac{\omega_\rmi{E}}{2}\left[\ImPart{\Big\{\mbox{Li}_2\left(\frac{2}{1+i\omega_\rmi{E}}\right)\Big\}}-\ImPart{\Big\{\mbox{Li}_2\left(\frac{2}{1-i\omega_\rmi{E}}\right)\Big\}}\right] \, .
\eeqa}

\section{Branch cut discontinuities}\label{sec:BC_Disc}

In this appendix, we give more details on the computations related to the branch cut discontinuities of the Landau damping contributions to the exact one-loop HTLpt pressure, i.e.\,\,related to eqs.~(\ref{General_Tgluon_contribution_after_collapse_Ld}),~(\ref{General_Lgluon_contribution_after_collapse_Ld}), and~(\ref{General_Quark_contribution_after_collapse_Ld}), as well as to eq.~(\ref{expf2}). First of all, we explicitly write the discontinuity of the hypergeometric function which is involved in the relevant $\arctan$ functions. To this end, let us define a notation, namely the superscripts $\oplus/\ominus$, which will denote a function approaching the real axis from above/below, i.e.\,\,for which the complex variable $\omega$ satisfies $\ImPart{\omega}\rightarrow 0^{\pm}$.

Then, provided that ${\rm {Re}}\,(k)>{\rm {Re}}\,(\omega)$, which is the case along the Landau damping cuts, we write the branch cut discontinuity of the hypergeometric function as
\beqa
\disc{}_2F_1\left(\mbox{\small $\frac{1}{2},1;\frac{3}{2}-\epsilon;\frac{k^2}{\omega^2}$}\right)&\equiv&{}_2F_1^{\oplus}\left(\mbox{\small $\frac{1}{2},1;\frac{3}{2}-\epsilon;\frac{k^2}{\omega^2}$}\right)-{}_2F_1^{\ominus}\left(\mbox{\small $\frac{1}{2},1;\frac{3}{2}-\epsilon;\frac{k^2}{\omega^2}$}\right) \, ,
\eeqa
with
{\allowdisplaybreaks
\beqa
{}_2F_1^{\oplus}\left(\mbox{\small $\frac{1}{2},1;\frac{3}{2}-\epsilon;\frac{k^2}{\omega^2}$}\right)&\equiv& {}_2F_1\left(\mbox{\small $\frac{1}{2},1;\frac{3}{2}-\epsilon;\frac{k^2}{\omega^2}$}\right) \, , \label{Disc_Plus} \\
{}_2F_1^{\ominus}\left(\mbox{\small $\frac{1}{2},1;\frac{3}{2}-\epsilon;\frac{k^2}{\omega^2}$}\right)&\equiv& e^{2i\pi\epsilon} \, {}_2F_1\left(\mbox{\small $\frac{1}{2},1;\frac{3}{2}-\epsilon;\frac{k^2}{\omega^2}$}\right) \nonumber \\
&+& \frac{2i\pi \, e^{i\pi\epsilon} \, \Gamma\left(\frac{3}{2}-\epsilon\right)}{\Gamma\left(1-\epsilon\right)\Gamma\left(\frac{1}{2}-\epsilon\right)\Gamma\left(1+\epsilon\right)}\,\,{}_2F_1\left(\mbox{\small $\frac{1}{2},1;1+\epsilon;\frac{\omega^2-k^2}{\omega^2}$}\right) \, . \label{Disc_Minus} 
\eeqa}
We also notice that in the limit $\epsilon\rightarrow 0$, these functions reduce to
{\allowdisplaybreaks
\beqa
{}_2F_1^{\oplus}\left(\mbox{\small $\frac{1}{2},1;\frac{3}{2};\frac{k^2}{\omega^2}$}\right)&=& \frac{\omega}{k}\,\arctanh\left(\frac{k}{\omega}\right)=\frac{\omega}{k}\,\arctanh\left(\frac{\omega}{k}\right)-\frac{i\pi\,\omega}{2k} \, \nonumber \\
&=& \frac{\omega}{2\,k}\,\log\left(\frac{k+\omega}{k-\omega}\right)-\frac{i\pi\,\omega}{2k} \, , \label{Disc_Plus_3d} \\
{}_2F_1^{\ominus}\left(\mbox{\small $\frac{1}{2},1;\frac{3}{2};\frac{k^2}{\omega^2}$}\right)&=& \frac{\omega}{k}\,\arctanh\left(\frac{k}{\omega}\right) +\frac{i\pi\,\omega}{k}=\frac{\omega}{k}\,\arctanh\left(\frac{\omega}{k}\right)+\frac{i\pi\,\omega}{2k} \, \nonumber \\
&=& \frac{\omega}{2\,k}\,\log\left(\frac{k+\omega}{k-\omega}\right)+\frac{i\pi\,\omega}{2k} \, ,\label{Disc_Minus_3d}
\eeqa}
\hspace{-0.12cm}where in the last two equalities for each function, we have made use of the fact that $k>\omega$, making the logarithm real valued.

Now, let us explicitly write down the discontinuities of the Landau damping angles in $d=3-2\epsilon$ dimensions. The discontinuity in eq.~(\ref{General_Tgluon_contribution_after_collapse_Ld}) reads
{\allowdisplaybreaks
\beqa
\disc\phi_{\rmi{T},\epsilon}&\equiv&\disc\arctan\left[\frac{\frac{m_\rmi{D}^2}{2-2\epsilon}\frac{k^2-\omega^2}{k^2} \ \ImPart{\left\{_2F_1\left(\mbox{$\frac{1}{2},1;\frac{3}{2}-\epsilon;\frac{k^2}{\omega^2}$}\right)\right\}}}{k^2-\omega^2+\frac{m_\rmi{D}^2}{2-2\epsilon} \frac{\omega^2}{k^2} \left[ 1 + \frac{k^2-\omega^2}{\omega^2} \ \RePart{\left\{_2F_1\left(\mbox{$\frac{1}{2},1;\frac{3}{2}-\epsilon;\frac{k^2}{\omega^2}$}\right)\right\}}\right]}\right] \, \nonumber \\
&=&\arctan\left[\frac{\frac{m_\rmi{D}^2}{2-2\epsilon}\frac{k^2-\omega^2}{k^2} \ \ImPart{\left\{_2F_1^{\oplus}\left(\mbox{$\frac{1}{2},1;\frac{3}{2}-\epsilon;\frac{k^2}{\omega^2}$}\right)\right\}}}{k^2-\omega^2+\frac{m_\rmi{D}^2}{2-2\epsilon} \frac{\omega^2}{k^2} \left[ 1 + \frac{k^2-\omega^2}{\omega^2} \ \RePart{\left\{_2F_1^{\oplus}\left(\mbox{$\frac{1}{2},1;\frac{3}{2}-\epsilon;\frac{k^2}{\omega^2}$}\right)\right\}}\right]}\right] \, \nonumber \\
&-&\arctan\left[\frac{\frac{m_\rmi{D}^2}{2-2\epsilon}\frac{k^2-\omega^2}{k^2} \ \ImPart{\left\{_2F_1^{\ominus}\left(\mbox{$\frac{1}{2},1;\frac{3}{2}-\epsilon;\frac{k^2}{\omega^2}$}\right)\right\}}}{k^2-\omega^2+\frac{m_\rmi{D}^2}{2-2\epsilon} \frac{\omega^2}{k^2} \left[ 1 + \frac{k^2-\omega^2}{\omega^2} \ \RePart{\left\{_2F_1^{\ominus}\left(\mbox{$\frac{1}{2},1;\frac{3}{2}-\epsilon;\frac{k^2}{\omega^2}$}\right)\right\}}\right]}\right] \, .
\eeqa}
\hspace{-0.12cm}In the same way, one can explicitly write the discontinuity in eq.~(\ref{General_Lgluon_contribution_after_collapse_Ld}) as
{\allowdisplaybreaks
\beqa
\disc\phi_{\rmi{L},\epsilon}&\equiv&\disc\arctan\left[\frac{m_\rmi{D}^2 \ \ImPart{\left\{_2F_1\left(\mbox{$\frac{1}{2},1;\frac{3}{2}-\epsilon;\frac{k^2}{\omega^2}$}\right)\right\}}}{k^2 + m_\rmi{D}^2 - m_\rmi{D}^2 \ \RePart{\left\{_2F_1\left(\mbox{$\frac{1}{2},1;\frac{3}{2}-\epsilon;\frac{k^2}{\omega^2}$}\right)\right\}}}\right] \, \nonumber \\
&=&\arctan\left[\frac{m_\rmi{D}^2 \ \ImPart{\left\{_2F_1^{\oplus}\left(\mbox{$\frac{1}{2},1;\frac{3}{2}-\epsilon;\frac{k^2}{\omega^2}$}\right)\right\}}}{k^2 + m_\rmi{D}^2 - m_\rmi{D}^2 \ \RePart{\left\{_2F_1^{\oplus}\left(\mbox{$\frac{1}{2},1;\frac{3}{2}-\epsilon;\frac{k^2}{\omega^2}$}\right)\right\}}}\right] \, \nonumber \\
&-&\arctan\left[\frac{m_\rmi{D}^2 \ \ImPart{\left\{_2F_1^{\ominus}\left(\mbox{$\frac{1}{2},1;\frac{3}{2}-\epsilon;\frac{k^2}{\omega^2}$}\right)\right\}}}{k^2 + m_\rmi{D}^2 - m_\rmi{D}^2 \ \RePart{\left\{_2F_1^{\ominus}\left(\mbox{$\frac{1}{2},1;\frac{3}{2}-\epsilon;\frac{k^2}{\omega^2}$}\right)\right\}}}\right] \, ,
\eeqa}
\hspace{-0.12cm}and the discontinuity in eq.~(\ref{General_Quark_contribution_after_collapse_Ld}) as
\beqa
\disc\theta_\rmi{q$_f$,$\epsilon$}&\equiv&\disc\arctan\left[\Xi_\rmi{$f$,$\epsilon$}\right]=\arctan\left[\Xi^{\oplus}_\rmi{$f$,$\epsilon$}\right]-\arctan\left[\Xi^{\ominus}_\rmi{$f$,$\epsilon$}\right] \, ,
\eeqa
\hspace{-0.12cm}with $\Xi^{\oplus/\ominus}_\rmi{$f$,$\epsilon$}$ defined by
{\allowdisplaybreaks
\beqa
&&\Xi^{\oplus/\ominus}_\rmi{$f$,$\epsilon$}\equiv \\
&&\frac{\frac{m_{q_{\mbox{\tiny$f$}}}^4}{k^2} \left[2 \ \ImPart{\left\{_2F_1^{\oplus/\ominus}\left(\mbox{$\frac{1}{2},1;\frac{3}{2}-\epsilon;\frac{k^2}{\omega^2}$}\right)\right\}}+\frac{k^2-\omega^2}{\omega^2} \ \ImPart{\left\{_2F_1^{\oplus/\ominus}\left(\mbox{$\frac{1}{2},1;\frac{3}{2}-\epsilon;\frac{k^2}{\omega^2}$}\right)^2\right\}}\right]}{k^2-\omega^2+2 m_{q_{\mbox{\tiny$f$}}}^2+\frac{m_{q_{\mbox{\tiny$f$}}}^4}{k^2}\,\left[1-2\,\RePart{\left\{_2F_1^{\oplus/\ominus}\left(\mbox{$\frac{1}{2},1;\frac{3}{2}-\epsilon;\frac{k^2}{\omega^2}$}\right)\right\}}-\frac{k^2-\omega^2}{\omega^2}\,\RePart{\left\{_2F_1^{\oplus/\ominus}\left(\mbox{$\frac{1}{2},1;\frac{3}{2}-\epsilon;\frac{k^2}{\omega^2}$}\right)^2\right\}}\right]} \, . \nonumber
\eeqa}

Consequently, with the help of eqs.~(\ref{Disc_Plus_3d}) and~(\ref{Disc_Minus_3d}), we see that in the limit of three dimensions, i.e. for $\epsilon\rightarrow 0$, these discontinuities reduce to
{\allowdisplaybreaks
\beqa
\disc\phi_{\rmi{T},\epsilon}&{\stackrel{\epsilon\rightarrow 0}{\longrightarrow}}&-2\,\arctan\left[\frac{\frac{\pi}{4}m_\rmi{D}^2\frac{\omega}{k^3}(k^2-\omega^2)}{k^2-\omega^2+\frac{m_\rmi{D}^2}{2}\frac{\omega^2}{k^2}\Big[1+\frac{k^2-\omega^2}{2k\omega}\log\left(\frac{k+\omega}{k-\omega}\right)\Big]}\right] \, , \\
\disc\phi_{\rmi{T},\epsilon}&{\stackrel{\epsilon\rightarrow 0}{\longrightarrow}}&-2\,\arctan\left[\frac{\frac{\pi}{2}m_\rmi{D}^2\frac{\omega}{k}}{k^2+m_\rmi{D}^2\Big[1-\frac{\omega}{2k}\log\left(\frac{k+\omega}{k-\omega}\right)\Big]}\right] \, , \\
\disc\theta_\rmi{q$_f$,$\epsilon$}&{\stackrel{\epsilon\rightarrow 0}{\longrightarrow}}&-2\, \\
&\times&\arctan\left[\frac{\frac{\pi m_\rmi{q$_f$}^4}{k^2}\Big[\frac{\omega}{k}+\frac{k^2-\omega^2}{2k^2}\log\left(\frac{k+\omega}{k-\omega}\right)\Big]}{k^2-\omega^2+2m_\rmi{q$_f$}^2+\frac{m_\rmi{q$_f$}^4}{k^2}\bigg[1-\frac{\omega}{k}\log\Big(\frac{k+\omega}{k-\omega}\Big)-\frac{k^2-\omega^2}{4k^2}\bigg[\log\Big(\frac{k+\omega}{k-\omega}\Big)^2-\pi^2\bigg]\bigg]}\right] \, . \nonumber
\eeqa}
\hspace{-0.12cm}In addition, since the discontinuity and the imaginary part commute with each other, we can rewrite eq.~(\ref{expf2}) as
\beq
\disc \ImPart{\left\{_2F_1\left(\mbox{$\frac{1}{2},1;\frac{3}{2}-\epsilon;\frac{k^2}{\omega^2}$}\right)\right\}}=\ImPart{\disc\left\{_2F_1\left(\mbox{$\frac{1}{2},1;\frac{3}{2}-\epsilon;\frac{k^2}{\omega^2}$}\right)\right\}} \, .
\eeq
Therefore, using eqs.~(\ref{Disc_Plus}) and~(\ref{Disc_Minus}), it is easy to see that once we expand this expression around $k\rightarrow\infty$, with $\omega$ fixed, we get
{\allowdisplaybreaks
\beqa
\disc \ImPart{\left\{_2F_1\left(\mbox{$\frac{1}{2},1;\frac{3}{2}-\epsilon;\frac{k^2}{\omega^2}$}\right)\right\}} &\underset{\underset{\mbox{\tiny $\omega$ fixed}}{k\longrightarrow \infty}}{=}& -\,4\,\, \frac{\Gamma\left(\frac{3}{2}-\epsilon\right)\,\Gamma\left(\frac{3}{2}\right)}{\Gamma\left(1-\epsilon\right)}\,\frac{\omega}{k}+{\cal O}\left(\frac{\omega^3}{k^3}\right) \; .
\eeqa}

\section{Mass expansion of the one-loop HTLpt pressure}\label{sec:Mass_expansion}

In this appendix, we will expand the one-loop HTLpt pressure in powers of $m_\rmi{D}/T\sim g$ and $m_\rmi{q$_f$}/T\sim g$, and compute the needed sum-integrals.

\subsection{Separation of scales}

At zero quark chemical potentials, there are two momentum scales in the sum-integrals, namely the hard scale $2\pi T$, and the soft scale given by $m_\rmi{D}$ as well as $m_\rmi{q$_f$}$. The hard region encompasses all fermionic momenta $K=((2n+1)\pi T,{\bf k})$ as well as bosonic momenta $K=(2n\pi T,{\bf k})$ with $n\neq0$ and even the $n=0$ mode when $k$ is of order $T$. The soft region on the other hand corresponds to bosonic momenta with $n=0$ and $k$ at most of order $g\,T$. At finite chemical potentials, this picture clearly remains unaltered, as the new hard scale $\mu_f$ only enters explicitly the fermionic momenta, or the bosonic ones via the soft scale $m_\rmi{D}$.

The soft scale contributions to the HTLpt pressure reads
\beqa
p_\rmi{HTLpt}^\rmi{\tiny High-T,(s)}&=&-(2-2\epsilon)\frac{d_\rmi{A}\,T}{2}\int_{\bf k}\log(k^2)-\frac{d_\rmi{A}\,T}{2}\int_{\bf k}\log\Big(k^2+m_\rmi{D}^2\Big) \;,
\label{softp}
\eeqa
where we have used $\Pi_\rmi{T}(0,{\bf k})=0$ and $\Pi_\rmi{L}(0,{\bf k})=m_\rmi{D}^2$. The first integral here vanishes in dimensional regularization, as it is scale free. The second integral is dominated by momenta of order $m_\rmi{D}$, and directly yields the soft contribution to the pressure~\cite{braaten1}.

Next, we move on to the hard scale contributions. Assuming $T$ to be large enough so that the ratios $m_\rmi{D}/T$ and $m_\rmi{q$_f$}/T$ are parametrically small, we straightforwardly expand the encountered sum-integrals in powers of these quantities. This yields
{\allowdisplaybreaks
\beqa
p_\rmi{HTLpt}^\rmi{\tiny High-T,(h)}&=&-(1-\epsilon)\,d_\rmi{A}\,\sumint_{K}\log\left(K^2\right)+2\,N_\rmi{c}\,\sum_f\ \sumint_{\{{K\}}}\log\left(K^2\right) \nonumber \\
&-&\frac{d_\rmi{A}\,m_\rmi{D}^2}{2}\,\sumint_{K}\frac{1}{K^2}+4\,N_\rmi{c}\,\sum_f m_\rmi{q$_f$}^2\,\sumint_{\{K\}}\frac{1}{K^2} \nonumber \\
&+&\frac{d_\rmi{A}\,m_\rmi{D}^4}{8-8\epsilon}\,\sumint_{K}\left[\frac{1}{\left(K^{2}\right)^{2}}-\frac{2}{k^2\,K^2}-(6-4\epsilon)\frac{{\cal T}_\rmi{K}}{\left(k^{2}\right)^{2}}+\frac{2\,{\cal T}_\rmi{K}}{k^2\,K^2}+(3-2\epsilon)\frac{({\cal T}_\rmi{K})^2}{\left(k^{2}\right)^{2}}\right] \nonumber \\
&-&2\,N_\rmi{c}\,\sum_f\,m_\rmi{q$_f$}^4\,\sumint_{\{K\}}\left[\frac{2}{\left(K^{2}\right)^{2}}-\frac{1}{k^2\,K^2}+\frac{2\,\widetilde{{\cal T}}_\rmi{K}}{k^2\,K^2}-\frac{(\widetilde{{\cal T}}_\rmi{K})^2}{k^2\,\left(\widetilde{\omega}_n-i\mu_f\right)^2}\right] \,,
\label{hardp}
\eeqa}
\hspace{-0.12cm}implying that the $m/T$-\,expansion of the one-loop HTLpt pressure is formally given by the sum of eqs.~(\ref{softp}) and (\ref{hardp}) as well as the counterterm $\Delta p$ of eq.~(\ref{Counter_term}), as
\beq
p^\rmi{\tiny High-$T$}_\rmi{HTLpt} \equiv p_\rmi{HTLpt}^\rmi{\tiny High-T,(s)}\,+\,p_\rmi{HTLpt}^\rmi{\tiny High-T,(h)}\,+\,\Delta p\;.
\eeq

\subsection{HTL master sum-integrals}\label{sec:HTLpt_High-T_sum-integrals}

Next, we will provide details of the evaluation of the one-loop sum-integrals that appear in the mass expansion of the pressure above. The most general forms of one-loop HTLpt sum-integrals encountered at any order of the mass expansion are
\beqa
\mathcal{I}^{u,m}_{w,\,l}&\equiv&\sumint_{K}\left[\frac{\left(\omega_n\right)^u\,({\cal T}_\rmi{K})^m}{\left(k^{2}\right)^{w}\,\left(K^{2}\right)^{l}}\right] \label{Master_Sum_Int_B} \, , \\
\widetilde{\mathcal{I}}^{u,m}_{w,\,l}&\equiv&\sumint_{\{K\}}\left[\frac{\left(\widetilde{\omega}_n-i\mu_f\right)^u\,(\widetilde{{\cal T}}_\rmi{K})^m}{\left(k^{2}\right)^{w}\,\left(K^{2}\right)^{l}}\right] \label{Master_Sum_Int_F} \, ,
\eeqa
where the HTL functions ${\cal T}_\rmi{K}$ and $\widetilde{{\cal T}}_\rmi{K}$ are defined in eqs.~(\ref{HTL_function_B}) and~(\ref{HTL_function_F}). 

In the case $m=0$, the sum-integrals simplify significantly, as the HTL functions do not appear in the integrands. They can then be straightforwardly evaluated by first integrating over spatial momenta in $3-2\epsilon$ dimensions, thus expressing the resulting sum in terms of the Riemann Zeta function or its generalized form. Consequently, we obtain
{\allowdisplaybreaks
\beqa
\mathcal{I}^{u,0}_{w,\,l}&=&e^{\gamma_\rmi{\tiny E}\epsilon}\left(\frac{\lmsb}{4\pi T}\right)^{2\epsilon}\left[\frac{\left(2\pi T\right)^{4+u-2(l+w)}}{(2\pi)^{3}}\right]\left[\frac{\Gamma\left(\frac{3}{2}-\epsilon-w\right)\Gamma\left(\epsilon-\frac{3}{2}+w+l\right)\Gamma\left(1-\epsilon\right)}{\Gamma\left(2-2\epsilon\right)\Gamma(l)} \right] \nonumber \\
& & \ \ \ \ \ \ \ \ \ \ \ \ \ \ \ \ \ \ \ \ \ \times\bigg[\big(1+(-1)^u\big)\,\zeta\Big(2\epsilon-3-u+2l+2w\Big)\bigg] \, , \\
\widetilde{\mathcal{I}}^{u,0}_{w,\,l}&=&e^{\gamma_\rmi{\tiny E}\epsilon}\left(\frac{\lmsb}{4\pi T}\right)^{2\epsilon}\left[\frac{\left(2\pi T\right)^{4+u-2(l+w)}}{(2\pi)^{3}}\right]\left[\frac{\Gamma\left(\frac{3}{2}-\epsilon-w\right)\Gamma\left(\epsilon-\frac{3}{2}+w+l\right)\Gamma\left(1-\epsilon\right)}{\Gamma\left(2-2\epsilon\right)\Gamma(l)} \right] \nonumber \\
& & \ \ \ \ \ \ \ \ \ \ \ \ \ \ \ \ \ \ \ \ \ \times\left[\zeta\Big(2\epsilon-3-u+2l+2w\,;\frac{1}{2}-\frac{i\mu_f}{2\pi T}\Big) \right. \nonumber \\
& & \ \ \ \ \ \ \ \ \ \ \ \ \ \ \ \ \ \ \ \ \ \ \ \ \ \ \ \ \ \ \left.+\,(-1)^{u}\,\,\zeta\Big(2\epsilon-3-u+2l+2w\,;\frac{1}{2}+\frac{i\mu_f}{2\pi T}\Big)\right] \,.
\eeqa}
\hspace{-0.12cm}Finally, the specific sum-integrals needed read
{\allowdisplaybreaks
\beqa
\sumint_{K}\log(K^2)&=&-\frac{\pi^2T^4}{45}\left(\frac{\bar{\Lambda}}{4\pi T}\right)^{2\epsilon}\bigg[1+{\cal O}(\epsilon)\bigg]\;,\\
\sumint_{\{K\}}\log(K^2)&=&\frac{7\pi^2}{360}\left(\frac{\bar{\Lambda}}{4\pi T}\right)^{2\epsilon}\left[T^4+\frac{30\mu_f^2T^2}{7\pi^2}+\frac{15\mu_f^4}{7\pi^4}+{\cal O}(\epsilon)\right]\;,\\
{\cal I}^{0,0}_{0,1}&=&\frac{1}{12}\left(\frac{\bar{\Lambda}}{4\pi T}\right)^{2\epsilon}\bigg[T^2+{\cal O}(\epsilon)\bigg]\;,\\
{\cal \widetilde{I}}^{0,0}_{0,1}&=&-\frac{1}{24}\left(\frac{\bar{\Lambda}}{4\pi T}\right)^{2\epsilon}\left[T^2+\frac{3\mu_f^2}{\pi^2}+{\cal O}(\epsilon)\right]\\
{\cal I}^{0,0}_{0,2}&=&\frac{1}{(4\pi)^2}\left(\frac{\bar{\Lambda}}{4\pi T}\right)^{2\epsilon}\left[\frac{1}{\epsilon}+2\gamma_E+{\cal O}(\epsilon)\right]\;,\\
{\cal \widetilde{I}}^{0,0}_{0,2}&=&\frac{1}{(4\pi)^2}\left(\frac{\bar{\Lambda}}{4\pi T}\right)^{2\epsilon}\left[\frac{1}{\epsilon}-\Psi\left(\frac{1}{2}+\frac{i\mu_f}{2\pi T}\right)-\Psi\left(\frac{1}{2}-\frac{i\mu_f}{2\pi T}\right)+{\cal O}(\epsilon)\right]\,,\\
{\cal I}^{0,0}_{1,1}&=&\frac{2}{(4\pi)^2}\left(\frac{\bar{\Lambda}}{4\pi T}\right)^{2\epsilon}\left[\frac{1}{\epsilon}+2\gamma_E+2+{\cal O}(\epsilon)\right]\;,\\
{\cal \widetilde{I}}^{0,0}_{1,1}&=&\frac{2}{(4\pi)^2}\left(\frac{\bar{\Lambda}}{4\pi T}\right)^{2\epsilon}\left[\frac{1}{\epsilon}+2-\Psi\left(\frac{1}{2}+\frac{i\mu_f}{2\pi T}\right)-\Psi\left(\frac{1}{2}-\frac{i\mu_f}{2\pi T}\right)+{\cal O}(\epsilon)\right]\;.\ \ \ \ \ \ \ \ \ 
\eeqa}

For $m\neq0$, the master sum-integrals are harder to evaluate, with the difficulty being the angular average in the HTL functions, given by eqs.~(\ref{HTL_function_B}) and~(\ref{HTL_function_F}). The HTL functions give rise to terms like $1/[(\omega_n^2+k^2)(\omega_n^2+c^2k^2)]$, and in this case one cannot simply rescale the variable $k$. However, one can use a simple decomposition to separate the denominator which is $c$-dependent from the $c$-independent one,
\beq
\frac{1}{(\omega_n^2+k^2)(\omega_n^2+c^2k^2)}=\frac{1}{k^2\,(c^2-1)}\left[\frac{1}{(\omega_n^2+k^2)}-\frac{1}{(\omega_n^2+c^2k^2)}\right]\;,
\label{relrescale}
\eeq
after which  it is easy to rescale each term. Based on this observation, we have developed an iterative and systematic algorithm which for a given $m\neq0$ allows for analytic representations of the solutions to the sum-integrals~(\ref{Master_Sum_Int_B}) and~(\ref{Master_Sum_Int_F}) in terms of sum-integrals with $m=0$, providing that $l$ is a positive integer. 

In order to obtain the pressure to order $g^5$ in the mass expansion, we need the master integrals for $m=1$ and $m=2$, of which we will here consider the $m=1$ case. Here, the difficult term in the master sum-integrals is proportional to $1/[(\omega_n^2+k^2)^l(\omega_n^2+c^2k^2)]$. Using eq.~(\ref{relrescale}) $l$ times, we can write this term as
\beq
\frac{1}{(\omega_n^2+k^2)^l(\omega_n^2+c^2k^2)}=\frac{(1-c^2)^{-l}}{(k^2)^l(\omega_n^2+c^2k^2)}-\sum_{r=1}^l\left[\frac{(1-c^2)^{-r}}{(k^2)^r\,(\omega_n^2+k^2)^{l-r+1}}\right] \, .
\eeq
The master integrals in eqs.~(\ref{Master_Sum_Int_B}) and ~(\ref{Master_Sum_Int_F}) with $m=1$ can then, after a convenient rescaling $k\rightarrow k/c$, be written in the form
\beq
\mathbfcal{I}^{u,1}_{w,\,l}=\mathcal{J}_{w,l}\,\,\mathbfcal{I}^{u+2,0}_{w+l,\,1}-\sum_{r=1}^l\left[\mathcal{J}_{r}\,\,\mathbfcal{I}^{u+2,0}_{w+r,\,l-r+1}\right] \, ,
\eeq
where the symbol $\mathbfcal{I}$ stands either for the bosonic sum-integral $\mathcal{I}$ or the fermionic one $\widetilde{\mathcal{I}}$, and where we have defined
\beqa
\mathcal{J}_{w,l}&\equiv&\frac{\Gamma\left(\frac{3}{2}-\epsilon\right)}{\Gamma\left(\frac{3}{2}\right)\Gamma\left(1-\epsilon\right)}\int^{1}_{0}\kern-0.5em\mathop{{\rm d}\!}\nolimits c \ \frac{c^{2\epsilon-3+2(l+w)}}{(1-c^2)^{l+\epsilon}}=\frac{\Gamma\left(\frac{3}{2}-\epsilon\right)\Gamma\left(1-\epsilon-l\right)\Gamma\left(\epsilon-1+l+w\right)}{\Gamma\left(\frac{1}{2}\right)\,\Gamma\left(w\right)\Gamma\left(1-\epsilon\right)} \, , \ \ \ \ \ \ \ \ \ 
\eeqa
and
\beq
\mathcal{J}_{r}\equiv\frac{\Gamma\left(\frac{3}{2}-\epsilon\right)}{\Gamma\left(\frac{3}{2}\right)\Gamma\left(1-\epsilon\right)}\int^{1}_{0}\kern-0.5em\mathop{{\rm d}\!}\nolimits c \ (1-c^2)^{-(r+\epsilon)}=\frac{\Gamma\left(\frac{3}{2}-\epsilon\right)\Gamma\left(1-\epsilon-r\right)}{\Gamma\left(1-\epsilon\right)\Gamma\left(\frac{3}{2}-\epsilon-r\right)} \, .
\eeq
The specific sum-integrals needed are finally
{\allowdisplaybreaks
\beqa
{\cal I}_{2,0}^{0,1}&=&-\frac{1}{(4\pi)^2}\left(\frac{\bar{\Lambda}}{4\pi T}\right)^{2\epsilon}\left[\frac{1}{\epsilon}+2\gamma_E+\log4+{\cal O}(\epsilon)\right]\;,\\
{\cal \widetilde{I}}_{1,1}^{0,1}&=&\frac{2}{(4\pi)^2}\left(\frac{\bar{\Lambda}}{4\pi T}\right)^{2\epsilon}\left[\log2\left\{\frac{1}{\epsilon}+\log2-\Psi\left(\frac{1}{2}+\frac{i\mu_f}{2\pi T}\right)-\Psi\left(\frac{1}{2}-\frac{i\mu_f}{2\pi T}\right)\right\}\right.\, \nonumber \\
&& \ \ \ \ \ \ \ \ \ \ \ \ \ \ \ \ \ \ \ \ \ \ \ \left.+\frac{\pi^2}{6}+{\cal O}(\epsilon)\right]\;,\\
{\cal I}_{1,1}^{0,1}&=&\frac{2}{(4\pi)^2}\left(\frac{\bar{\Lambda}}{4\pi T}\right)^{2\epsilon}\left[\frac{\log2}{\epsilon}+\frac{\pi^2}{6}+2\gamma_E\log2+\log^22+{\cal O}(\epsilon)\right]\;,\\
{\cal I}_{2,0}^{0,2}&=&-\frac{2}{3(4\pi)^2}\left(\frac{\bar{\Lambda}}{4\pi T}\right)^{2\epsilon}\left[\frac{1+2\log2}{\epsilon}+2\gamma_E\left(1+2\log2\right)-\frac{4}{3}+\frac{22}{3}\log2\right.\, \nonumber \\
&& \ \ \ \ \ \ \ \ \ \ \ \ \ \ \ \ \ \ \ \ \ \ \ \ \ \ +2\log^22+{\cal O}(\epsilon)\bigg]\;,\\
{\cal \widetilde{I}}_{1,0}^{\,-2,2}&=&\frac{4\log2}{(4\pi)^2}\left(\frac{\bar{\Lambda}}{4\pi T}\right)^{2\epsilon}\left[\frac{1}{\epsilon}+\log2-\Psi\left(\frac{1}{2}+\frac{i\mu_f}{2\pi T}\right)-\Psi\left(\frac{1}{2}-\frac{i\mu_f}{2\pi T}\right)+{\cal O}(\epsilon)\right] \; . \ \ \ \ \ \ 
\eeqa}

Finally, we would like to note that unfortunately the above method is not straightforwardly applicable to the case of two-loop HTL sum-integrals. We, however, suspect that the problem may be circumvented by means of Mellin-Barnes transformations~\cite{Sylvain2}.

%%%%%%%%%%%%%%%%%%%%%%%%%%%%%%%%%%%%%%%%%%%%%%%%%%%%%%%%%%%%%%%%%%%%%%%%%%%%%%%%%%%%%%%%%%%%%%%%%%%%%%%%%%%%%%%%%%%%%%%%%%%%%%%%%%%%%%%%%%%%%%%%%%%%%%
%%%%%%%%%%%%%%%%%%%%%%%%%%%%%%%%%%%%%%%%%%%%%%%%%%%%%%%%%%%%%%%%%%%%%%%%%%%%%%%%%%%%%%%%%%%%%%%%%%%%%%%%%%%%%%%%%%%%%%%%%%%%%%%%%%%%%%%%%%%%%%%%%%%%%%

%%% BIBLIOGRAPHY

%%% END OF THE DOCUMENT

\begin{thebibliography}{99}

\bibitem{Tannenbaum}
M.~J.~Tannenbaum,
\emph{Highlights from BNL-RHIC},
[arXiv:1201.5900].

\bibitem{Muller}
B.~M\"{u}ller, J.~Schukraft and B.~Wyslouch,
\emph{First Results from Pb+Pb collisions at the LHC},
\emph{Ann.\ Rev.\ Nucl.\ Part.\ Sci.}  {\bf 62} (2012) 361 [arXiv:1202.3233].

\bibitem{Satz}
H.~Satz,
\emph{The Quark-Gluon Plasma: A Short Introduction},
\emph{Nucl.\ Phys.\ A} {\bf 862-863} (2011) 4 [arXiv:1101.3937].

\bibitem{bazavov12}
A.~Bazavov et al.~(HotQCD Collaboration), 
\emph{Fluctuations and Correlations of net baryon number, electric charge, and strangeness: A comparison of lattice QCD results with the hadron resonance gas model},
\emph{Phys.\ Rev.\ D} {\bf 86} (2012) 034509 [arXiv:1203.0784].

\bibitem{bieleHighT}
A.~Bazavov, H.~-T.~Ding, P.~Hegde, F.~Karsch, C.~Miao, S.~Mukherjee, P.~Petreczky and C.~Schmidt et al.,
\emph{Quark number susceptibilities at high temperatures},
[arXiv:1309.2317].

\bibitem{biele1}
C.~Schmidt, 
\emph{QCD bulk thermodynamics and conserved charge fluctuations with HISQ fermions},
\emph{J.\ Phys.\ Conf.\ Ser.} {\bf 432} (2013) 012013v [arXiv:1212.4283].

\bibitem{biele2}
C.~Schmidt,  
\emph{Baryon number and charge fluctuations from lattice QCD},
\emph{Nucl.\ Phys.\ A} {\bf 904-905} (2013) 865c [arXiv:1212.4278].

\bibitem{wuppertalcharge}
S.~Bors\'{a}nyi, Z.~Fodor, S.~D.~Katz, S.~Krieg, C.~Ratti, and K.~Szab\'{o},
\emph{Fluctuations of conserved charges at finite temperature from lattice QCD},
\emph{JHEP} {\bf 01} (2012) 138 [arXiv:1112.4416].

\bibitem{chi4}
S.~Bors\'{a}nyi, 
\emph{Thermodynamics of the QCD transition from lattice},
\emph{Nucl.\ Phys.\ A} {\bf 904-905} (2013) 270c [arXiv:1210.6901].

\bibitem{HandsNc2_1}
S.~Hands, P.~Kenny, S.~Kim and J.~-I.~Skullerud,
\emph{Lattice Study of Dense Matter with Two Colors and Four Flavors},
\emph{Eur.\ Phys.\ J.\ A} {\bf 47} (2011) 60 [arXiv:1101.4961].

\bibitem{HandsNc2_2}
S.~Cotter, P.~Giudice, S.~Hands and J.~-I.~Skullerud,
\emph{Towards the phase diagram of dense two-color matter},
\emph{Phys.\ Rev.\ D} {\bf 87} (2013) 034507 [arXiv:1210.4496].

\bibitem{aleksi1}
A.~Vuorinen,
\emph{The Pressure of QCD at finite temperatures and chemical potentials},
\emph{Phys.\ Rev.\ D} {\bf 68} (2003) 054017 [hep-ph/0305183].

\bibitem{aleksi2}
A.~Vuorinen,
\emph{Quark number susceptibilities of hot QCD up to $g^6\,\ln\,g$},
\emph{Phys.\ Rev.\ D} {\bf 67} (2003) 074032 [hep-ph/0212283].

\bibitem{aleksi3}
A.~Vuorinen,
\emph{The Pressure of QCD at finite temperatures and chemical potentials},
\emph{Phys.\ Rev.\ D} {\bf 68} (2003) 054017 [hep-ph/0305183].

\bibitem{ipp3}
A.~Ipp, K.~Kajantie, A.~Rebhan and A.~Vuorinen,
\emph{The Pressure of deconfined QCD for all temperatures and quark chemical potentials},
\emph{Phys.\ Rev.\ D} {\bf 74} (2006) 045016 [hep-ph/0604060].

\bibitem{blaizot1}
J.~-P.~Blaizot, E.~Iancu , and A.~Rebhan,
\emph{Quark number susceptibilities from HTL resummed thermodynamics},
\emph{Phys.\ Lett.\ B} {\bf 523} (2001) 143 [hep-ph/0110369].

\bibitem{blaizot2}
J.~-P.~Blaizot, E.~Iancu, and A.~Rebhan,
\emph{Comparing different hard thermal loop approaches to quark number susceptibilities}, 
\emph{Eur.\ Phys.\ J.\ C} {\bf 27} (2003) 433 [hep-ph/0206280].

\bibitem{mustafa1}
P.~Chakraborty, M.~G.~Mustafa, and M.~H.~Thoma,
\emph{Quark number susceptibility in hard thermal loop approximation},
\emph{Eur.\ Phys.\ J.\ C} {\bf 23} (2002) 591 [hep-ph/0111022].

\bibitem{mustafa2}
P.~Chakraborty, M.~G.~Mustafa, and M.~H.~Thoma,
\emph{Quark number susceptibility, thermodynamic sum rule, and hard thermal loop approximation},
\emph{Phys.\ Rev.\ D} {\bf 68} (2003) 085012 [hep-ph/0303009].

\bibitem{mustafa3}
N.~Haque, M.~G.~Mustafa, and M.~H.~Thoma,
\emph{Conserved Density Fluctuation and Temporal Correlation Function in HTL Perturbation Theory},
\emph{Phys.\ Rev.\ D} {\bf 84} (2011) 054009 [arXiv:1103.3394].

\bibitem{baierredlich}
R.~Baier and K.~Redlich,
\emph{Hard thermal loop resummed pressure of a degenerate quark gluon plasma},
\emph{Phys. Rev. Lett.} {\bf 84} 2100 (2000) [hep-ph/9908372].

\bibitem{JensMike1}
J.~O.~Andersen and M.~Strickland,
\emph{The Equation of state for dense QCD and quark stars},
\emph{Phys.\ Rev.\ D} {\bf 66} (2002) 105001 [hep-ph/0206196].

\bibitem{sylvain1}
J.~O.~Andersen, S.~Mogliacci, N.~Su, and A.~Vuorinen,
\emph{Quark number susceptibilities from resummed perturbation theory},
\emph{Phys.\ Rev.\ D} {\bf 87} (2013) 074003 [arXiv:1210.0912].

\bibitem{mikehaque}
N.~Haque, M.~G.~Mustafa, and M.~Strickland,
\emph{Quark Number Susceptibilities from Two-Loop Hard Thermal Loop 
Perturbation Theory},
\emph{JHEP} {\bf 07} (2013) 184 [arXiv:1302.3228].

\bibitem{Haque1}
N.~Haque, M.~G.~Mustafa and M.~Strickland,
\emph{Two-loop HTL pressure at finite temperature and chemical potential}, 
\emph{Phys.\ Rev.\ D} {\bf 87} (2013) 105007 [arXiv:1212.1797].

\bibitem{haque}
N.~Haque, J.~O.~Andersen, M.~G.~Mustafa, M.~Strickland, and N.~Su,
\emph{Three-loop HTLpt Pressure and Susceptibilities at Finite Temperature and Density},
[arXiv:1309.3968].

\bibitem{ipp1}
A.~Ipp and A.~K.~Rebhan,
\emph{Thermodynamics of large $N_f$ QCD at finite chemical potential},
\emph{JHEP} {\bf 06} (2003) 032 [hep-ph/0305030].

\bibitem{ipp2}
A.~Ipp, A.~K.~Rebhan, and A.~Vuorinen,
\emph{Perturbative QCD at nonzero chemical potential: Comparison with the large $N_f$ limit and apparent convergence},
\emph{Phys.\ Rev.\ D} {\bf 69} (2004) 077901 [hep-ph/0311200].

\bibitem{ads}
J.~Casalderrey-Solana and D.~Mateos,
\emph{Off-diagonal Flavour Susceptibilities from AdS/CFT},
\emph{JHEP} {\bf 08} (2012) 165 [arXiv:1202.2533].

\bibitem{PNJL1}
A.~Bhattacharyya, P.~Deb, A.~Lahiri and R.~Ray,
\emph{Susceptibilities with multi-quark interactions in PNJL model},
\emph{Phys.\ Rev.\ D} {\bf 82} (2010) 114028 [arXiv:1008.0768].

\bibitem{ModelForQCD2}
D.~-k.~He, X.~-x.~Ruan, Y.~Jiang, W.~-M.~Sun and H.~-S.~Zong,
\emph{A model study of quark-number susceptibility at finite chemical potential and temperature},
\emph{Phys.\ Lett.\ B} {\bf 680} (2009) 432.

\bibitem{PNJL3}
C.~Ratti, S.~Roessner and W.~Weise,
\emph{Quark number susceptibilities: Lattice QCD versus PNJL model},
\emph{Phys.\ Lett.\ B} {\bf 649} (2007) 57 [hep-ph/0701091].

\bibitem{Jinfeng}
S.~Shi and J.~Liao,
\emph{Conserved Charge Fluctuations and Susceptibilities in Strongly Interacting Matter},
\emph{JHEP} {\bf 1306} (2013) 104 [arXiv:1304.7752].

\bibitem{motspt}
J.~O Andersen and M.~Strickland, 
\emph{Mass expansions of screened perturbation theory},
\emph{Phys.\ Rev.\ D} {\bf 64} (2001) 105012 [hep-ph/0105214].

\bibitem{Seipt1} 
D.~Seipt, M.~Bluhm and B.~Kampfer,
\emph{Quark mass dependence of thermal excitations in QCD in one-loop approximation},
\emph{J.\ Phys.\ G} {\bf 36} 045003 (2009) [arXiv:0810.3803].
  
\bibitem{blaizot3}
J.~-P.~Blaizot, E.~Iancu and A.~Rebhan,
\emph{Thermodynamics of the high temperature quark gluon plasma},
\emph{In Hwa, R.C. (ed.) et al.: Quark gluon plasma}, 60-122 [hep-ph/0303185].

\bibitem{kraemmer1}
U.~Kraemmer and A.~Rebhan,
\emph{Advances in perturbative thermal field theory},
\emph{Rept.\ Prog.\ Phys.} {\bf 67} (2004) 351 [hep-ph/0310337].

\bibitem{JensMike2}
J.~O.~Andersen and M.~Strickland,
\emph{Resummation in hot field theories},
\emph{Annals Phys.} {\bf 317} (2005) 281 [hep-ph/0404164].

\bibitem{Jens2}
J.~O.~Andersen, E.~Braaten and M.~Strickland,
\emph{Hard thermal loop resummation of the thermodynamics of a hot gluon plasma},
\emph{Phys.\ Rev.\ D} {\bf 61} (2000) 014017 [hep-ph/9905337].
 
\bibitem{Jens3}
J.~O.~Andersen, E.~Braaten and M.~Strickland,
\emph{Hard thermal loop resummation of the free energy of a hot quark-gluon plasma}, 
\emph{Phys.\ Rev.\ D} {\bf 61} (2000) 074016 [hep-ph/9908323].

\bibitem{Kajantie1}
K.~Kajantie, M.~Laine, K.~Rummukainen and M.~E.~Shaposhnikov,
\emph{Generic rules for high temperature dimensional reduction and their application to the standard model},
\emph{Nucl.\ Phys.\ B} {\bf 458} (1996) 90 [hep-ph/9508379].

\bibitem{braaten2}
E.~Braaten and A.~Nieto,
\emph{Effective field theory approach to high temperature thermodynamics},
\emph{Phys.\ Rev.\ D} {\bf 51} (1995) 6990 [hep-ph/9501375].

\bibitem{Jens1}
J.~O.~Andersen, L.~E.~Leganger, M.~Strickland, and N.~Su,
\emph{Three-loop HTL QCD thermodynamics},
\emph{JHEP} {\bf 08} (2011) 053 [arXiv:1103.2528].
 
\bibitem{aleksi4}
A.~Vuorinen and L.~G.~Yaffe,
\emph{Z(3)-symmetric effective theory for SU(3) Yang-Mills theory at high temperature},
\emph{Phys.\ Rev.\ D} {\bf 74} (2006) 025011 [hep-ph/0604100].

\bibitem{deforcrand1}
P.~de~Forcrand, A.~Kurkela and A.~Vuorinen,
\emph{Center-Symmetric Effective Theory for High-Temperature SU(2) Yang-Mills Theory},
\emph{Phys.\ Rev.\ D} {\bf 77} (2008) 125014 [arXiv:0801.1566].

\bibitem{zhang1}
T.~Zhang, T.~Brauner, A.~Kurkela and A.~Vuorinen,
\emph{Two-color QCD via dimensional reduction},
\emph{JHEP} {\bf 02} (2012) 139 [arXiv:1112.2983].

\bibitem{kajantie3}
K.~Kajantie, M.~Laine, K.~Rummukainen and Y.~Schr\"{o}der,
\emph{The Pressure of hot QCD up to $g^6 \ln(1/g)$},
\emph{Phys.\ Rev.\ D} {\bf 67} (2003) 105008 [hep-ph/0211321].

\bibitem{direnzo1}
F.~Di Renzo, M.~Laine, V.~Miccio, Y.~Schr\"{o}der and C.~Torrero,
\emph{The Leading non-perturbative coefficient in the weak-coupling expansion of hot QCD pressure},
\emph{JHEP} {\bf 07} (2006) 026 [hep-ph/0605042].

\bibitem{gynther1}
A.~Gynther, A.~Kurkela and A.~Vuorinen,
\emph{The $N_f^3$ $g^6$ term in the pressure of hot QCD},
\emph{Phys.\ Rev.\ D} {\bf 80} (2009) 096002 [arXiv:0909.3521].

\bibitem{York}
Y.~Schr\"{o}der,
\emph{A fresh look on three-loop sum-integrals},
\emph{JHEP} {\bf 1208} (2012) 095 [arXiv:1207.5666].

\bibitem{mikkoyork}
M.~Laine and Y.~Schr\"{o}der,
\emph{Quark mass thresholds in QCD thermodynamics},
\emph{Phys.\ Rev.\ D} {\bf 73} (2006) 085009 [hep-ph/0603048].

\bibitem{Hart}
A.~Hart, M.~Laine and O.~Philipsen,
\emph{Static correlation lengths in QCD at high temperatures and finite densities},
\emph{Nucl.\ Phys.\ B} {\bf 586} (2000) 443 [hep-ph/0004060].

\bibitem{Moeller}
J.~M\"{o}ller and Y.~Schr\"{o}der,
\emph{Three-loop matching coefficients for hot QCD: Reduction and gauge independence},
\emph{JHEP} {\bf 1208} (2012) 025 [arXiv:1207.1309].

\bibitem{blaizot4}
J.~-P.~Blaizot, E.~Iancu and A.~Rebhan,
\emph{On the apparent convergence of perturbative QCD at high temperature},
\emph{Phys.\ Rev.\ D} {\bf 68} (2003) 025011 [hep-ph/0303045].

\bibitem{htlpt2loop}
J.~O.~Andersen, E.~Braaten, E.~Petitgirard, and M.~Strickland, 
\emph{HTL perturbation theory to two loops},
\emph{Phys.\ Rev.\ D} {\bf 66} (2002) 085016 [hep-ph/0205085].

\bibitem{Borsanyi_delta_P}
S.~Bors\'{a}nyi, G.~Endr\H{o}di, Z.~Fodor, S.~D.~Katz, S.~Krieg, C.~Ratti and K.~K.~Szab\'{o},
\emph{QCD equation of state at nonzero chemical potential: continuum results with physical quark masses at order $\mu^2$},
\emph{JHEP} {\bf 08} (2012) 053 [arXiv:1204.6710]; the error bars of the data
are from private communication.

\bibitem{handsu6u2}
C.~R.~Allton, M.~Doring, S.~Ejiri, S.~J.~Hands, O.~Kaczmarek, F.~Karsch, E.~Laermann and K.~Redlich,
\emph{Thermodynamics of two flavor QCD to sixth order in quark chemical potential},
\emph{Phys.\ Rev.\ D} {\bf 71} (2005) 054508 [hep-lat/0501030].

\bibitem{gupta}
R.~V.~Gavai, S.~Gupta and P.~Majumdar,
\emph{Susceptibilities and screening masses in two flavor QCD},
\emph{Phys.\ Rev.\ D} {\bf 65} (2002) 054506 [hep-lat/0110032].

\bibitem{kajantie4}
K.~Kajantie, M.~Laine, K.~Rummukainen, M.~E.~Shaposhnikov,
\emph{3-D SU(N) + adjoint Higgs theory and finite temperature QCD},
\emph{Nucl.\ Phys.\ B} {\bf 503} (1997) 357 [hep-ph/9704416].

\bibitem{runningalpha}
A.~Bazavov, N.~Brambilla, X.~Garcia, P.~Petreczky, J.~Soto, and A.~Vairo,
\emph{Determination of $\alpha_s$ from the QCD static energy},
\emph{Phys.\ Rev.\ D} {\bf 86} (2012) 11403 [arXiv:1205.6155].

\bibitem{Bazavov13}
A.~Bazavov, H.~-T.~Ding, P.~Hegde, O.~Kaczmarek, F.~Karsch, E.~Laermann, Y.~Maezawa and S.~Mukherjee et al.,
\emph{Strangeness at High Temperatures: From Hadrons to Quarks},
\emph{Phys.\ Rev.\ Lett.} {\bf 111} (2013) 082301 [arXiv:1304.7220].

\bibitem{htlptforth}
N.~Haque, J.~O.~Andersen, M.~G.~Mustafa, M.~Strickland, and N.~Su, in preparation.

\bibitem{Peikert}
F.~Karsch, E.~Laermann and A.~Peikert,
\emph{The Pressure in two flavor, (2+1)-flavor and three flavor QCD},
\emph{Phys.\ Lett.\ B} {\bf 478} (2000) 447 [hep-lat/0002003].

\bibitem{Peikert2}
A.~Peikert,
PhD dissertation: \emph{QCD thermodynamics with $2+1$ quark flavours in lattice simulations},
Bielefeld, May 2000.

\bibitem{AbsoluteScale}
F.~Karsch, E.~Laermann and A.~Peikert,
\emph{Quark mass and flavor dependence of the QCD phase transition},
\emph{Nucl.\ Phys.\ B} {\bf 605} (2001) 579 [hep-lat/0012023].

\bibitem{mathfile}
\href{http://www.physik.uni-bielefeld.de/~vuorinen/DREoS.nb}{www.physik.uni-bielefeld.de/$\sim$vuorinen/DREoS.nb}

\bibitem{braaten1}
E.~Braaten and E.~Petitgirard,
\emph{Solution to the three loop $\Phi$-derivable approximation for massless scalar thermodynamics}, 
\emph{Phys.\ Rev.\ D} {\bf 65} 085039 (2002) [hep-ph/0107118].

\bibitem{Sylvain2}
I.~Kondrashuk, S.~Mogliacci, and Y.~Schr\"{o}der,
in preparation.

\end{thebibliography}
\end{document}